\documentclass[letterpaper]{article}

\usepackage[T1]{fontenc}

\usepackage{geometry}
\usepackage{setspace}

\usepackage[style=chem-acs, articletitle=true]{biblatex}
\usepackage{graphicx}
\graphicspath{{figures/}} % Tell LaTeX to look for images in the figures/ folder
\usepackage{float}
\newfloat{scheme}{htbp}{los}
\floatname{scheme}{Scheme}
\floatname{chart}{Chart}
\newfloat{graph}{htbp}{loh}

\usepackage[version = 4]{mhchem} % Formulas using \ce{}
\usepackage{bm}% bold math
\usepackage{enumerate}
\usepackage{booktabs}
\usepackage{tabularx}
\usepackage{subcaption}
\usepackage{makecell}
\usepackage{threeparttable}
\usepackage{multirow}
\usepackage{svg, epstopdf}
\usepackage{float}

\usepackage[page]{appendix}
\usepackage{hyperref}% add hypertext capabilities
\hypersetup{
  pdftitle={Energy eigenstates of electrons, magnons and phonons in Fe3O4 (magnetite), MnFe2O4 (Jacobsite), and mixed Mn-Zn ferrite},
  pdfauthor={Deepak Dhariwal; Michael von Spakovsky; William T. Reynolds Jr.},
  colorlinks=true,
  citecolor=blue
}
\DeclareMathOperator{\Tr}{{Tr}}

\usepackage{authblk}
\author[1]{Deepak Dhariwal*}
\affil[1]{\small Department of Materials Science and Engineering, Virginia Tech, Blacksburg, Virginia, USA 24060}
\author[2]{Michael R. von Spakovsky}
\affil[2]{\small Department of Mechanical Engineering, Virginia Tech, Blacksburg, Virginia, USA 24060}
\author[1]{William T. Reynolds Jr.}

\title{Energy eigenstates of electrons, magnons, and phonons in \ce{Fe3O4}~(magnetite), \ce{MnFe2O4}~(jacobsite), and mixed \ce{Mn-Zn} ferrite}
\date{\normalfont *Email: deepak20@vt.edu}

\begin{document}

\maketitle

\begin{abstract}
We report first-principles calculations of the electronic structure, magnon excitations, and phonons in magnetite (\ce{Fe3O4}), jacobsite (\ce{MnFe2O4}), and mixed manganese--zinc ferrites \ce{(Mn_{x},Zn_{1-x})Fe2O4} for representative compositions ($0\le x \le 1$) and A/B-site cation arrangements. Electronic structures are computed using density functional theory (DFT) augmented by rotationally invariant DFT+$U$+$J$, with on-site Hubbard and Hund's parameters, $U$ and $J$, respectively, determined self-consistently by spin-polarized linear-response perturbations of the chosen correlated subspaces (including, where applied, the ligand $2p$ subspace). A classical Heisenberg spin Hamiltonian is parameterized by mapping DFT+$U$+$J$ total energies for multiple collinear spin configurations onto nearest-neighbor exchange couplings, which are then used to obtain magnon dispersions and magnon densities of states within linear spin-wave theory. Phonon spectra and densities of states are obtained from finite-displacement force constants and dynamical matrices computed on the same DFT+$U$+$J$-relaxed structures. Overall, the workflow provides a consistent, composition- and configuration-aware route to electronic, vibrational, and magnetic excitation spectra across the Mn/Zn ferrite space.
\end{abstract}

\section*{Keywords}

spinel ferrites; DFT+$U$+$J$ relaxed structures; Hubbard parameter, Hund's exchange parameter; linear spin wave theory (LSWT); density of states

\section{Introduction}\label{sec:intro}

Ferrites or ferrimagnetic oxides (\ce{MeO}\,$\cdot$\,\ce{Fe2O3}, Me = Mn, Fe, Co, Ni, Zn, etc.) underpin many high-frequency magnetic components such as inductors, transformers, and coupled electronic devices because their large resistivity, tunable ferrimagnetism, and chemical robustness mitigate conduction losses while sustaining useful permeability at higher switching frequencies and elevated operating temperatures ~\cite{harrisModernFerritesVolume2023, somiyaHandbookAdvancedCeramics2013}.  Within this class, spinel ferrites are especially attractive since they accommodate mixed-valence transition metals on crystallographically distinct tetrahedral (A sites) and octahedral (B sites) sublattices, allowing composition and cation distribution to be engineered for specific operating windows. For comparison purposes, the focus in this study is on a set of different chemistries, namely, \ce{Fe3O4} (magnetite), \ce{MnFe2O4} (Mn ferrite or jacobsite), and \ce{(Mn_{x}, Zn_{1-x})Fe2O4} for $x=0.5$ (mixed Mn--Zn ferrite). These are chosen because they share the spinel motif with magnetite yet differ systematically in their cation chemistry and magnetic behavior (e.g., non-magnetic \ce{Zn^{2+}} on A sites versus \ce{Mn^{2+/3+}} or \ce{Fe^{2+/3+}} on A/B sites) ~\cite{harrisModernMicrowaveFerrites2012}.

\ce{Fe3O4} serves as the archetypal ferrite, providing a well-characterized reference point for electronic structure and magnetic order. Stoichiometric \ce{Fe3O4} has a Curie temperature of 858 K (585 $^\circ$C) and saturation magnetization (at 20 $^\circ$C) of 480 kA/m~\cite{cullityIntroductionMagneticMaterials2008, smitFerritesPhysicalProperties1959}. It is both an n-type and a p-type semiconductor. The electrical conductivity of magnetite between $10^2 - 10^3\, \Omega^{-1} {\rm cm}^{-1}$ is almost metallic \cite{zieseMagnetoresistanceGrainBoundaries2002,cornellIronOxidesStructure2003}. Building on this, \ce{MnFe2O4} in its bulk or near‐bulk stoichiometric form, has a Curie temperature in the range of 570-600 K (300-400 $^\circ$C), saturation magnetization in some cases of 450-550 kA/m)~\cite{cullityIntroductionMagneticMaterials2008} though nanoparticles and non-stoichiometric variants tend to reduce these values~\cite{balajiMagneticPropertiesMnFe2O42002}. The \ce{MnFe2O4} is far better insulator than magnetite with its electrical  conductivity between $10^{-2}-10^{-4}\, \Omega^{-1} {\rm cm}^{-1}$ depending on the processing route \cite{hakamiStructuralDielectricMagnetic2024}. At the technological frontier, many Mn--Zn mixed ferrites show an initial relative permeability ($\mu_i$) between 5,000 and 15,000 at frequencies between 10-100 kHz and have low core losses. However, this type of performance is highly sensitive to composition, grain size, resistivity, and fabrication method~\cite{JFEFerrite2022, heSoftMagneticMaterials2023, thakurReviewMnZnFerrites2020}.

The energy losses in an inductor under an externally applied alternating electromagnetic field arise from transport phenomena.  The time-alternating field drives the inductor material out of equilibrium and induces flows of charge, magnetization, and thermal energy. These flows dissipate energy in ways that depend upon frequency and give rise to phenomena such as ferromagnetic resonance~\cite{srivastavaAngleDependenceFerromagnetic1999} and temperature-dependent permeability~\cite{sunTemperatureFrequencyCharacteristics2011}. Since each type of flow dissipates energy, the overall energy dissipation is coupled. To capture this, the electron, phonon, and magnon structures presented here will be utilized (in a subsequent contribution) by with the steepest-entropy-ascent quantum thermodynamic (SEAQT) formalism~\cite{berettaSteepestEntropyAscent2014, liGeneralizedThermodynamicRelations2016, liSteepestentropyascentQuantumThermodynamic2016, liSteepestentropyascentModelMesoscopic2018, liSteepestEntropyAscent2018, wordenPredictingCoupledElectron2024} to model the non-equilibrium transport that gives rise to hysteresis loss over an electromagnetic cycle. The SEAQT formalism employs an equation of motion for a system defined by a Hamiltonian operator that includes the electron, phonon, and magnon structures (i.e., so called energy eigenstructures) of the given material. The system state is represented by the density operator, and the dissipation operator of the equation of motion is derived using the steepest-entropy-ascent principle. Dissipative losses arise from entropy generation, which is manifested in the SEAQT framework by energy redistribution among the available energy eigenstates of the material. For a magnetic ferrite, the eigenstates are described by its electron, phonon, and magnon density of states (DOS), which constitute the energy eigenstructure.

The central question of the present contribution is how does the configuration of cations, i.e., which species occupy A versus B sites and in what proportions, modify the electronic, vibrational, and spin-wave energy eigenvalues and degeneracies (i.e., DOS) of a given material? Empirical proxies such as tabulated material conductivities, permeability, and structural parameters~\cite{thakurReviewMnZnFerrites2020, smitFerritesPhysicalProperties1959} are often insufficiently transferable across compositions and A/B site arrangements because: (i) mixed valence and exchange interactions change with cationic distribution, altering the carrier density and magnetic coupling~\cite{santos-carballalFirstprinciplesStudyInversion2015}; (ii) oxygen sub-lattice relaxations subtly change metal-oxygen bond lengths and angles, which modify 3{\emph d}--2{\emph p} orbital overlap, super-exchange strength, and, thus, the hybridization pattern that shapes the DOS ~\cite{naveasFirstPrinciplesCalculationsMagnetite2023a}; and (iii) alloying with Mn and Zn not only varies the number of available carriers but also changes their orbital character and the spectrum of excitations~\cite{sakuraiCationDistributionValence2008}. Consequently, reliable loss predictions for a given material chemistry and cation arrangement require microscopic, composition-aware spectra rather than generic property values. This motivates the use of computational quantum chemistry methods such as DFT and allied post-DFT calculations to compute the DOS of electrons, phonons, and magnons for the relevant material systems.

Concretely, mutually self-consistent DOS sets for \ce{Fe3O4}, \ce{MnFe2O4}, and mixed \ce{(Mn_{x}, Zn_{1-x})Fe2O4} for $x=0.5$ are constructed and compared. ``Mutually self-consistent'' here indicates that for each material system: (1) the electronic structure and DOS are solved self-consistently with the same exchange-correlation family of functionals, compatible on-site electron-electron correlation strategy, $k$-point resolution, and convergence thresholds; (2) the magnon DOS (spin-excitations) are derived from Heisenberg exchange parameters mapped from the same electronic ground states and analyzed within a common spin-Hamiltonian framework; and (3) the phonon DOS is obtained from the corresponding electronic structures using a single, consistent force-constant workflow.

Despite its central role in modern electronic-structure modeling~\cite{beckePerspectiveFiftyYears2014}, approximate-DFT can struggle to describe the electronic structure and magnetic properties of transition-metal oxides such as iron oxides~\cite{mengWhenDensityFunctional2016}. This difficulty is closely linked to the localized nature of the Fe $3d$ states and the substantial self-interaction error and delocalization-based errors of commonly used exchange-correlation functionals, including the local density approximation (LDA) and the generalized gradient approximation (GGA)~\cite{pavariniCorrelatedElectronsModels2012}. More fundamentally, these errors reflect the challenge of capturing exchanges and correlations accurately using only density-based approximations and not explicitly solving the many-body Schr\"odinger equation.

To address this, DFT+U has over the past couple of decades become a widely adopted compromise between accuracy and computational cost by augmenting conventional functionals with an on-site correction that better represents Coulomb interactions among localized electrons ~\cite{anisimovFirstprinciplesCalculationsElectronic1997}. More recently, several extensions and refinements of DFT+U have been developed with the goal of further correcting for static correlation effects and delocalization errors~\cite{bajajMolecularDFT+UTransferable2021, linscottRoleSpinCalculation2018}. Within this broader context, the Hubbard model remains a useful conceptual framework for rationalizing the physics of correlated transition metal compounds~\cite{streltsovOrbitalPhysicsTransition2017}, while a growing body of work has emphasized that Hund's exchange parameter, J, can be essential for capturing phenomena such as Jahn-Teller distortions, emergent intra-atomic exchange, and Kondo-like behavior~\cite{streltsovOrbitalPhysicsTransition2017, georgesStrongCorrelationsHunds2013}. These considerations motivate going beyond the simplified $U_{\rm eff} = U - J$ treatment in DFT+U\textsubscript{eff}~\cite{dudarevElectronenergylossSpectraStructural1998} to treating $J$ explicitly on an equal footing with $U$ as a separate exchange term rather than folding it into an effective parameter. The result is DFT+$U$+$J$.

The challenge of DFT+$U$+$J$ type functionals is that the $U$ and $J$ parameters must be specified before a calculation can be performed~\cite{leiriacampojrExtendedDFTMethod2010}. Because predicted energetics, electronic structure, and magnetic ordering can change substantially with these choices, obtaining reliable parameter values is essential. Another challenge is the limited transferability of $U$ and $J$. Numerous studies have shown that these parameters are very sensitive to the local chemical environment and depend on the details of computational setup. For example, the choice of pseudopotentials and the definition of correlated subspace (i.e., site occupation projection scheme) can significantly alter the computed $U$ values~\cite{linscottRoleSpinCalculation2018, wangLocalProjectionDensity2016}. As a result, $U$ (and by extension $J$, which is often somewhat less environment-sensitive) cannot be treated as a universal material constant that can be tabulated, but instead must be determined on a case-by-case basis~\cite{mooreHighthroughputDeterminationHubbard2024}.

These challenges can be addressed by computing $U$ and $J$ from first principles. Here a linear response (LR) method~\cite{cococcioniLinearResponseApproach2005a} as opposed to a constrained random phase approximation (cRPA) method~\cite{vaugierHubbardHundExchange2012}is adopted. The former offers a favorable balance between accuracy and computational cost for the large number of configurations considered, while the latter is typically more expensive and, therefore, less compatible for high-throughput workflows. The LR method introduced by Cococcioni and coworkers~\cite{cococcioniLinearResponseApproach2005a} is based on the idea that self-interaction error and delocalization manifest in the curvature of the total energy with respect to the occupation of localized orbitals. In this framework $U$ and $J$ are obtained directly from the response of orbital and occupations to small on-site perturbations, yielding a procedure that is: (i) systematic, because it quantifies the corrective interaction implied by the underlying DFT functional, and (ii) predictive, because it relies only on DFT calculations rather than empirical fitting.

The following sections provide details of the approach used here, which consists of first computing the $U$ and $J$ onsite parameters self-consistently from first principles followed by geometry optimization using the converged $U$ and $J$ parameters. This unified computational pipeline ensures that spectral differences reflect chemistry and cation configuration, rather than methodological artifacts. Within each composition, representative A/B cation distributions (e.g., site preferences) are examined as well to separate chemistry effects from configurational effects. The outcome is a comparative spectral map comprised of electronic, magnonic, and phononic eigenstates across Fe--, Mn--, and Mn--Zn spinel ferrites and across plausible A/B arrangements. The present study is therefore intended as a controlled comparison within a common cubic spinel reference framework across all compositions and A/B-site arrangements, rather than as an exhaustive search over all lower-symmetry distortions that could give rise to competing crystal structures. It is intended to provide a mutually self-consistent computation of the electronic, magnonic, and phononic spectra of technologically important ferrite compositions using established tools: {\tt VASP}-based DFT+$U$+$J$, linear-response evaluation of $U$ and $J$, collinear energy mapping to a Heisenberg model, {\tt SpinW}-based linear spin-wave calculations, and finite-displacement phonons with {\tt Phonopy}.

\section{Computational Procedure}\label{sec:comp-details}

\begin{figure}
\centering
\includegraphics[scale=0.4]{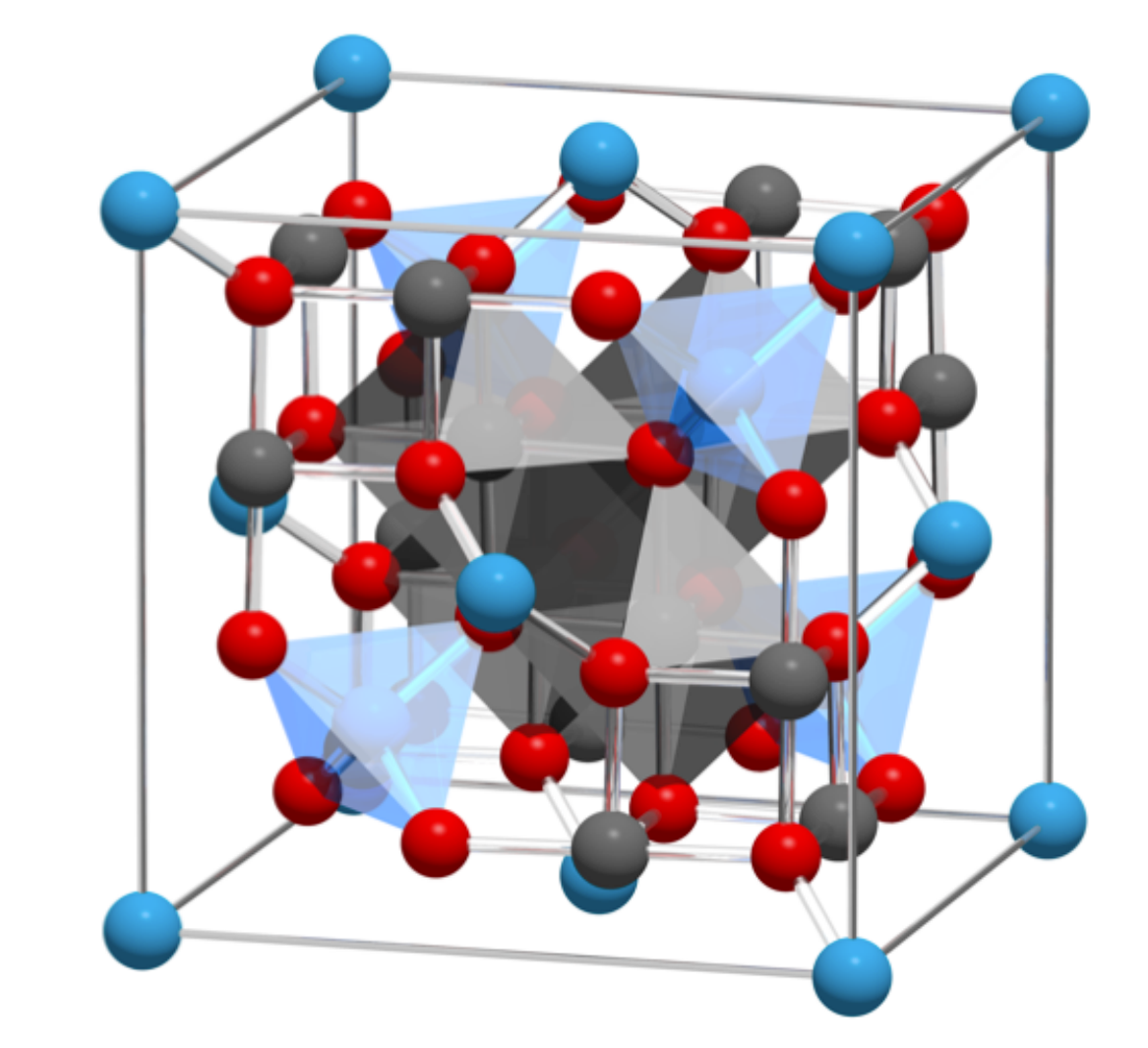}
\caption{\small \em Full unit cell of a spinel ferrite. Tetrahedral cations are shown in blue, octahedral cations in gray, and oxygen anions in red. The atomic radii are not shown to scale. The glass bonds show the nearest neighbor 
% chemical and/or magnetic 
interactions.} 
\label{fig:figure43}
\end{figure}

Electronic-structure, magnetic, and interatomic-force calculations are performed within DFT using the Vienna \emph{Ab-initio} Simulation Package ({\tt VASP})\cite{kresseInitioMolecularDynamics1993, kresseTheoryCrystalStructures1994, kresseEfficiencyAbinitioTotal1996, kresseEfficientIterativeSchemes1996, kresseUltrasoftPseudopotentialsProjector1999}. Exchanges and correlations are treated using the Perdew-Burke-Ernzerhof (PBE) generalized gradient approximation~\cite{perdewGeneralizedGradientApproximation1996a}. Higher-level functionals such as SCAN or HSE could provide useful independent benchmarks for selected cases, but such benchmarking lies beyond the scope of the present comparative LR-based DFT+$U$+$J$ study. Core-valence interactions are described with the projector augmented-wave (PAW) method, treating Fe $3d\,4s$ and O $2s\,2p$ as valence states. Wavefunctions are expanded in a plane-wave basis with a kinetic-energy cutoff of 500~eV. Brillouin-zone integrations use a $10\times10\times10$ Monkhorst-Pack \emph{k}-point mesh~\cite{monkhorstSpecialPointsBrillouinzone1976}. Prior to the linear-response workflow, convergence tests were carried out for the kinetic-energy cutoff over the range of $450-520$ eV and for Monkhorst-Pack meshes from $7 \times 7 \times 7$ to $11 \times 11 \times 11$, using total-energy differences as the convergence metric. On this basis, a cutoff of 500~eV and a $10 \times 10 \times 10$ k-point mesh were adopted for the production calculations reported here. Electronic self-consistency is converged to $10^{-7}$~eV, using Gaussian smearing (0.005~eV). For the linear-response calculations used to evaluate correlated-subspace occupations, the k-point sampling was further refined to $15 \times 15 \times 15$. Total densities of states are evaluated with the Bl\"ochl tetrahedron method. Structural relaxation is carried out for the conventional cubic (spinel) unit cell (space group $Fd\bar{3}m$, No.\ 227) containing 56 atoms (8 formula units) shown in \autoref{fig:figure43}. The cell shape is constrained to remain cubic, while oxygen internal coordinates are relaxed within the retained cubic-reference symmetry. Cations are held at ideal crystallographic positions to preserve the target space group. In all production calculations, symmetry was retained in \texttt{VASP} with \texttt{ISYM=2}. This constrained structural treatment was chosen deliberately to preserve a common cubic spinel crystallographic reference across all compositions and cation arrangements, while still allowing the oxygen sub-lattice to relax self-consistently within that framework. The resulting relaxed oxygen coordinates, together with the associated distributions of Me–O bond lengths and Me–O–Me bond angles, provide the local measure of tetrahedral and octahedral distortion used throughout this work. This choice does not constitute a full search over all lower-symmetry structural minima; rather, it defines the common crystallographic reference within which chemistry- and configuration-driven spectral trends are compared. Ionic relaxations employ a conjugate-gradient algorithm and are terminated when the maximum Hellmann-Feynman force falls below $10^{-6}$~eV~\AA$^{-1}$.

Spin-polarized collinear calculations are performed to represent ferrimagnetic ordering with antiparallel alignment imposed between tetrahedral (A) and octahedral (B) sub-lattices as an initial condition. Site-resolved charges and magnetic moments are obtained from a Bader partitioning of the converged densities rather than from integration within fixed atom-centered spheres. In Bader analysis, real space is divided into atomic basins bounded by zero-flux surfaces of the charge-density gradient~\cite{baderAtomsMoleculesQuantum1994}. Bader charges are computed by integrating the (PAW-reconstructed) electron density over each Bader basin, and Bader magnetic moments by integrating the spin magnetization density over the same basins (``Bader magnetization'')~\cite{henkelmanFastRobustAlgorithm2006,tangGridbasedBaderAnalysis2009,yuAccurateEfficientAlgorithm2011}. Spin-orbit interactions are not included.

\subsection{The Hubbard functional for correlated electrons}\label{subsec:hubbard}

The DFT+$U$+$J$ approach is a corrective extension of a base (semi-)local DFT functional in which an explicit on-site interaction term is added for a chosen set of localized orbitals (the ``Hubbard subspaces'') and a corresponding double-counting contribution is subtracted~\cite{pavariniCorrelatedElectronsModels2012}. The resulting total-energy functional can be written as

\begin{equation}
\begin{aligned}
E_{\rm DFT+U+J} \left[\rho,\{\bm{n}_{\iota}^{\sigma}\}\right]\, & =\, E_{\rm DFT}[\rho] + E_{\rm Hub}\left[\{\bm{n}_{\iota}^{\sigma}\}\right]
- E_{\rm dc}\left[\{n_{\iota}^{\sigma}\}\right] \\[5pt]
& \equiv\, E_{\rm DFT}[\rho] + E_{U,J}\left[\{\bm{n}_{\iota}^{\sigma}\}\right],
\end{aligned}
\end{equation}

% for arXiv (two column format)
%\begin{equation}
%\begin{aligned}
%& E_{\rm DFT+$U$+$J$} \left[\rho,\{\bm{n}_{\iota}^{\sigma}\}\right]\\ 
%
%& \hspace{2cm} =\, E_{\rm DFT}[\rho] + E_{\rm Hub}\left[\{\bm{n}_{\iota}^{\sigma}\}\right]
%- E_{\rm dc}\left[\{n_{\iota}^{\sigma}\}\right] \\[5pt]
%
%& \hspace{2cm} \equiv\, E_{\rm DFT}[\rho] + E_{U,J}\left[\{\bm{n}_{\iota}^{\sigma}\}\right],
%\end{aligned}
%\end{equation}

where $\iota$ indexes the Hubbard sites (typically atomic sites), $\sigma\in\{\uparrow,\downarrow\}$, and the spin-resolved on-site occupation matrices are

\begin{equation}
\left(\bm{n}_{\iota}^{\sigma}\right)_{mm'}
= \left\langle \phi_{\iota m}\,\middle|\, \hat{\rho}^{\sigma}\,\middle|\,\phi_{\iota m'} \right\rangle.
\end{equation}

Here $\{\phi_{\iota m}\}$ denotes a set of localized orbitals spanning the correlated subspace on site $\iota$ (e.g., the transition-metal $3d$ manifold), and $m$ labels the orbital quantum numbers within that subspace. The corresponding spin- and site-resolved occupations are given by 

\begin{equation}
n_{\iota}^{\sigma}=\Tr\,\bm{n}_{\iota}^{\sigma}, \qquad
n_{\iota}=n_{\iota}^{\uparrow}+n_{\iota}^{\downarrow}.
\end{equation}

To capture anisotropic Coulomb and exchange interactions in a rotationally invariant manner, the DFT+$U$+$J$ functional introduced by Liechtenstein \emph{et al.} is used~\cite{liechtensteinDensityfunctionalTheoryStrong1995a}. In this formulation, the Hubbard interaction energy can be expressed in terms of Coulomb matrix elements projected onto the localized-orbital basis such that

\begin{equation}
    \begin{aligned}
    E_{\rm Hub}\, & =\, \frac{1}{2} \sum_{\lbrace m \rbrace, \iota, \sigma} \langle m, m'' \lvert V_{ee} \rvert m', m''' \rangle (n_{\iota}^{\sigma})_{m m'} (n_{\iota}^{-\sigma})_{m'' m'''}\, \\ 
    & \hspace{0.4cm} +\, \frac{1}{2} \sum_{\lbrace m \rbrace, \iota, \sigma}  \Big\{ \langle m, m'' \lvert V_{ee} \rvert m', m''' \rangle\, \\
    & \hspace{0.4cm} -\, \langle m, m'' \lvert V_{ee} \rvert m''', m' \rangle  \Big\}\ (n_{\iota}^{\sigma})_{m m'} (n_{\iota}^{\sigma})_{m'' m'''}
    \end{aligned}
\end{equation}

where the $\langle \cdot \lvert V_{ee} \rvert \cdot \rangle$ are the Coulomb integrals within the correlated subspace. The double-counting term approximately removes the portion of these interactions already included in the base XC functional. In the fully localized limit (FLL) form commonly used for insulating/ionic regimes,

\begin{equation}
E_{\rm dc}
=\sum_{\iota}\frac{U_{\iota}}{2}\,n_{\iota}(n_{\iota}-1)
+\sum_{\iota,\sigma}\frac{J_{\iota}}{2}\,n_{\iota}^{\sigma}(n_{\iota}^{\sigma}-1),
\end{equation}

so that the corrective functional is parameterized by $U_{\iota}$ and $J_{\iota}$~\cite{himmetogluHubbardcorrectedDFTEnergy2014,liechtensteinDensityfunctionalTheoryStrong1995a,pavariniCorrelatedElectronsModels2012}.

\subsection{Spin-polarized linear response}\label{subsec:spin-pol-LR}

In the LR approach, on-site interaction parameters are obtained by applying small localized perturbations to the correlated subspaces and measuring the induced changes in their occupations~\cite{cococcioniLinearResponseApproach2005a}. For a set of Hubbard sites $\{\iota\}$, the site projector is defined as

\begin{equation}
    \hat{P}_{\iota}=\sum_{m}\,|\phi_{\iota m}\rangle\langle\phi_{\iota m}|,
\end{equation}

and an external on-site potential of the form

\begin{equation}
\hat{V}=\sum_{\iota} v_{\iota}\,\hat{P}_{\iota}.
\end{equation}

is applied. The constrained energy associated with these perturbations can be written as

\begin{equation}
    E[\{v_{\iota}\}]
    =\min_{\rho}\left\{E[\rho]+\sum_{\iota} v_{\iota}\,n_{\iota}\right\},
\end{equation}

where $n_{\iota}=n_{\iota}^{\uparrow}+n_{\iota}^{\downarrow}$ is the total occupation of the correlated subspace on site $\iota$.
The central quantities are the (charge-channel) response matrices

\begin{equation}
    \chi_{\iota\iota'}=\frac{\partial n_{\iota}}{\partial v_{\iota'}},
    \qquad
    \chi^{(0)}_{\iota\iota'}=\left.\frac{\partial n_{\iota}}{\partial v_{\iota'}}\right|_{\rm unscreened},
\end{equation}

where $\chi$ is the fully self-consistent (screened) response and $\chi^{(0)}$ is the unscreened response, which is typically evaluated from the non-self-consistent step before Hartree+XC screening fully develops. The Hubbard parameter associated with site $\iota$ is then obtained from the difference between the inverse unscreened and screened responses such that

\begin{equation}\label{eq:U-equation}
U_{\iota}=\left(\left[\chi^{(0)}\right]^{-1}-\chi^{-1}\right)_{\iota\iota}.
\end{equation}

Note that {\tt VASP} adopts an opposite sign convention for $\chi$. The description here adopts the sign convention used by the foundational LR literature~\cite{cococcioniLinearResponseApproach2005a}. The above procedure yields the charge-channel interaction $U$, because the perturbation shifts the two spin channels equally ($v_{\iota}^{\uparrow}=v_{\iota}^{\downarrow}$) and the measured response is the change in total occupation $n_{\iota}$.

To obtain $J$, a \emph{spin-dependent} (exchange-field-like) perturbation that splits the two spin channels with opposite sign is applied so that

\begin{equation}
    \hat{V}^{\sigma}=\sum_{\iota} v_{\iota}^{\sigma}\,\hat{P}_{\iota},
    \qquad
    v_{\iota}^{\uparrow}=+b_{\iota},\ \ v_{\iota}^{\downarrow}=-b_{\iota}.
\end{equation}

In this case, the natural response variable is the local subspace magnetization expressed as

\begin{equation}
    m_{\iota}=n_{\iota}^{\uparrow}-n_{\iota}^{\downarrow},
\end{equation}

and the \emph{spin-channel} response matrices are given by

\begin{equation}
    \chi^{m}_{\iota\iota'}=\frac{\partial m_{\iota}}{\partial b_{\iota'}},
    \qquad
    \left(\chi^{m}_{0}\right)_{\iota\iota'}=\left.\frac{\partial m_{\iota}}{\partial b_{\iota'}}\right|_{\rm unscreened}.
\end{equation}

The Hund's parameter on site $\iota$ is then obtained in direct analogy with \autoref{eq:U-equation} as
\begin{equation}\label{eq:J-equation}
    J_{\iota}\, =\, - \left(\left[\chi^{m}_{0}\right]^{-1}-\left[\chi^{m}\right]^{-1}\right)_{\iota\iota},
\end{equation}
which isolates the screened on-site exchange interaction associated with spin polarization within the correlated subspace. Note that there is an additional negative sign for the expression of $J$ as originally proposed~\cite{lambertEvaluationFirstprinciplesHubbard2024,linscottRoleSpinCalculation2018,  vaugierHubbardHundExchange2012,macenultyOptimizationStrategiesDeveloped2023}. \\

\subsection{Calculation of \emph{U} and \emph{J}}\label{subsec:U-J-proc}

The site- and species-resolved $U$ and $J$ parameters are determined self-consistently for each ferrite composition and cation arrangement considered in this work using an iterative geometry-parameter feedback procedure. Starting from an initial structure (experimental lattice constant when available, otherwise a reasonable literature/relaxed starting point), a structural relaxation is first performed using the base GGA functional without Hubbard corrections. Linear-response calculations are then carried out to extract the on-site interaction parameters following the method outlined in \autoref{subsec:spin-pol-LR}. The resulting $\{U_{\iota},J_{\iota}\}$ values are subsequently used in a new relaxation employing the rotationally invariant DFT+$U$+$J$ functional (Lichtenstein form) as implemented in {\tt VASP} by setting {\tt LDAUTYPE=3}. This cycle (relaxation $\rightarrow$ LR extraction of $\{U,J\}$ $\rightarrow$ updated DFT+$U$+$J$ relaxation) is repeated until successive iterations agree within $10^{-3}$ \AA\ in the lattice parameter, $10^{-2}\ \mu_B$ in the site magnetic moments, and  $10^{-3}$~eV  in the LR-derived $U$ and $J$ values. All production calculations in this iterative workflow were performed with symmetry retained (\texttt{ISYM=2}) in order to achieve stable and reproducible convergence across the full set of ferrite configurations. As an auxiliary robustness check, additional relaxation tests were carried out for \ce{Fe3O4}. In a fixed-cubic-cell, all-ions-relaxed test, the internal atomic coordinates moved away from the intended cubic-spinel reference, including departures from the ideal tetrahedral/octahedral-site geometry. In a separate full cell + ionic relaxation, the structure moved further from the targeted cubic reference, including distortion of the cell shape itself. We also attempted symmetry-off calculations, but these frequently became numerically unstable or failed to converge, especially within the cyclic relaxation/linear-response workflow. Thus, they were not adopted as a viable production strategy for the present study.

In the LR step, the converged ground state is perturbed by small on-site potentials applied to the localized Hubbard subspaces associated with each symmetry-inequivalent atom type present in the configuration (e.g., tetrahedral-site cations, octahedral-site cations, and oxygen). For the charge-channel response used to determine $U$, equal perturbations are applied to both spin channels, $v_{\iota}^{\uparrow}=v_{\iota}^{\downarrow}=\pm\alpha$, and changes in the total subspace occupancy $n_{\iota}=n_{\iota}^{\uparrow}+n_{\iota}^{\downarrow}$ are recorded. For the spin-channel response used to determine $J$, opposite-sign perturbations are applied to the two spin channels, $v_{\iota}^{\uparrow}=+\beta$ and $v_{\iota}^{\downarrow}=-\beta$, and the induced change in subspace magnetization $m_{\iota}=n_{\iota}^{\uparrow}-n_{\iota}^{\downarrow}$ is evaluated. For each inequivalent site/species, a set of 20 equally spaced perturbation amplitudes spanning $\pm 0.1$~eV is used to construct the LR response matrices and extract the corresponding $U_{\iota}$ and $J_{\iota}$ \cite{macenultyOptimizationStrategiesDeveloped2023}.

\subsection{The Heisenberg model}\label{subsec:heisenberg}

To describe magnetic excitations and extract exchange parameters from first-principles energetics, an effective Heisenberg-Dirac-van Vleck (HDvV) type spin Hamiltonian is adopted, which assumes that magnetism can be represented by a set of localized moments interacting pairwise~\cite{heisenbergZurTheorieFerromagnetismus1928,singhStoryMagnetismHeisenberg2018}. In its general bilinear form, it can be written as

\begin{equation}\label{eq:Heisenberg-tensor}
    \hat{\mathcal{H}}=-\sum_{i\neq j}\hat{\bm S}_i^{\top}\,\bm J_{ij}\,\hat{\bm S}_j
\end{equation}

where $\hat{\bm S}_i$ is the spin operator at magnetic site $i$ and $\bm J_{ij}$ is the (generally anisotropic) exchange tensor of rank 2. In the present work, single-ion anisotropy and Dzyaloshinskii-Moriya interactions are not retained. The exchange tensor is, therefore, reduced to an isotropic scalar $J_{ij}$~\cite{liechtensteinLocalSpinDensity1987}. For collinear configurations, it is convenient to map total energies onto a classical (or, alternatively, mean-field) version of the HDvV Hamiltonian (denoted by $H$) by replacing spin operators with classical vectors $\bm S_i=S_i\bm e_i$ of fixed magnitude $S_i$ and direction $\bm e_i$. The result is

\begin{equation}\label{eq:Heisenberg-classical}
    H = -\sum_{i\neq j} J_{ij}\,\bm S_i^{\top}\bm S_j
\end{equation}

This energy-mapping strategy of evaluating DFT energies for several chosen spin configurations and fitting the parameters of an effective spin Hamiltonian is widely used and is complementary to approaches based on infinitesimal spin rotations (or, magnetic force theorem)~\cite{liechtensteinLocalSpinDensity1987,katsnelsonFirstprinciplesCalculationsMagnetic2000,xiangMagneticPropertiesEnergymapping2013}. The present model is restricted to nearest-neighbor (NN) couplings on the spinel lattice, and the corresponding effective NN exchange constants used in subsequent spin-wave calculations are reported.

\subsection{Determining the exchange coupling constants}\label{subsec:exchange-calculations}

In insulating and semiconducting spinel ferrites, the dominant magnetic interactions are typically \emph{superexchange} couplings mediated by O $2p$ states along metal-oxygen-metal exchange paths rather than direct $d$-$d$ overlap. The relative strengths and signs of these couplings are commonly rationalized using the Goodenough-Kanamori-Anderson rules, and for many spinel ferrites, the A-B coupling is the strongest and antiferromagnetic, producing ferrimagnetic order~\cite{harrisGoodenoughKanamoriAnderson2022,srivastavaExchangeConstantsSpinel1979}. Motivated by classic experimental analyses of spinel ferrites~\cite{glasser1963spin}, the nearest-neighbor exchange network is parameterized here by three effective constants: $J_{AB}$ for nearest-neighbor bonds between tetrahedral (A) and octahedral (B) sub-lattices, $J_{AA}$ for nearest-neighbor A-A bonds on the diamond sub-lattice, and $J_{BB}$ for nearest-neighbor B-B bonds on the pyrochlore sub-lattice.

\paragraph{Nearest-neighbor Hamiltonian on the spinel lattice:}
Let $A$ and $B$ denote the sets of A- and B-sub-lattice sites in the crystallographic cell used for the mapping. Only \emph{magnetic} cations are included in the sums. Non-magnetic Zn is naturally handled by setting $S_i=0$ on Zn sites (so bonds involving Zn contribute zero). For collinear reference states, the Ising variables $\sigma_i=\pm 1$ that encode whether the moment on site $i$ is parallel ($+1$) or antiparallel ($-1$) to a chosen global axis are introduced. The NN Heisenberg Hamiltonian for mapping is then

\begin{equation}\label{eq:NN-Heis-spinel}
    \begin{aligned}
    H
    = & -J_{AB}\sum_{\langle i\in A,\,j\in B\rangle} \bm{S}_i^{\top}\bm{S}_j\,\sigma_i\sigma_j\,
    -\, J_{AA}\sum_{\langle i,i'\in A\rangle} \bm{S}_i^{\top}\bm{S}_{i'}\,\sigma_i\sigma_{i'}\, 
    -\, J_{BB}\sum_{\langle j,j'\in B\rangle} \bm{S}_j^{\top}\bm{S}_{j'}\,\sigma_j\sigma_{j'}
    \end{aligned}
\end{equation}

% for arXiv (two column)
%\begin{equation}\label{eq:NN-Heis-spinel}
%    \begin{aligned}
%    H
%    = -J_{AB}\sum_{\langle i\in A,\,j\in B\rangle} \bm{S}_i^{\top}\bm{S}_j\,\sigma_i\sigma_j\, \\
%    -\, J_{AA}\sum_{\langle i,i'\in A\rangle} \bm{S}_i^{\top}\bm{S}_{i'}\,\sigma_i\sigma_{i'}\, \\
%    -\, J_{BB}\sum_{\langle j,j'\in B\rangle} \bm{S}_j^{\top}\bm{S}_{j'}\,\sigma_j\sigma_{j'}
%    \end{aligned}
%\end{equation}

where $\langle\cdots\rangle$ denotes sums over nearest-neighbor bonds in the spinel lattice. With the sign convention used in \autoref{eq:NN-Heis-spinel}, $J>0$ favors ferromagnetic alignment, and $J<0$ favors antiferromagnetic alignment.

\paragraph{Bond-weighted sums valid for Fe$_3$O$_4$, Mn ferrite, and \ce{Mn-Zn} ferrite:}
For a given chemical configuration (including the specific A/B occupation pattern), the bond-weighted NN sums are defines as

\begin{equation}\label{eq:Wdefs}
    \begin{aligned}
    W_{AB}\, &= \sum_{\langle i\in A,\,j\in B\rangle} S_i\, S_j \\[5pt]
    W_{AA}\, &= \sum_{\langle i,i'\in A\rangle} S_i\, S_{i'} \\[5pt]
    W_{BB}\, &= \sum_{\langle j,j'\in B\rangle} S_j\, S_{j'}
    \end{aligned}
\end{equation}

These quantities reduce to the familiar ``bond counts times sub-lattice moments'' when all magnetic sites on a sub-lattice share the same magnitude where, for example, Fe$_3$O$_4$ treated with a uniform $S_A$ and $S_B$ yields $W_{AB}=N_{AB}S_AS_B$, $W_{AA}=N_{AA}S_A^2$, $W_{BB}=N_{BB}S_B^2$, with $N_{AB}=12$, $N_{AA}=2$, $N_{BB}=6$ per formula unit for the spinel NN topology~\cite{bercoffExchangeConstantsTransfer1997}.

For mixed-cation cases, \autoref{eq:Wdefs} automatically incorporates the correct weighting. Equivalently, the A-B weight may be written as a sum over pair types. For MnFe$_2$O$_4$, this results in
\begin{equation}\label{eq:WABpairtypes}
    W_{AB}
    = \sum_{s\in\{\mathrm{Fe},\mathrm{Mn}\}}\sum_{t\in\{\mathrm{Fe},\mathrm{Mn}\}}
    N_{AB}^{(s_A,t_B)}\,S_{s_A}S_{t_B},
\end{equation}
and for \ce{Mn-Zn} ferrites the same form applies with $t\in\{\mathrm{Fe},\mathrm{Mn},\mathrm{Zn}\}$ but $S_{\mathrm{Zn}}=0$ so that Zn-containing bonds do not contribute. Analogous pair-type decompositions hold for $W_{AA}$ and $W_{BB}$.

\paragraph{Energy mapping and closed-form expressions for $J_{AB}$, $J_{AA}$, and $J_{BB}$:}
Collinear DFT total energies are computed on the same relaxed structure for a set of reference spin states and  energy \emph{differences} are mapped onto \autoref{eq:NN-Heis-spinel} (energy-mapping analysis)~\cite{xiangMagneticPropertiesEnergymapping2013,ciofiniMappingManyelectronGeneralised2005,liechtensteinLocalSpinDensity1987}. The four reference states used are

\begin{enumerate}[(i)]
    \item ferrimagnetic $\mathrm{FiM}$ (A$\uparrow$, B$\downarrow$);
    \item ferromagnetic $\mathrm{FM}$ (A$\uparrow$, B$\uparrow$);
    \item an A-sublattice antiferromagnetic state $\mathrm{A}_{\mathrm{AF}}$ (A N\'eel order on the diamond net; B is kept collinear);
    \item a B-sublattice antiferromagnetic state $\mathrm{B}_{\mathrm{AF}}$ (collinear ``2-up/2-down'' arrangement on each pyrochlore tetrahedron; A is kept collinear).
\end{enumerate}

For any reference state $k$, the bond-correlation sums are defined as

\begin{equation}\label{eq:Cdefs}
    \begin{aligned}
    C_{AB}^{(k)} & =\sum_{\langle i\in A,\,j\in B\rangle} S_iS_j\,\sigma_i^{(k)}\sigma_j^{(k)}, \\[5pt]
    C_{AA}^{(k)} & =\sum_{\langle i,i'\in A\rangle} S_iS_{i'}\,\sigma_i^{(k)}\sigma_{i'}^{(k)}, \\[5pt]
    C_{BB}^{(k)} & =\sum_{\langle j,j'\in B\rangle} S_jS_{j'}\,\sigma_j^{(k)}\sigma_{j'}^{(k)}
    \end{aligned}
\end{equation}

The DFT energy differences relative to $\mathrm{FiM}$ then satisfy

\begin{equation}\label{eq:DeltaEgeneral}
    \begin{aligned}
    \Delta E_k & \equiv E_k - E_{\mathrm{FiM}} \\[5pt]
    &= -J_{AB}\left(C_{AB}^{(k)}-C_{AB}^{(\mathrm{FiM})}\right)\,
    -J_{AA}\left(C_{AA}^{(k)}-C_{AA}^{(\mathrm{FiM})}\right)\, 
    -J_{BB}\left(C_{BB}^{(k)}-C_{BB}^{(\mathrm{FiM})}\right)
    \end{aligned}
\end{equation}

% for arXiv (two column)
%\begin{equation}\label{eq:DeltaEgeneral}
%    \begin{aligned}
%    \Delta E_k & \equiv E_k - E_{\mathrm{FiM}} \\[5pt]
%
%    &= -J_{AB}\left(C_{AB}^{(k)}-C_{AB}^{(\mathrm{FiM})}\right)\, \\
%    & \hspace{1cm} -J_{AA}\left(C_{AA}^{(k)}-C_{AA}^{(\mathrm{FiM})}\right)\, \\
%    & \hspace{1cm} -J_{BB}\left(C_{BB}^{(k)}-C_{BB}^{(\mathrm{FiM})}\right)
%    \end{aligned}
%\end{equation}

This form is completely general and remains valid for Fe$_3$O$_4$, MnFe$_2$O$_4$, and \ce{Mn-Zn} ferrites because the chemical configuration enters only through the site magnitudes $\{S_i\}$ and the bond lists.

A particularly simple closed form exists for $J_{AB}$ using $\mathrm{FiM}$ and $\mathrm{FM}$ because these two states have identical A-A and B-B correlations, while every A-B bond changes sign. Since $C_{AB}^{(\mathrm{FM})}=+W_{AB}$ and $C_{AB}^{(\mathrm{FiM})}=-W_{AB}$,

\begin{equation}\label{eq:JAB_general}
    J_{AB}
    = -\,\frac{E_{\mathrm{FM}}-E_{\mathrm{FiM}}}{2\,W_{AB}} .
\end{equation}

The remaining constants are obtained from $\mathrm{A}_{\mathrm{AF}}$ and $\mathrm{B}_{\mathrm{AF}}$. For $\mathrm{A}_{\mathrm{AF}}$ the A-A nearest-neighbor bonds on the diamond net flip sign so that $C_{AA}^{(\mathrm{A_{AF}})}=-W_{AA}$, while the A-B correlation $C_{AB}^{(\mathrm{A_{AF}})}$ is evaluated from \autoref{eq:Cdefs} for the imposed collinear pattern. This is important in mixed-cation cases where perfect cancellation is not guaranteed by symmetry alone. Substituting into \autoref{eq:DeltaEgeneral} yields

\begin{equation}\label{eq:JAA_general}
    J_{AA}
    = \frac{\Delta E_{\mathrm{A_{AF}}} + J_{AB}\left(C_{AB}^{(\mathrm{A_{AF}})}+W_{AB}\right)}{2\,W_{AA}} .
\end{equation}

Similarly, for $\mathrm{B}_{\mathrm{AF}}$ the B-B correlations change and are evaluated explicitly from \autoref{eq:Cdefs}. The result is

\begin{equation}\label{eq:JBB_general}
J_{BB}
= -\,\frac{\Delta E_{\mathrm{B_{AF}}} + J_{AB}\left(C_{AB}^{(\mathrm{B_{AF}})}+W_{AB}\right)}{C_{BB}^{(\mathrm{B_{AF}})}-W_{BB}} .
\end{equation}
When all B-site moments have equal magnitude and the ideal ``2-up/2-down'' pattern is used, such as in \ce{Fe3O4}, $C_{BB}^{(\mathrm{B_{AF}})}=-(1/3)W_{BB}$, so that \autoref{eq:JBB_general} reduces to,

\begin{equation}
    J_{BB}=\frac{3}{4}\,\frac{\Delta E_{\mathrm{B_{AF}}}+J_{AB}W_{AB}}{W_{BB}}
\end{equation} 

In MnFe$_2$O$_4$ and \ce{Mn-Zn} ferrites, the general form of \autoref{eq:JBB_general} is retained and $C_{BB}^{(\mathrm{B_{AF}})}$ (and any residual $C_{AB}^{(\mathrm{B_{AF}})}$) is evaluated directly from the imposed collinear configuration.

Finally, it is emphasized that the NN three-parameter model produces \emph{effective} exchange constants that are appropriate for the chosen lattice, composition, and cation arrangement. If further-neighbor interactions are non-negligible, the fitted NN constants should be interpreted as renormalized values within the reduced model~\cite{xiangMagneticPropertiesEnergymapping2013}. To test this point explicitly, a shell-resolved beyond-NN benchmark for cubic \ce{Fe3O4} is presented in the \autoref{app:second-NN-Fe3O4}, where second-shell couplings within the A-B, B-B, and A-A pair families are extracted. The resulting $J_{AB}^{(2)}$, $J_{BB}^{(2)}$ and $J_{AA}^{(2)}$ terms are substantially smaller than their first-shell counterparts, indicating that the dominant exchange hierarchy captured by the main NN model remains intact, while longer-range terms provide comparatively modest corrections.

\subsection{Magnon calculations}\label{subsec:magnons}
Magnon (spin-wave) spectra are computed within linear spin-wave theory (LSWT) using the NN isotropic Heisenberg Hamiltonian on the spinel lattice described in the previous section. LSWT describes small transverse fluctuations about a long-range ordered reference state by mapping spin operators to bosons via the Holstein-Primakoff transformation and retaining only the quadratic (harmonic) terms, i.e., the leading order in a $1/S$ expansion~\cite{holsteinFieldDependenceIntrinsic1940,auerbachInteractingElectronsQuantum1994}. This approximation is appropriate here because the goal is to quantify how changes in cation configuration modify the \emph{harmonic} magnon eigenvalue spectrum and its DOS.

The exchange constants $\{J_{AB},J_{AA},J_{BB}\}$ are taken directly from the DFT energy-mapping procedure described in the previous section and, therefore, correspond to the same relaxed crystal structure and collinear ferrimagnetic alignment (moments parallel within each sub-lattice and antiparallel between A and B). The mapping yields \emph{effective} nearest-neighbor exchanges for each specific composition and the A/B occupation pattern. Non-magnetic Zn is naturally accommodated in the subsequent spin-wave calculation by assigning a zero moment on Zn sites.

Spin-wave calculations are performed with {\tt SpinW}~\cite{tothLinearSpinWave2015a}. The crystallographic lattice and magnetic basis are constructed from the DFT-optimized conventional spinel cell to ensure consistency with the structural and vibrational models. Exchange parameters are supplied in meV using the opposite sign convention and normalization adopted in the Heisenberg mapping for how these are defined within the {\tt SpinW} code. Unless stated otherwise, Land\'e $g$ factors are set to $g=2.0$ for all magnetic cations.

In LSWT and its implementation in {\tt SpinW}, the quadratic bosonic Hamiltonian in reciprocal space is obtained by Fourier transforming the exchange network and then diagonalized at each wavevector by a bosonic Bogoliubov transformation~\cite{tothLinearSpinWave2015a,colpaDiagonalizationQuadraticBoson1978}. Single-ion anisotropy and Dzyaloshinskii-Moriya interactions are neglected in the present work. This is consistent with the ``soft-ferrite'' character of \ce{Mn-Zn} ferrites (nearly zero magnetocrystalline anisotropy) and with the general expectation that antisymmetric exchange is typically a weak correction compared with the dominant symmetric exchange in bulk materials~\cite{thakurReviewMnZnFerrites2020,camleyConsequencesDzyaloshinskiiMoriyaInteraction2023}. Where relevant to magnetite, spin-wave measurements provide a useful experimental context for the exchange-dominated spectra~\cite{mcqueeneyInvestigationPresenceCharge2006}.

Brillouin-zone integration is performed on a uniform $15\times 15\times 15$ $q$-point mesh, and convergence is checked against $10\times 10\times 10$ and $12\times 12\times 12$ meshes. The dominant peak positions change by less than 5~meV upon refinement. The discrete spectrum is broadened with a Gaussian of width 0.05~meV to obtain a smooth mDOS. Reducing the width to 0.01~meV does not materially change the peak locations or the integrated spectral weight.

\subsection{Phonon calculations}\label{subsec:phonons}

The phonon DOS for each material is computed using the finite-displacement (supercell) method as implemented in {\tt Phonopy}, with interatomic force constants obtained from first-principles Hellmann-Feynman forces computed in {\tt VASP}~\cite{togoFirstprinciplesCalculationsFerroelastic2008,togoFirstPrinciplesPhonon2015a,parlinskiFirstPrinciplesDeterminationSoft1997}. In this approach, small symmetry-adapted displacements are applied to crystallographically inequivalent atoms in a supercell. The resulting forces are used to assemble the real-space force-constant matrix from which the dynamical matrix is constructed and diagonalized to obtain phonon frequencies throughout the Brillouin zone~\cite{togoFirstPrinciplesPhonon2015a,parlinskiFirstPrinciplesDeterminationSoft1997}.

For each configuration, a $2\times 2\times 2$ supercell of the conventional spinel cell (448 atoms) is built and each symmetry-inequivalent atom displaced by $\pm 0.01$~\AA\ along Cartesian directions. Using the same crystallographic reference and A/B occupation pattern as in the magnetic model ensures that differences in phonon spectra reflect the chemistry and cation configuration rather than changes in the underlying lattice description. Forces are evaluated for each displaced structure using the converged LR parameters ($U$ and $J$) and a $3\times 3\times 3$ $k$-point mesh for the supercell. The resulting force constants are symmetrized in {\tt Phonopy} and used to compute the phonon DOS on a $15\times 15\times 15$ $q$-mesh. No imaginary frequencies are observed for the relaxed structures, and the phonon frequencies converge to within 0.05~meV with respect to the Brillouin-zone sampling and displacement amplitude.

\section{Discussion}\label{sec:discussion}

This section organizes the outputs of the unified workflow (\autoref{sec:comp-details}) and frames them in a consistent way for cross-composition and cross-configuration comparison. The focus is on three spinel ferrites relevant to soft-magnetic applications: \ce{Fe3O4}, \ce{MnFe2O4}, and \ce{(Mn_{0.5},Zn_{0.5})Fe2O4}. Rather than surveying the full compositional and configurational space, the computational setup is fixed and the comparison restricted to a compact, representative set of cation arrangements so that observed trends can be traced to chemistry and local coordination rather than to methodological differences.

\subsection{Structural properties}

\begin{figure}[t]
\centering
\includegraphics[scale=0.6]{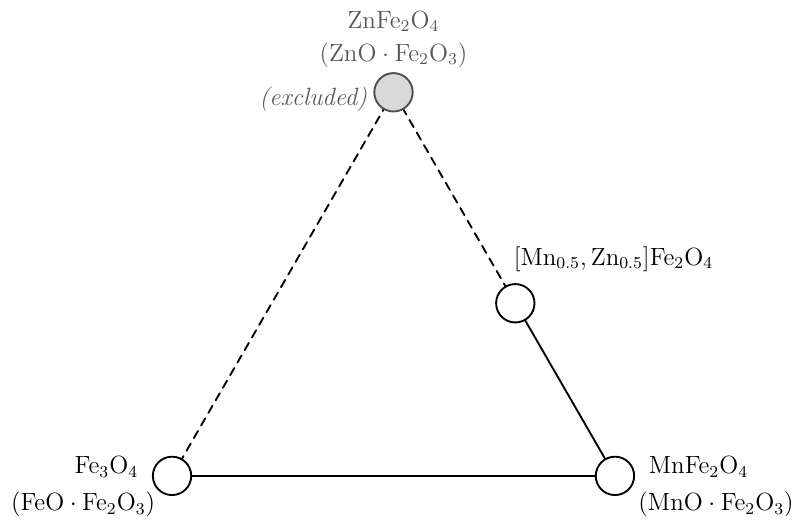}
\caption{\small \em Pseudo-ternary map with markers indicating studied compositions. The dashed segments indicate directions toward the excluded \ce{ZnFe2O4} end member.}
\label{fig:figure2}
\end{figure}

To organize the chemistry, the pseudo-ternary map in \autoref{fig:figure2} is used. Its three vertices are the spinel end members written as \ce{MeFe2O4} (equivalently \ce{MeO}$\cdot$\ce{Fe2O3}) formula units: \ce{Fe3O4} (\ce{FeO}$\cdot$\ce{Fe2O3}), \ce{MnFe2O4} (\ce{MnO}$\cdot$\ce{Fe2O3}), and \ce{ZnFe2O4} (\ce{ZnO}$\cdot$\ce{Fe2O3}). Strictly speaking, the compounds studied are spinels of the \ce{MeFe2O4} family, not literal mixtures of monoxides, but this triangular framing provides a useful compass for the chemistry where \ce{Fe3O4} sits near the Fe-rich vertex, \ce{MnFe2O4} steps along the \ce{Fe} $\leftrightarrow$ \ce{Mn} edge, and \ce{(Mn_{0.5}, Zn_{0.5})Fe2O4} marks a deliberate excursion toward the \ce{Zn} corner with a fixed 1:1 \ce{Mn/Zn} ratio. This triangle is used as an organizing motif rather than a thermodynamic phase diagram. It visualizes substitution paths without implying that every interior point is sampled or even synthetically accessible. Throughout, only the discrete compositions and cation configurations explicitly computed are compared, because site preference and cation inversion constrain which distributions are physically relevant. Here, ``inversion'' refers to the redistribution of cations between the tetrahedral (A) and octahedral (B) sub-lattices relative to the \emph{normal} spinel arrangement and is conveniently quantified by an inversion parameter $\delta$ (fraction of the nominal A-site cations occupying B sites), with $\delta=0$ for a normal spinel and $\delta=1$ for a fully inverse spinel.

Within each composition, a small set of cation distributions over tetrahedral (A) and octahedral (B) sub-lattices is considered to separate chemical substitution effects from configurational effects. \ce{Fe3O4} serves as the canonical reference spinel, including the inequivalent B-site electronic character captured by the relaxed cubic cell. For \ce{MnFe2O4}, a low-inversion limit (\ce{Mn} predominantly on A, $\delta \approx 0$) is compared against a more mixed distribution of Mn among A and B sites to reveal how inversion reshapes local coordination and exchange pathways. For \ce{(Mn_{0.5},Zn_{0.5})Fe2O4}, three symmetry-distinct arrangements consistent with \ce{Zn}’s strong tendency toward A-site occupancy are examined~\cite{bohraNanostructuredZnFe2O4Exotic2021,bockDelithiationSpinelFerrites2020,anantharamanMagneticPropertiesUltrafine1998}, while varying how \ce{Mn} is partitioned between A and B sites. A fully inverse \ce{MnFe2O4} limit with Mn placed exclusively on B sites was also probed in auxiliary calculations, but it was not retained because, within the same self-consistent LR-based DFT+$U$+$J$ workflow and cubic-reference constraints used throughout this study, it yielded abnormally large LR parameters and unstable SCF convergence, preventing a controlled comparison on the same methodological footing as the retained cases.

Although \ce{ZnFe2O4} is the natural third end member of the map in \autoref{fig:figure2}, it is not treated as a primary target in this work. In the normal spinel limit, \ce{Zn^2+} occupies A sites and carries no magnetic moment, while \ce{Fe^3+} resides on B sites. This suppresses the dominant A-B superexchange channel that underpins ferrimagnetic spinels such as \ce{Fe3O4} and \ce{MnFe2O4}\cite{bohraNanostructuredZnFe2O4Exotic2021}. As a consequence, the magnetic response of \ce{ZnFe2O4} is unusually sensitive to inversion and defect chemistry and is governed largely by B-B interactions~\cite{rodrigueztorresEvidenceDefectinducedFerromagnetism2011}. Here, instead, the focus is on compositions where \ce{Zn} is introduced alongside magnetic cations (\ce{Mn}, \ce{Fe}) so that the A-B exchange network remains active and the extracted exchange constants and magnon spectra stay directly comparable across the set.

\newcommand{\UJ}[2]{\makecell[l]{{\em U} = #1\\{\em J\,} = #2}}

\begin{table*}[t]
%\centering
    \caption{Linear-response results: on-site Hubbard $U$ and Hund's exchange $J$ parameters for each inequivalent site. ``50/50 Mn on A \& B'' means half of the Mn on A sites and half of the Mn on B sites.}
    \label{tab:lr-UJ}
    \renewcommand\arraystretch{2}
    \setlength{\tabcolsep}{1.8pt}

%==================== Fe3O4 ====================%
    \begin{subtable}[t]{\textwidth}
    \centering
    \caption{\ce{Fe3O4}}
    \label{tab:lr-UJ-fe3o4}
    \begin{tabular}{|c|c|c|c|}
    \hline
    \textbf{Configuration} & \textbf{Fe-A} & \textbf{Fe-B} & \textbf{O} \\
    \hline
    Prototype
      & \UJ{4.491}{$-0.488$}
      & \UJ{5.235}{$-1.316$}
      & \UJ{10.146}{$1.209$}
    \\ \hline
    \end{tabular}
    \end{subtable}
    \hfill\\[5pt]
%==================== MnFe2O4 ====================%
    \begin{subtable}[t]{\textwidth}
    \centering
    \caption{\ce{MnFe2O4}}
    \label{tab:lr-UJ-mnfe2o4}
    \begin{tabular}{|c|c|c|c|c|c|}
    \hline
    \textbf{Configuration} & \textbf{Mn-A} & \textbf{Mn-B} & \textbf{Fe-A} & \textbf{Fe-B} & \textbf{O} \\
    \hline
    \makecell{Config.\,1\\(all Mn on A)}
      & \UJ{5.968}{0.203}
      & N/A
      & N/A
      & \UJ{4.974}{$-0.547$} 
      & \UJ{9.881}{$1.072$}
    \\ \hline
    \makecell{Config.\,2\\(50/50 Mn on A \& B)}
      & \UJ{5.637}{$0.335$}
      & \UJ{5.640}{$0.529$}
      & \UJ{5.617}{$-2.740$}
      & \UJ{5.069}{$-1.060$}
      & \UJ{10.083}{$1.152$}
    \\ \hline
    \end{tabular}
    \end{subtable}
    \hfill\\[5pt]
%==================== (Mn,Zn)Fe2O4 ====================%
    \begin{subtable}[t]{\textwidth}
    \centering
    \caption{\ce{(Mn_{0.5},Zn_{0.5})Fe2O4}}
    \label{tab:lr-UJ-mnznfe2o4}
    \begin{tabular}{|c|c|c|c|c|c|c|}
    \hline
    \textbf{Configuration} & \textbf{Zn-A} & \textbf{Mn-A} & \textbf{Mn-B} & \textbf{Fe-A} & \textbf{Fe-B} & \textbf{O} \\
    \hline
    \makecell{Config.\,1\\ (all Zn \& Mn on A)}
      & \UJ{2.017}{$0.013$}
      & \UJ{4.075}{$1.344$}
      & N/A
      & N/A
      & \UJ{5.171}{$-1.824$}
      & \UJ{10.126}{$1.181$}
    \\ \hline
    \makecell{Config.\,2\\ (all Zn on A, all Mn on B))}
      & \UJ{2.645}{0.287}
      & N/A
      & \UJ{5.701}{$0.466$}
      & \UJ{4.866}{$-0.849$}
      & \UJ{4.982}{$-0.042$}
      & \UJ{10.037}{$1.121$}
    \\ \hline
    \makecell{Config.\,3\\ (all Zn on A, 50/50 Mn on A \& B)}
      & \UJ{3.670}{$0.981$}
      & \UJ{5.788}{$0.318$}
      & \UJ{6.701}{$0.188$}
      & \UJ{4.708}{$-0.443$}
      & \UJ{5.029}{$-0.758$}
      & \UJ{10.068}{$1.383$}
    \\ \hline
    \end{tabular}
    \end{subtable}

\end{table*}

\autoref{tab:lr-UJ} reports the LR interaction parameters used in the DFT+$U$+$J$ description, resolved by species and by A/B sub-lattice. These parameters are obtained self-consistently alongside geometry convergence. Starting from a trial parameter set, the cubic cell is relaxed under the imposed structural constraints (fixed cation framework with oxygen internal coordinates allowed to be adjusted). The LR is then performed on that relaxed state to update $(U,J)$ for each inequivalent subspace, and the relaxation is repeated using the updated values. This relaxation-LR cycle is continued until both the structural descriptors (e.g., \emph{a, u} for geometry) and the LR parameters change negligibly between successive iterations. With that closed loop and identical numerical settings for every case, trends in \autoref{tab:lr-UJ} can be interpreted as environmental screening trends tied to local coordination, rather than as artifacts of inconsistent parametrization. 

To place the values in \autoref{tab:lr-UJ} in context, it is useful to compare them with representative Hubbard parameter choices used previously in the ferrite literature. For \ce{Fe3O4}, older DFT+U studies commonly adopted a single \ce{Fe} $U_{\rm eff}$ of about 3.8 eV without distinguishing between A- and B-site Fe \cite{santos-carballalFirstprinciplesStudyInversion2015, yuFe3O4SurfaceElectronic2012}, whereas more recent self-consistent studies have reported distinct \ce{Fe_{oct}} and \ce{Fe_{tet}} values and, in some cases, additional intersite $V$ corrections \cite{naveasFirstPrinciplesCalculationsMagnetite2023a}. In that sense, the present Fe-A and Fe-B values fall within the broad scale already used for magnetite, while also retaining the site dependence that is absent from fixed-$U_{\rm eff}$  treatments. For \ce{MnFe2O4}, the representative literature is less standardized and is dominated by fixed species-based $U$ or $U_{\rm eff}$ choices (3.9 eV for Mn and 4.5 eV for Fe \cite{huangCationMagneticOrders2013}), typically without A/B-site differentiation and without explicit $J$ \cite{santos-carballalFirstprinciplesStudyInversion2015}. The self-consistent values reported here lie in the same few-eV range, but additionally resolve the dependence on cation sublattice and local configuration. For the mixed \ce{(Mn_{0.5},Zn_{0.5})Fe2O4} case, published Hubbard parameter choices are much sparser and, where available, are again usually given as fixed $U_{\rm eff}$ values rather than as site-resolved $U+J$ sets. One recent DFT+U study on \ce{(Mn_{0.5},Zn_{0.5})Fe2O4} explicitly uses $U_{\rm eff}$ = 4.3 eV for Fe, 3.5 eV for Mn and 6.0 eV for Zn \cite{liTailoringStructuralElectronic2025}. The comparison to prior work should, therefore, be read as a robustness context for the scale and chemical trends of the present LR parameters, not as a one-to-one validation against a unique canonical parameter set. It is also important to note that explicit published $J$ values are much less common in the representative ferrite studies surveyed here than fixed $U_{\rm eff}$ values or, in some recent cases, $U+V$ parameterizations. This is one reason why direct experimental validation of isolated $U$ and $J$ numbers is not possible in a one-to-one sense: these are effective parameters of a chosen computational framework rather than directly measurable observables. Experimental comparison is instead most meaningful at the level of the resulting lattice parameters, local moments, exchange trends, and spectral features generated by the parameterization. In this context, it is important to note that the reported structural descriptors, exchange constants, phonons, and DOS all correspond to the converged self-consistent LR-based DFT+$U$+$J$ workflow described in \autoref{subsec:U-J-proc}, rather than to the initial GGA-relaxed structures used only to initialize that iterative procedure.

The inclusion of oxygen in the LR analysis is intentional. In spinel ferrites, the electronic structure and magnetic superexchange are controlled by Fe/Mn\,(3\emph{d}) -- O\,(2\emph{p}) hybridization. Oxygen is, therefore, the dominant screening and mediation channel for the correlated \emph{d} subspaces and exchange pathways. Performing LR on the O\,($2p$) subspace provides a consistent way to quantify how the ligand network participates in screening within the same projector definition used for the cations~\cite{lambertEvaluationFirstprinciplesHubbard2024,linscottRoleSpinCalculation2018,mooreHighthroughputDeterminationHubbard2024}. In \autoref{tab:lr-UJ}, $U_{\rm Oxygen}$ is comparatively large (near 10 eV across all compositions) and varies only weakly from one configuration to another. In the present context, this should be read less as a `standalone oxygen correlation strength' and more as a stability indicator of the ligand screening environment under a fixed LR definition, namely, that the ligand framework remains broadly comparable across the series, while most configuration sensitivity enters via which cations occupy A and B sites and how that redistributes \emph{d}--\emph{p} hybridization locally~\cite{chaiMinimumTrackingLinear2024,gebauerOxygenVacanciesZirconia2023}.

A key point for interpreting \autoref{tab:lr-UJ} is that the LR-extracted $J$ is an effective on-site exchange parameter within the chosen Hubbard functional and projector subspace, not a direct measurement of the bare atomic Hund's coupling. Within LR, $U$ and $J$ are obtained by fitting the curvature of the total energy with respect to controlled perturbations of on-site occupations (and, depending on the implementation, spin-resolved occupations). In strongly covalent or highly screened environments, particularly for \ce{Fe}-$d$ subspaces hybridized with O-$p$, this effective $J$ can become negative without implying a literal reversal of Hund's rule on the atom. Rather, a negative value indicates that, in the effective subspace used by the functional, the exchange-like contribution that would normally be represented by a positive $J$ is already accounted for (or over-screened) by the underlying DFT description and screening so that the LR-consistent corrective term would enter with the opposite sign to reproduce the response curvature. Practically, readers should interpret $U$ and $J$ as a paired parameterization of the on-site correction. In that sense, a negative $J$ tends to increase the net on-site penalty relative to using $U$ alone within the same functional form. In the remainder of this section, attributing atomic `Hund's rule' meaning to the sign of $J$ is, therefore, avoided, and instead how the self-consistent parameter set varies with A/B coordination and Mn/Zn substitution is tracked as is how those changes correlate with the DOS and exchange constants.

The structural descriptors in \autoref{tab:struct-props} summarize both the global geometry and the local coordination landscape of each relaxed 56-atom cubic cell. In these relaxations, the cation framework is held fixed at the chosen A/B occupations, while the oxygen sub-lattice is allowed to relax within the imposed symmetry so that the lattice parameter $a$ and oxygen internal parameter $u$ self-consistently accommodate the cation chemistry and site distribution. Because the relaxed oxygen positions generate a set of Me--O bonds that are generally not all symmetry-equivalent (especially in chemically mixed or partially inverted configurations), each Me--O distance is reported in a statistical form, namely, the mean $\pm$ standard deviation taken over all corresponding Me--O bonds of that species and sub-lattice within the cell. In this representation, the mean captures the typical bond scale relevant to orbital overlap and polyhedral geometry, while the spread quantifies the degree of tetrahedral/octahedral distortion induced by the fixed cation arrangement and the oxygen relaxation. These structural metrics are converged together with the LR parameters through the same relaxation-LR cycle described above so the reported bond-length statistics and the final $(U,J)$ values are mutually consistent descriptors of the same relaxed state. It is important to interpret these structural descriptors in the context of the reference model used here. The constrained relaxations preserve the cubic spinel framework and the prescribed A/B-site occupations, while the relaxed oxygen coordinates and the resulting bond-length distributions quantify the local oxygen-sublattice accommodation within that framework. This representation captures oxygen-driven local distortion on a common crystallographic basis, but it does not exhaust the full distortion space accessible to real materials. Consistent with this limitation, auxiliary \ce{Fe3O4} tests in which the constraints were released drove the structure away from the targeted cubic reference. Accordingly, the present results should be read as properties of the cubic-reference spinel model adopted in this work.

% Table: Structural + magnetic + charge descriptors
\begin{table*}
    \centering
    \caption{Structural and local descriptors for the relaxed spinel cells studied in this work. For each configuration, the lattice parameter $a$ and oxygen internal parameter $u$ are reported, while the bond lengths $d$ of Me-O bond (with Me = Fe, Mn, Zn), magnetic moments $\mu$, and Bader charges $|e|$ are listed in a site/species-resolved form.}
    \label{tab:struct-props}
    \renewcommand\arraystretch{1.15}
    \setlength{\tabcolsep}{3.5pt}
    \begin{tabular}{|c|c|c|c|c|c|}
    \hline
    \textbf{Compound} &
    \textbf{$a$ (\AA)} &
    \textbf{$u$} &
    \textbf{d (\AA)} &
    \textbf{$\mu$ ($\mu_B$)} &
    \textbf{$|e|$} \\
    \hline

    \multicolumn{6}{|l|}{} \\
    \multicolumn{6}{|l|}{\makecell[l]{\textbf{\ce{Fe3O4}}\\ \ }} \\
    \hline
    Prototype &
    $8.391$ &
    $0.2567$ &
    \begin{tabular}[c]{@{}l@{}}
    Fe\textsubscript{A}--O: $1.8347 \pm 0.035$\\
    Fe\textsubscript{B}--O: $2.0653 \pm 0.055$\\
    \end{tabular}
    &
    \begin{tabular}[c]{@{}l@{}}
    Fe\textsubscript{A}: $4.225\, (\uparrow)$ \\
    Fe\textsubscript{B}: $3.812/4.101\, (\downarrow)$ \\
    \end{tabular}
    &
    \begin{tabular}[c]{@{}l@{}}
    Fe\textsubscript{A}: $2.344$ \\
    Fe\textsubscript{B}: $1.991/2.075$ \\
    O: $-1.623$ \\
    \end{tabular}\\
    \hline

    \multicolumn{6}{|l|}{} \\
    \multicolumn{6}{|l|}{\makecell[l]{\textbf{\ce{MnFe2O4}}\\ \ }} \\
    \hline
    \makecell{Config.\,1\\(all Mn on A)} &
    $8.513$ &
    $0.2629$ &
    \begin{tabular}[c]{@{}l@{}}
    Mn\textsubscript{A}--O: $1.9051\, \pm$ 1.7e$-5$ \\
    Fe\textsubscript{B}--O: $2.0680\, \pm$ 1.2e$-5$ \\
    \end{tabular}
    &
    \begin{tabular}[c]{@{}l@{}}
    Mn\textsubscript{A}: $4.650\, (\uparrow)$\\
    Fe\textsubscript{B}: $4.322\, (\downarrow)$\\
    \end{tabular}
    &
    \begin{tabular}[c]{@{}l@{}}
    Mn\textsubscript{A}: $1.581$ \\
    Fe\textsubscript{B}: $1.994$ \\
    O: $-1.429$ \\
    \end{tabular}\\
    \hline

    \makecell{Config.\,2\\(50/50 Mn on A \& B)} &
    $8.520$ &
    $0.2615$ &
    \begin{tabular}[c]{@{}l@{}}
    Mn\textsubscript{A}--O: $2.0463 \pm 0.016$\\
    Fe\textsubscript{A}--O: $1.9232 \pm 0.065$\\
    Mn\textsubscript{B}--O: $2.0687 \pm 0.096$\\
    Fe\textsubscript{B}--O: $2.0439 \pm 0.058$\\
    \end{tabular}
    &
    \begin{tabular}[c]{@{}l@{}}
    Mn\textsubscript{A}: $4.641\, (\uparrow)$\\
    Fe\textsubscript{A}: $4.554\, (\uparrow)$ \\
    Mn\textsubscript{B}: $4.948\, (\downarrow)$ \\
    Fe\textsubscript{B}: $4.378/3.815\, (\downarrow)$ \\
    \end{tabular}
    &
    \begin{tabular}[c]{@{}l@{}}
    Mn\textsubscript{A}: $1.646$ \\
    Fe\textsubscript{A}: $1.907$ \\
    Mn\textsubscript{B}: $1.487$ \\
    Fe\textsubscript{B}: $1.932/1.688$ \\
    O: $-1.401$ \\
    \end{tabular}\\
    \hline

    \multicolumn{6}{|l|}{} \\
    \multicolumn{6}{|l|}{\makecell[l]{\textbf{\ce{(Mn_{0.5},Zn_{0.5})Fe2O4}}\\ \ }} \\
    \hline
    \makecell{Config.\,1\\ (all Zn \& Mn on A)} &
    $8.411$ &
    $0.2582$ &
    \begin{tabular}[c]{@{}l@{}}
    Zn\textsubscript{A}--O: $1.9890 \pm 0.023$ \\
    Mn\textsubscript{A}--O: $1.7893 \pm 0.001$ \\
    Fe\textsubscript{B}--O: $2.0670 \pm 0.066$ \\
    \end{tabular}
    &
    \begin{tabular}[c]{@{}l@{}}
    Mn\textsubscript{A}: $4.585\, (\uparrow)$ \\
    Fe\textsubscript{B}: $4.774\, (\downarrow)$ \\
    \end{tabular}
    &
    \begin{tabular}[c]{@{}l@{}}
    Zn\textsubscript{A}: $1.202$ \\
    Mn\textsubscript{A}: $1.593$ \\
    Fe\textsubscript{B}: $1.976$ \\
    O: $-1.397$ \\
    \end{tabular}\\
    \hline
    \makecell{Config.\,2\\ (all Zn on A, \\all Mn on B)} &
    8.448 &
    0.2587 &
    \begin{tabular}[c]{@{}l@{}}
    Zn\textsubscript{A}--O: $1.9821 \pm 0.019$ \\
    Fe\textsubscript{A}--O: $1.8981 \pm 0.042$ \\
    Mn\textsubscript{B}--O: $2.0352 \pm 0.086$ \\
    Fe\textsubscript{B}--O: $2.0561 \pm 0.051$ \\
    \end{tabular}
    &
    \begin{tabular}[c]{@{}l@{}}
    Fe\textsubscript{A}: $4.265\, (\uparrow)$ \\
    Mn\textsubscript{B}: $4.644\, (\downarrow)$ \\
    Fe\textsubscript{B}: $4.298/3.758\, (\downarrow)$ \\
    \end{tabular}
    &
    \begin{tabular}[c]{@{}l@{}}
    Zn\textsubscript{A}: $1.213$ \\
    Fe\textsubscript{A}: $1.839$ \\
    Mn\textsubscript{B}: $1.522$ \\
    Fe\textsubscript{B}: $1.884/1.624$ \\
    O: $-1.456$ \\
    \end{tabular}\\
    \hline
    \makecell{Config.\,3\\ (all Zn on A, \\50/50 Mn on A \& B)} &
    $8.414$ &
    $0.2612$ &
    \begin{tabular}[c]{@{}l@{}}
    Zn\textsubscript{A}--O: $1.9818 \pm 0.017$ \\
    Mn\textsubscript{A}--O: $2.0318 \pm 0.018$ \\
    Fe\textsubscript{A}--O: $1.9467 \pm 0.052$ \\
    Mn\textsubscript{B}--O: $2.0421 \pm 0.077$ \\
    Fe\textsubscript{B}--O: $2.0299 \pm 0.037$ \\
    \end{tabular}
    &
    \begin{tabular}[c]{@{}l@{}}
    Mn\textsubscript{A}: $4.629\, (\uparrow)$ \\
    Fe\textsubscript{A}: $4.237\, (\uparrow)$ \\
    Mn\textsubscript{B}: $4.651\, (\downarrow)$ \\
    Fe\textsubscript{B}: $4.353\, (\downarrow)$ \\
    \end{tabular}
    &
    \begin{tabular}[c]{@{}l@{}}
    Zn\textsubscript{A}: $1.211$ \\
    Mn\textsubscript{A}: $1.626$ \\
    Fe\textsubscript{A}: $1.852$ \\
    Mn\textsubscript{B}: $1.591$ \\
    Fe\textsubscript{B}: $1.891$ \\
    O: $-1.422$ \\
    \end{tabular}\\
    \hline

    \end{tabular}
\end{table*}

Several trends emerge directly from \autoref{tab:struct-props}. Substituting Mn into the spinel expands the lattice relative to \ce{Fe3O4}: $a$ increases from 8.391~\AA\ in \ce{Fe3O4} to $8.51$-$8.52$~\AA\ in \ce{MnFe2O4}, accompanied by an increase in $u\  (0.2567\, \to\, 0.261$--$0.263)$, indicating a modified oxygen framework and, hence,  an altered Me--O--Me geometry, consistent with a more expanded cation-oxygen skeleton. The ordered \ce{MnFe2O4} configuration with Mn exclusively on A sites yields essentially single-valued Mn\textsubscript{A}--O and Fe\textsubscript{B}--O distances (spreads $\sim 10^{-5}$ ~\AA), which is a direct consequence of the constrained relaxation. With a uniform cation environment and symmetry-preserving oxygen relaxation, the oxygen sub-lattice converges to nearly equivalent coordination polyhedra. By contrast, for the mixed A/B configuration, the Me--O distributions broaden substantially (Mn\textsubscript{A}--O: $2.0687 \pm 0.096$~\AA; and Fe\textsubscript{B}--O: $1.9232 \pm 0.065$~\AA), reflecting oxygen-sub-lattice distortions induced by chemical heterogeneity within the same coordination network. Across the Mn--Zn ferrite configurations, $a$ and $u$ shift modestly relative to \ce{Fe3O4}, but the bond-length spreads again consistent with how Mn is partitioned between A and B sites, signaling how the oxygen network accommodates competing site preferences. 

Bader charges are reported as well as a post-processing descriptor of charge redistribution. In Bader analysis, the total electron density is partitioned into atomic basins bounded by zero-flux surfaces in $\nabla \rho (\bm{r})$, and the integrated charge in each basin yields an `atomic' charge associated with that site. While Bader charges are not formal oxidation states and depend on the chosen density and partitioning scheme, they provide a consistent, geometry-sensitive proxy for how charge transfer and covalency trends evolve across compositions and configurations when the same workflow is used throughout~\cite{baderAtomsMoleculesQuantum1994}. In the present study they are particularly useful because the dominant physics (screening, \emph{d}--\emph{p} hybridization, and superexchange) depends on how electron density is shared between cations and oxygen.  Bader charges, therefore, complement the LR parameters by providing an independent, density-based view of how substitution and A/B occupancy shift the electronic environment.

In \ce{Fe3O4}, oxygen carries the most negative Bader charge in the series (O: $-1.623$), while the cations span a broader range (Fe(A): $2.344$; Fe(B): $1.99$ to $2.08$), consistent with non-equivalent Fe environments on the A and B sub-lattices. Upon Mn and/or Zn substitution, oxygen becomes systematically less negative (typically $\sim -1.40$ to $-1.46$), indicating a redistribution of charge density consistent with altered covalency and screening in the Me--O network under the same partitioning scheme. Zn carries the smallest positive Bader charge ($\sim 1.20$), consistent with its closed-shell character in the present bonding environment, while Mn sits near $\sim 1.5$ to 1.65. The Fe charges shift with site and configuration (e.g., Fe(A) $\sim 1.84$ to  1.91 in the Mn--Zn cases), reinforcing that A/B occupancy changes not only geometry but also the local electronic environment in a way that is consistent with the environment-dependent LR parameters reported in \autoref{tab:lr-UJ}. The magnetic moments remain in the high-spin range expected for Mn and Fe in these coordinations. Taken together, the bond-length statistics and Bader charges show that `inversion' is not merely a bookkeeping label, it induces measurable oxygen-sub-lattice distortions and charge redistribution that ultimately reshape exchange pathways.

% Table: NN Heisenberg exchange constants
\begin{table*}
    \centering
    \caption{Nearest-neighbor (NN) Heisenberg exchange constants for the relaxed spinel cells studied in this work. All $J_{ij}$ values are extracted by the same collinear energy-mapping protocol on the same 56-atom cubic cell for every compound and configuration. The convention of \autoref{eq:Heisenberg-classical} is used. Thus, $J>0$ favors ferromagnetic alignment and $J<0$ favors antiferromagnetic alignment. Where multiple species-resolved NN channels exist within the same exchange class, they are listed as multiple entries within the cell.}
    \label{tab:nn-exchange}
    \renewcommand\arraystretch{1.15}
    \setlength{\tabcolsep}{6pt}
    \begin{tabular}{|c|c|c|c|}
    \hline
    \textbf{Compound} &
    \textbf{$J_{AB}$ (meV)} &
    \textbf{$J_{BB}$ (meV)} &
    \textbf{$J_{AA}$ (meV)} \\
    \hline

    \multicolumn{4}{|l|}{} \\
    \multicolumn{4}{|l|}{\makecell[l]{\textbf{\ce{Fe3O4}}\\ \ }} \\
    \hline
    Prototype &
    \begin{tabular}{@{}l@{}}
    Fe\textsubscript{A}--Fe\textsubscript{B}: $-2.38$\\
    \end{tabular}
    &
    \begin{tabular}{@{}l@{}}
    Fe\textsubscript{B}--Fe\textsubscript{B}: $0.56$ \\
    \end{tabular}
    &
    \begin{tabular}{@{}l@{}}
    Fe\textsubscript{A}--Fe\textsubscript{A}: $-0.26$\\
    \end{tabular}
    \\
    \hline

    \multicolumn{4}{|l|}{} \\
    \multicolumn{4}{|l|}{\makecell[l]{\textbf{\ce{MnFe2O4}}\\ \ }} \\
    \hline
    \makecell{Config.\,1\\(all Mn on A)} &
    \begin{tabular}{@{}l@{}}
    Mn\textsubscript{A}--Fe\textsubscript{B}: $-1.88$\\
    \end{tabular}
    &
    \begin{tabular}{@{}l@{}}
    Fe\textsubscript{B}--Fe\textsubscript{B}: $1.18$\\
    \end{tabular}
    &
    \begin{tabular}{@{}l@{}}
    Mn\textsubscript{A}--Mn\textsubscript{A}: $0.74$ \\
    \end{tabular}

    \\
    \hline

    \makecell{Config.\,2\\(50/50 Mn on A \& B)} &
    \begin{tabular}{@{}l@{}}
    Mn\textsubscript{A}--Fe\textsubscript{B}: $-2.068$ \\
    Fe\textsubscript{A}--Mn\textsubscript{B}: $-1.541$ \\
    Fe\textsubscript{A}--Fe\textsubscript{B}: $-1.126$ \\
    Mn\textsubscript{A}--Mn\textsubscript{B}: $-1.730$ \\
    \end{tabular}
    &
    \begin{tabular}{@{}l@{}}
    Mn\textsubscript{B}--Fe\textsubscript{B}: $0.93$ \\
    Mn\textsubscript{B}--Mn\textsubscript{B}: $1.18$ \\
    Fe\textsubscript{B}--Fe\textsubscript{B}: $2.03$ \\
    \end{tabular}
    &
    \begin{tabular}{@{}l@{}}
    Mn\textsubscript{A}--Fe\textsubscript{A}: $-0.020$ \\
    Mn\textsubscript{A}--Mn\textsubscript{A}: $-0.061$ \\
    Fe\textsubscript{A}--Fe\textsubscript{A}:  $0.105$ \\
    \end{tabular}
    \\
    \hline

    \multicolumn{4}{|l|}{} \\
    \multicolumn{4}{|l|}{\makecell[l]{\textbf{\ce{(Mn_{0.5},Zn_{0.5})Fe2O4}}\\ \ }} \\
    \hline
    \makecell{Config.\,1\\ (all Zn \& Mn on A)} &
    \begin{tabular}{@{}l@{}}
    Mn\textsubscript{A}--Fe\textsubscript{B}: $-1.94$ \\
    \end{tabular}
    &
    \begin{tabular}{@{}l@{}}
    Fe\textsubscript{B}--Fe\textsubscript{B}: $2.04$ \\
    \end{tabular}
    &
    \begin{tabular}{@{}l@{}}
    Mn\textsubscript{A}--Mn\textsubscript{A}: $-0.004$ \\
    \end{tabular}
    \\
    \hline

    \makecell{Config.\,2\\ (all Zn on A, \\all Mn on B)} &
    \begin{tabular}{@{}l@{}}
    Fe\textsubscript{A}--Mn\textsubscript{B}: $-1.656$ \\
    Fe\textsubscript{A}--Fe\textsubscript{B}: $-1.382$ \\
    \end{tabular}
    &
    \begin{tabular}{@{}l@{}}
    Mn\textsubscript{B}--Fe\textsubscript{B}: $1.30$ \\
    Mn\textsubscript{B}--Mn\textsubscript{B}: $0.83$ \\
    Fe\textsubscript{B}--Fe\textsubscript{B}: $0.92$ \\
    \end{tabular}
    & 
    \begin{tabular}{@{}l@{}}
    Fe\textsubscript{A}--Fe\textsubscript{A}: $-0.07$ \\
    \end{tabular}
    \\
    \hline

    \makecell{Config.\,3\\ (all Zn on A, \\50/50 Mn on A \& B)} &
    \begin{tabular}{@{}l@{}}
    Mn\textsubscript{A}--Fe\textsubscript{B}: $-2.036$ \\
    Mn\textsubscript{A}--Mn\textsubscript{B}: $-1.247$ \\
    Fe\textsubscript{A}--Fe\textsubscript{B}: $-1.109$ \\
    Fe\textsubscript{A}--Mn\textsubscript{B}: $-1.444$ \\
    \end{tabular}
    &
    \begin{tabular}{@{}l@{}}
    Mn\textsubscript{B}--Fe\textsubscript{B}: $0.79$ \\
    Mn\textsubscript{B}--Mn\textsubscript{B}: $1.37$ \\
    Fe\textsubscript{B}--Fe\textsubscript{B}: $1.96$ \\
    \end{tabular}
    &
    \begin{tabular}{@{}l@{}}
    Mn\textsubscript{A}--Mn\textsubscript{A}: $-0.024$ \\
    Mn\textsubscript{A}--Fe\textsubscript{A}: $0.018$ \\
    Fe\textsubscript{A}--Fe\textsubscript{A}: $-0.008$ \\
    \end{tabular}
    \\
    \hline

    \end{tabular}
\end{table*}

\autoref{tab:nn-exchange} translates the relaxed local environments into nearest-neighbor Heisenberg couplings. To avoid ambiguities from differing normalizations, all exchange constants reported are extracted by the same collinear energy-mapping procedure applied to the same 56-atom cubic cell for every case, and the sign convention and units (meV) are kept identical throughout. Details of the mapping equations and alternative normalizations are deferred to \autoref{subsec:exchange-calculations}. Accordingly, the exchange constants reported below should be interpreted as exchange parameters of the common cubic-reference spinel model used throughout this study, rather than of a fully unconstrained low-symmetry structural minimum. Across every composition and configuration, the dominant interaction is the A--B channel, and it is consistently antiferromagnetic: $J_{AB}$ remains negative and of largest magnitude (e.g., $-2.38$ meV in \ce{Fe3O4}, $-1.88$ meV in the low-inversion \ce{MnFe2O4} configuration, and $\sim -1.38$ to $-2.04$ meV across the Mn--Zn configurations). This uniform sign and scale establishes the ferrimagnetic backbone of the series and provides a direct structural-magnetic link, namely, changes in which species occupy A and B sites primarily modulate the strength of the A--B exchange channel rather than its character. The same-sub-lattice interactions are smaller but not negligible and show stronger chemistry dependence. The B--B couplings are uniformly ferromagnetic in this dataset and can become comparable to $|J_{AB}|$ in several mixed cases (e.g., $J_{BB} = 2.03$ meV for Fe(B)--Fe(B) in mixed \ce{MnFe2O4}, and 2.04 meV in \ce{(Mn_{0.5},Zn_{0.5})Fe2O4} Config.~1), indicating that the B sub-lattice network stiffens substantially in some chemically/structurally distinct octahedral environments. In contrast, the A--A interactions are generally small in magnitude in mixed configurations (typically near zero with mixed signs), but become strongly configuration-dependent when A is occupied by a single magnetic species, as seen in the ferromagnetic Mn(A)--Mn(A) coupling ($J_{AA} = 0.74$ meV) in the low-inversion \ce{MnFe2O4} configuration. Where multiple species-resolved nearest-neighbor channels exist within the same exchange class, the table should be read channel-by-channel. The intra-class differences are precisely the fingerprint of local chemical heterogeneity in the fixed 56-atom cell.

The values in \autoref{tab:nn-exchange} are also broadly consistent with the experimental hierarchy long established for spinel ferrites. For cubic \ce{Fe3O4}, classic neutron-scattering measurements identified a dominant A--B exchange scale of about $J_{AB}\!\approx\!-2.3$ to $-2.4$ meV, together with a much smaller B--B contribution and little sensitivity to the A--A term \cite{osti_4763367, glasser1963spin} In that sense, the present \ce{Fe3O4} result, with $J_{AB}=-2.38$ meV and substantially smaller same-sublattice terms, is in very good agreement with the experimental spin-wave picture: the ferrimagnetic backbone is controlled primarily by antiferromagnetic A--B superexchange, while the A--A and B--B channels act as secondary corrections. For \ce{MnFe2O4}, the experimental microscopic literature is less complete, but it points in the same qualitative direction. Inelastic neutron scattering on acoustic magnons in \ce{MnFe2O4} resulted in effective $J_{AB}$ and $J_{BB}$ values derived directly from the measured spin-wave branches, providing experimental precedent for exchange-parameter extraction in manganese ferrite \cite{wegenerInelasticNeutronScattering1974}. Although those measurements do not furnish a universally adopted full $\{J_{AB},J_{AA},J_{BB}\}$ set comparable to the present configuration-resolved mapping, they support the same qualitative hierarchy found here: the dominant interaction is the antiferromagnetic A--B channel, whereas the same-sublattice couplings are weaker and more sensitive to the detailed cation arrangement. This trend is also consistent with broader experimental analyses of spinel ferrites based on temperature-dependent magnetization and susceptibility, which found that $|J_{AB}|$ generally exceeds $|J_{AA}|$ and $|J_{BB}|$ \cite{srivastavaExchangeConstantsSpinel1979}. For the mixed Mn--Zn ferrites, direct microscopic exchange constants for the exact composition/configuration studied here are much less standardized experimentally than for \ce{Fe3O4}, and the reported magnetic behavior is known to depend strongly on cation distribution, composition, and processing route. Accordingly, the present Mn--Zn values are best interpreted as a configuration-specific theoretical baseline, which preserve the experimentally expected dominance of the A--B superexchange channel while quantifying how Zn/Mn partitioning reorganizes the weaker A--A and B--B interactions within a common crystallographic and computational framework.

With the geometric descriptors and nearest-neighbor exchange constants established on a consistent footing, the resulting electronic, vibrational, and magnon spectra across the same set of relaxed 56-atom cubic cells are compared next. \\

\subsection{Electron density of states}

\begin{figure*}
\centering
\vspace{-2em}
\begin{subfigure}{0.45\textwidth}
    \includegraphics[width=\textwidth]{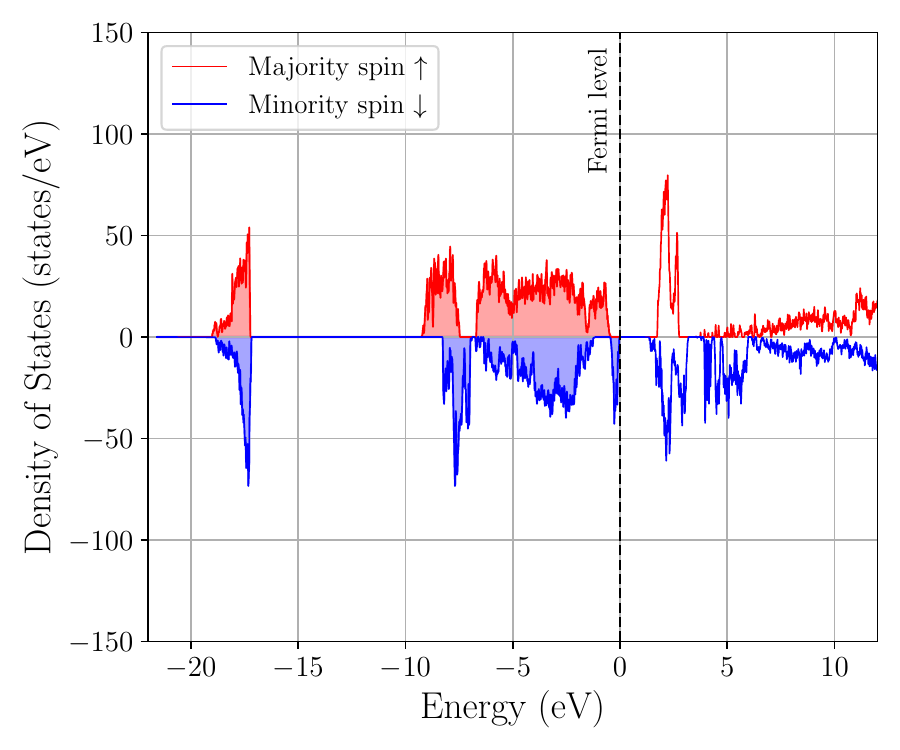}
    \caption{}\label{subfig:el_a}
\end{subfigure}%\hspace{1em}
\begin{subfigure}{0.45\textwidth}
    \includegraphics[width=\textwidth]{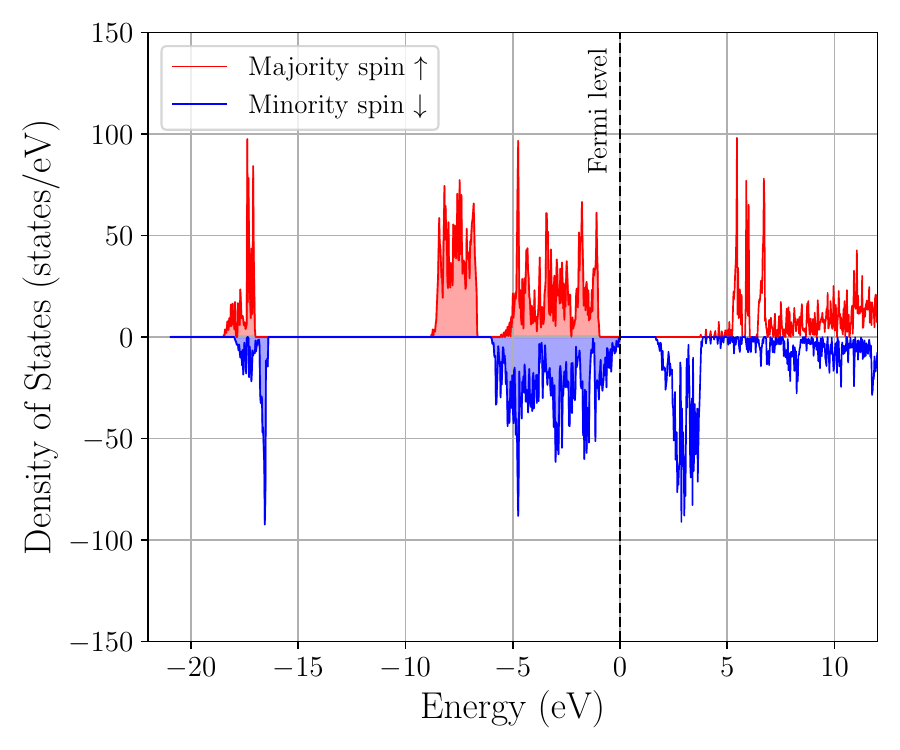}
    \caption{}\label{subfig:el_b}
\end{subfigure}\\
\vspace{0.5em}
\begin{subfigure}{0.45\textwidth}
    \includegraphics[width=\textwidth]{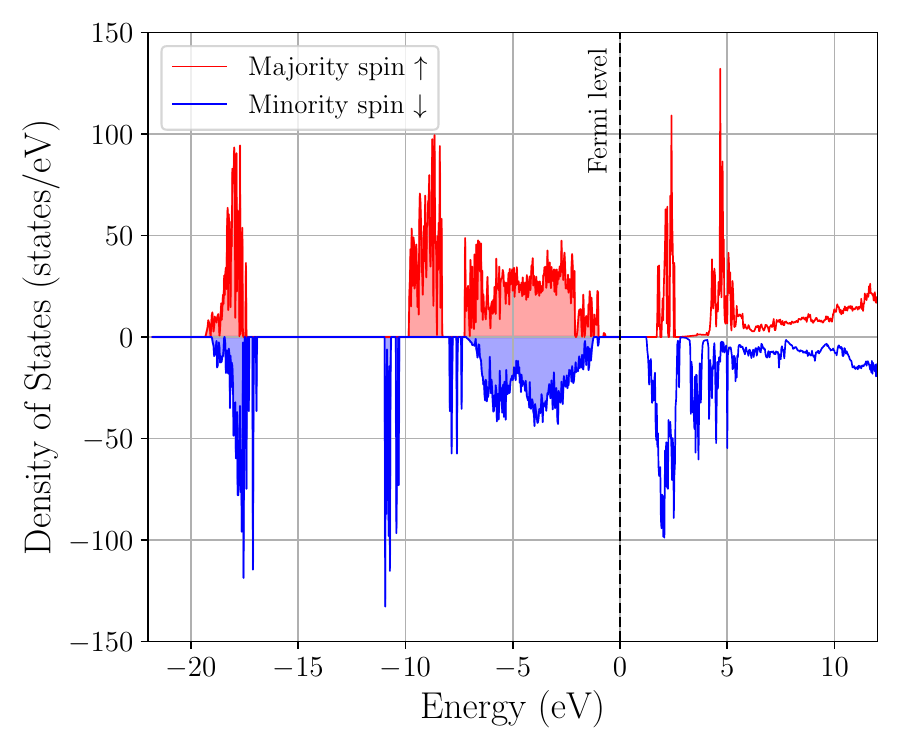}
    \caption{}\label{subfig:el_c}
\end{subfigure}%\hspace{2em}
\begin{subfigure}{0.45\textwidth}
    \includegraphics[width=\textwidth]{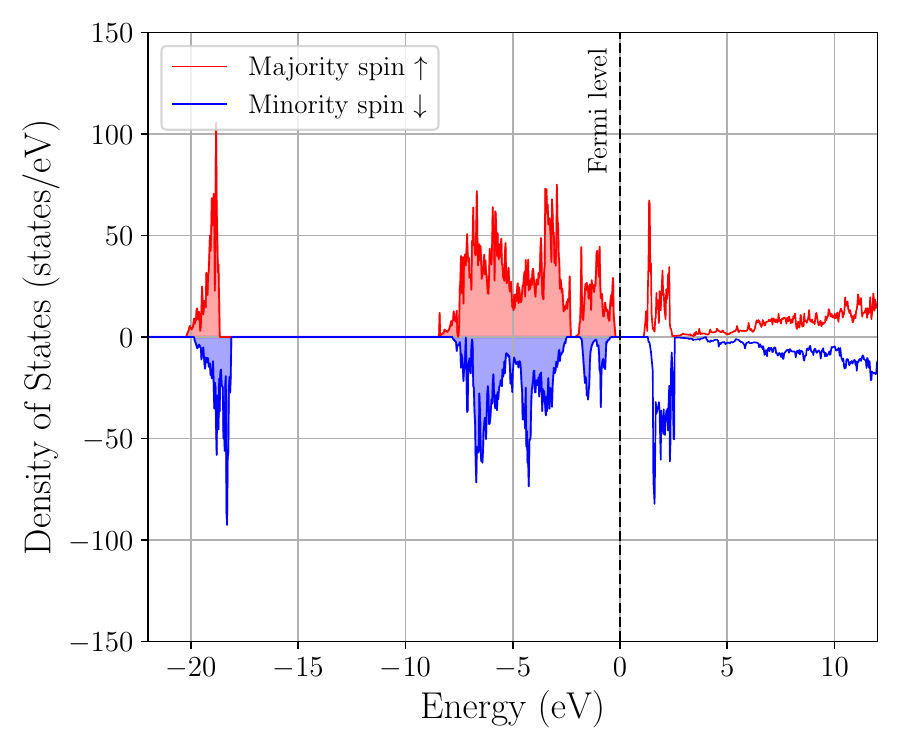}
    \caption{}\label{subfig:el_d}
\end{subfigure}\\
\vspace{0.5em}
\begin{subfigure}{0.45\textwidth}
    \includegraphics[width=\textwidth]{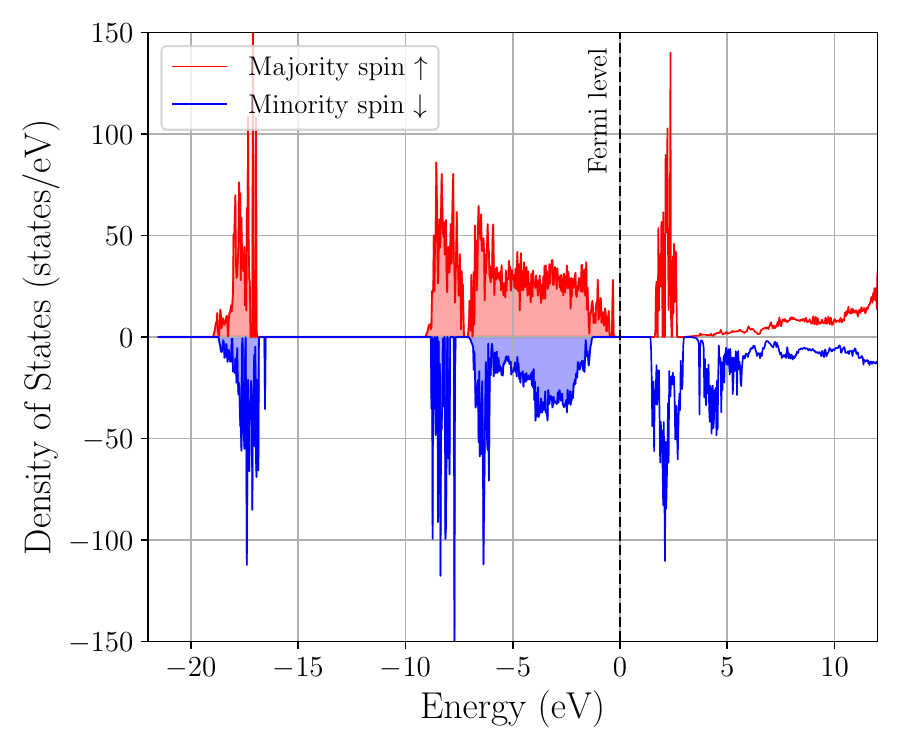}
    \caption{}\label{subfig:el_e}
\end{subfigure}%\hspace{2em}
\begin{subfigure}{0.45\textwidth}
    \includegraphics[width=\textwidth]{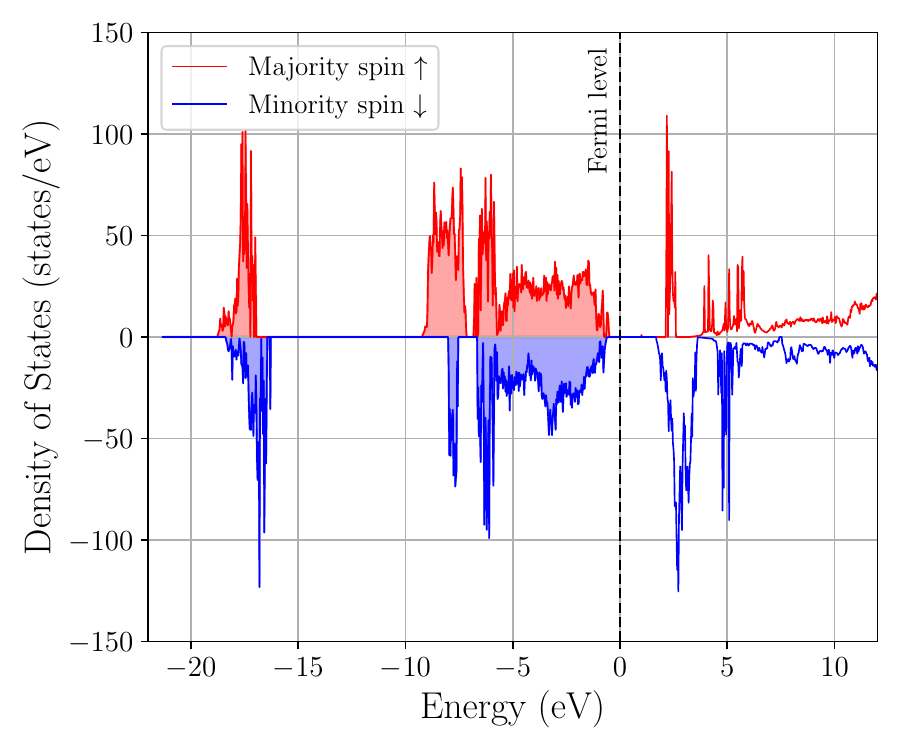}
    \caption{}\label{subfig:el_f}
\end{subfigure}

    \caption{\small \em Electron DOS for spinel ferrites with different cationic composition and configuration. (a) archetypal \ce{Fe3O4}, (b) \ce{MnFe2O4} Config.\,1 where all Mn atoms occupy A-sites of the spinel structure, (c) \ce{MnFe2O4} Config.\,2 where half of the Mn atoms occupy A-sites and the other half occupy B-sites, (d) \ce{(Mn_{0.5},Zn_{0.5})Fe2O4} Config.\,1 where all Zn \& Mn atoms occupy A-sites, (e) \ce{(Mn_{0.5},Zn_{0.5})Fe2O4} Config.\,2 where all Zn atoms occupy A-sites \& all Mn atoms occupy B-sites, (f) \ce{(Mn_{0.5},Zn_{0.5})Fe2O4} Config.\,3 where all Zn atoms occupy A-sites \& the Mn atoms occupy half A-sites and half B-sites.}  
\label{fig:figure3}
\end{figure*}

\begin{figure*}
    \centering
    \begin{subfigure}{0.45\textwidth}
        \includegraphics[width=\textwidth]{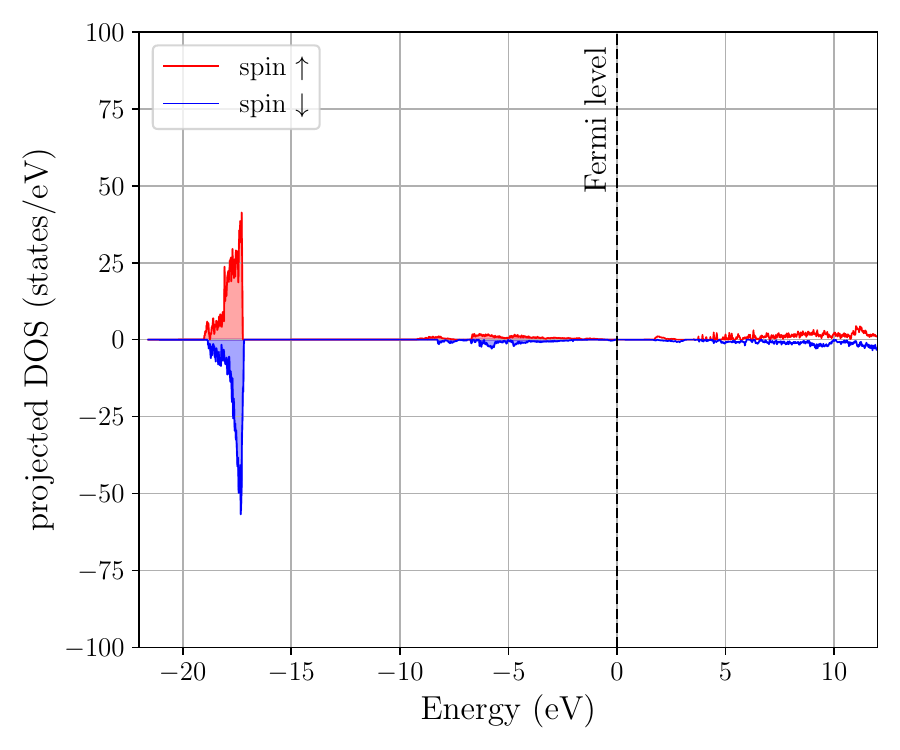}
        \caption{}\label{subfig:Fe3O4_s}
    \end{subfigure}\\
    \begin{subfigure}{0.45\textwidth}
        \includegraphics[width=\textwidth]{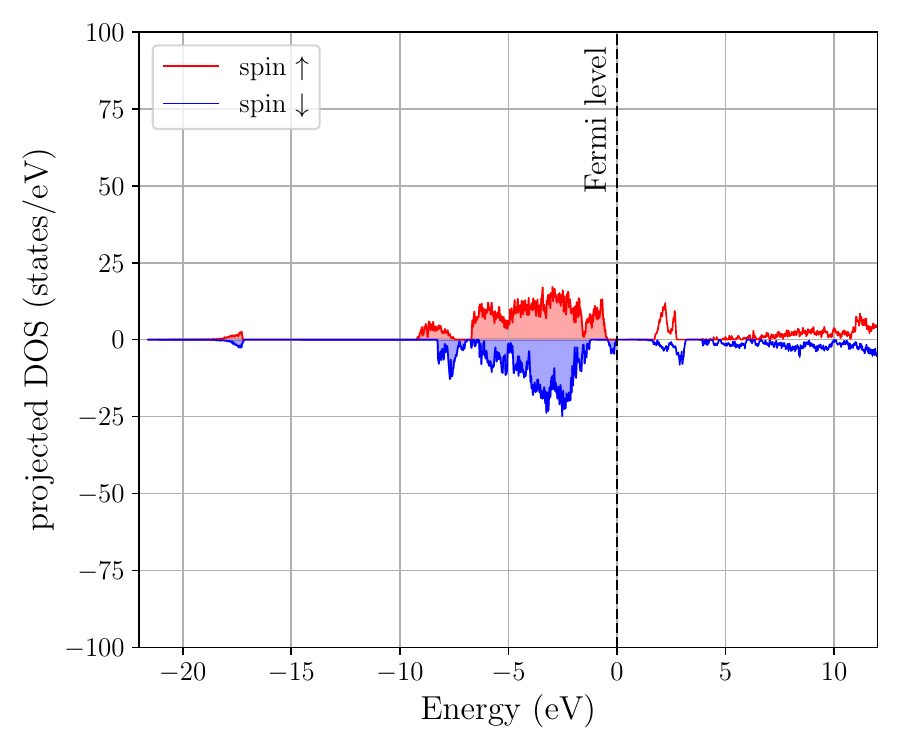}
        \caption{}\label{subfig:Fe3O4_p}
    \end{subfigure}\\
    \begin{subfigure}{0.45\textwidth}
        \includegraphics[width=\textwidth]{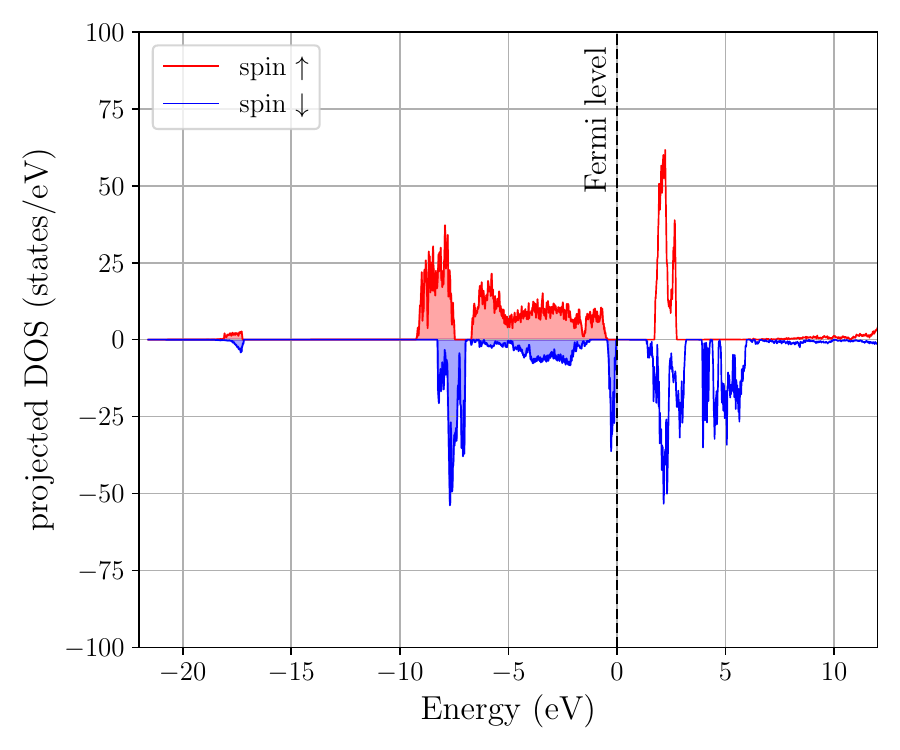}
        \caption{}\label{subfig:Fe3O4_d}
    \end{subfigure}
\caption{\small \em Orbital-projected electron DOS for \ce{Fe3O4}: (a) \(s\)-projected DOS, (b) \(p\)-projected DOS, and (c) \(d\)-projected DOS.}
\label{fig:figure3_1}
\end{figure*}

\begin{figure*}
    \centering
    \begin{subfigure}{0.45\textwidth}
        \includegraphics[width=\textwidth]{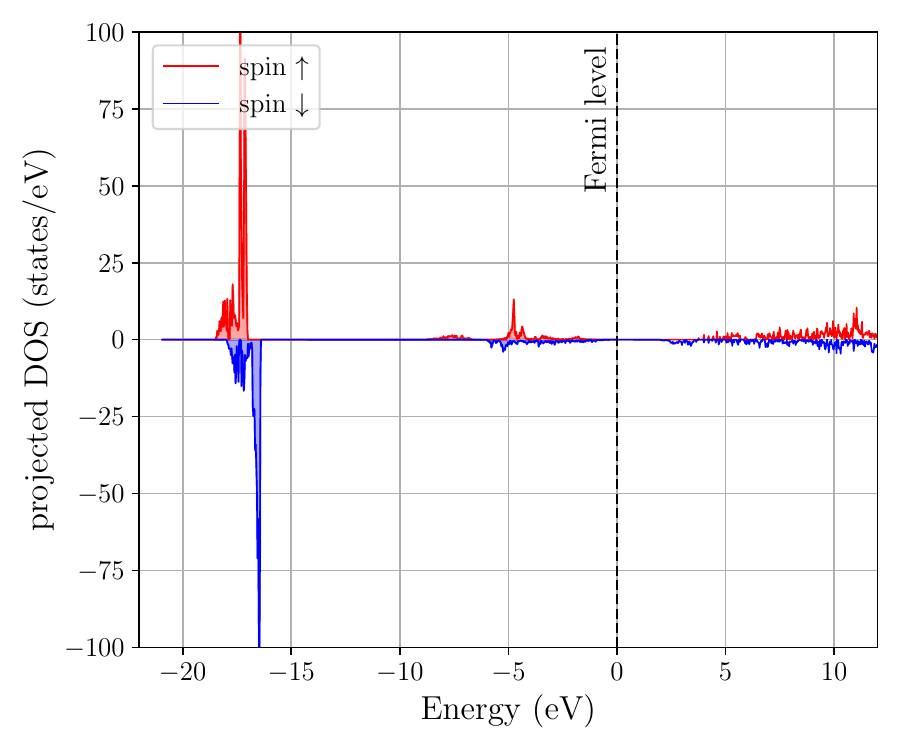}
        \caption{}\label{subfig:MnFe2O4_1_s}
    \end{subfigure}\\
    \begin{subfigure}{0.45\textwidth}
        \includegraphics[width=\textwidth]{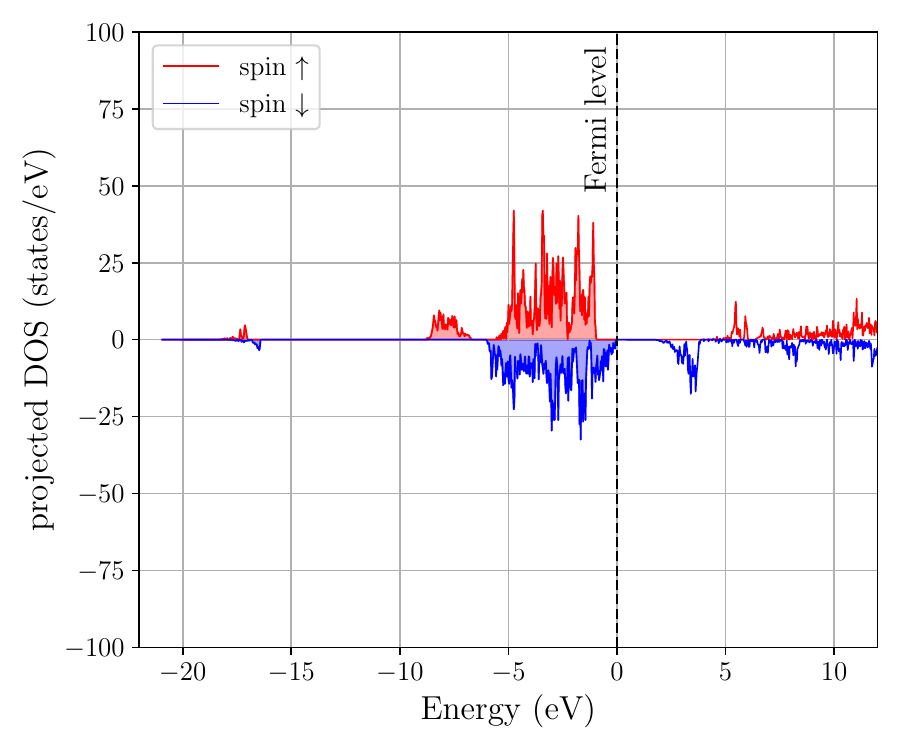}
        \caption{}\label{subfig:MnFe2O4_1_p}
    \end{subfigure}\\
    \begin{subfigure}{0.45\textwidth}
        \includegraphics[width=\textwidth]{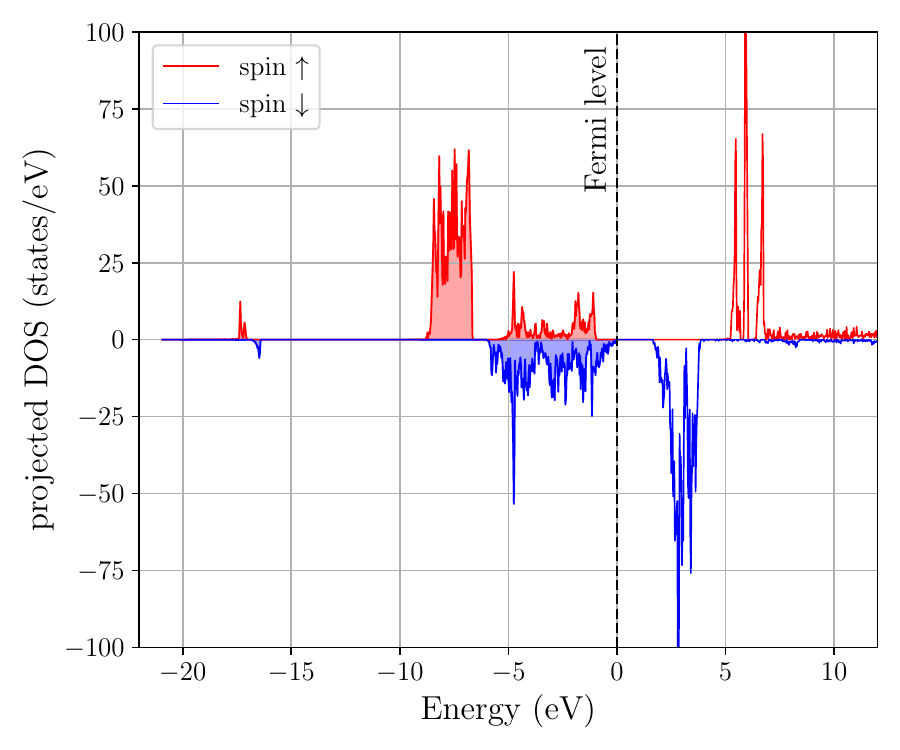}
        \caption{}\label{subfig:MnFe2O4_1_d}
    \end{subfigure}
\caption{\small \em Orbital-projected electron DOS for \ce{MnFe2O4}--Config.\,1: (a) \(s\)-projected DOS, (b) \(p\)-projected DOS, and (c) \(d\)-projected DOS.}
\label{fig:figure3_2}
\end{figure*}

\begin{figure*}
    \centering
    \begin{subfigure}{0.45\textwidth}
        \includegraphics[width=\textwidth]{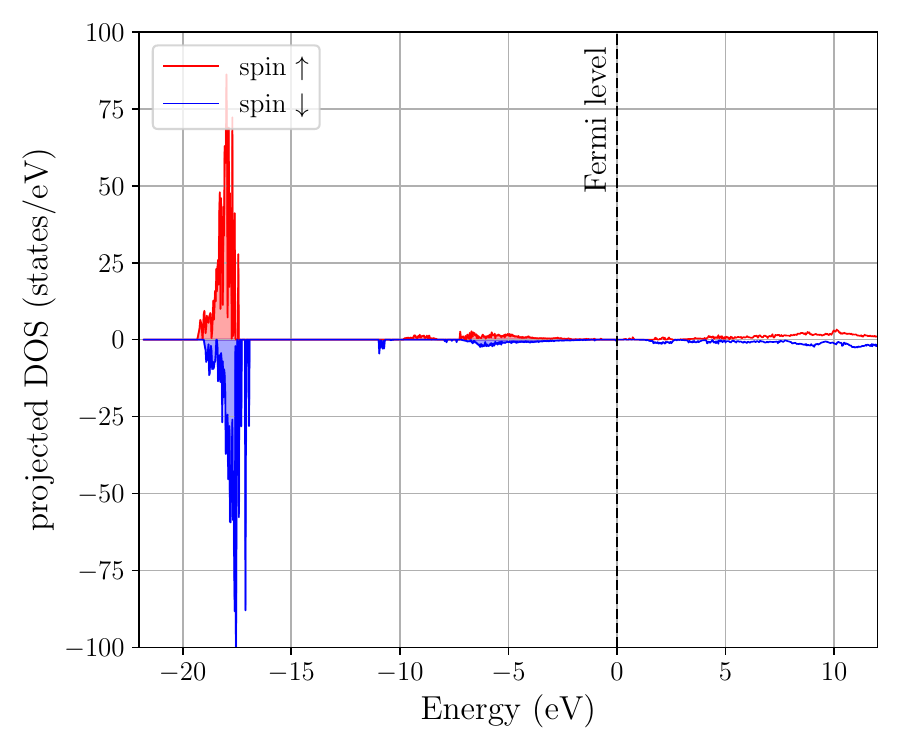}
        \caption{}\label{subfig:MnFe2O4_2_s}
    \end{subfigure}\\
    \begin{subfigure}{0.45\textwidth}
        \includegraphics[width=\textwidth]{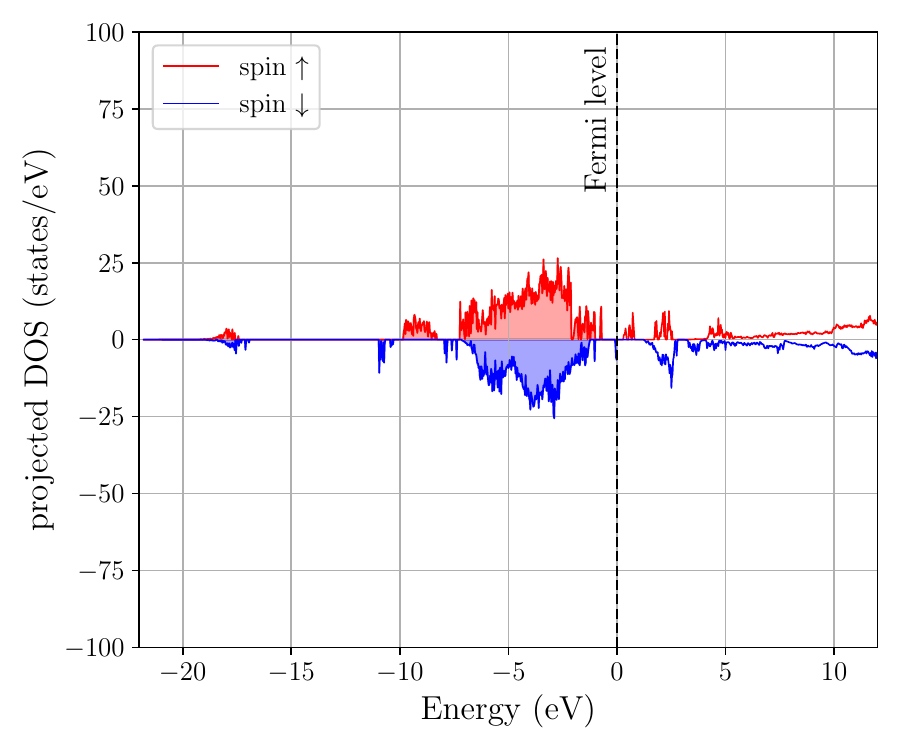}
        \caption{}\label{subfig:MnFe2O4_2_p}
    \end{subfigure}\\
    \begin{subfigure}{0.45\textwidth}
        \includegraphics[width=\textwidth]{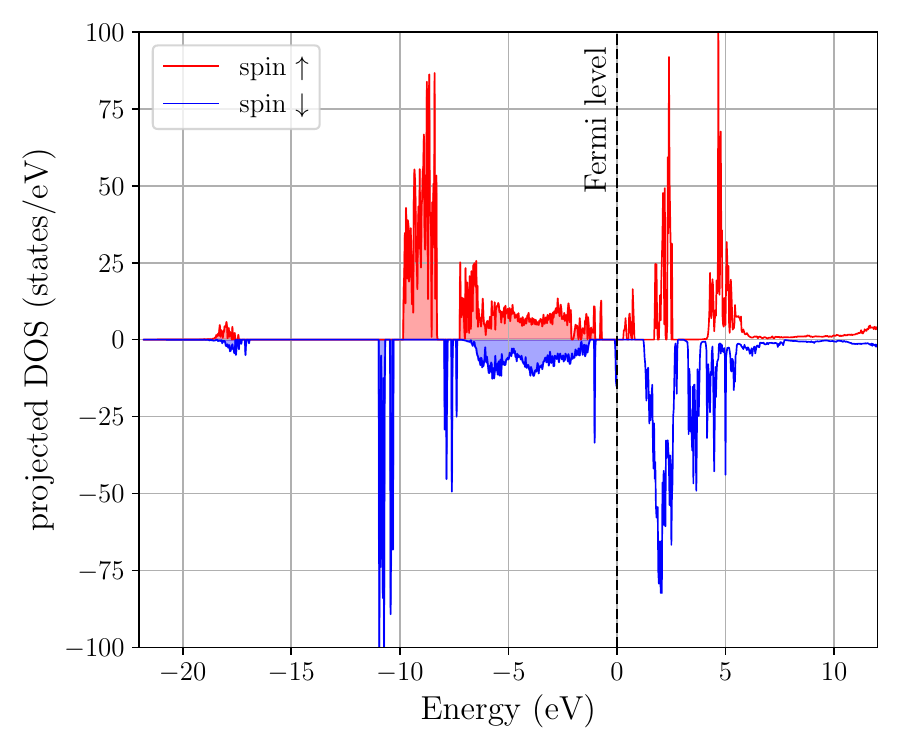}
        \caption{}\label{subfig:MnFe2O4_2_d}
    \end{subfigure}
\caption{\small \em Orbital-projected electron DOS for \ce{MnFe2O4}--Config.\,2: (a) \(s\)-projected DOS, (b) \(p\)-projected DOS, and (c) \(d\)-projected DOS.}
\label{fig:figure3_3}
\end{figure*}

\begin{figure*}
    \centering
    \begin{subfigure}{0.45\textwidth}
        \includegraphics[width=\textwidth]{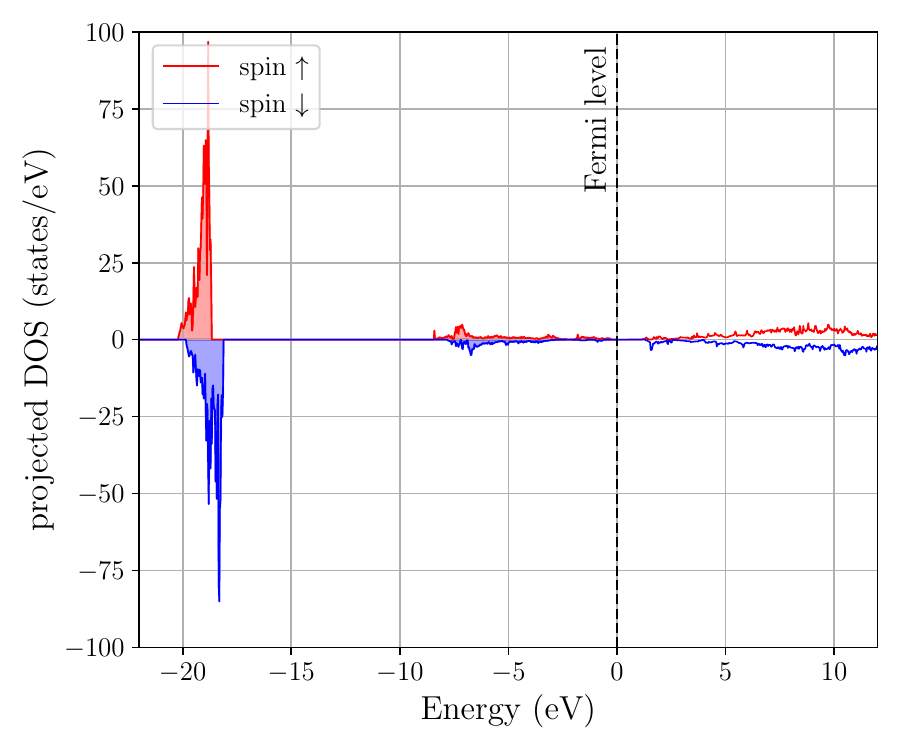}
        \caption{}\label{subfig:MZF_A_s}
    \end{subfigure}\\
    \begin{subfigure}{0.45\textwidth}
        \includegraphics[width=\textwidth]{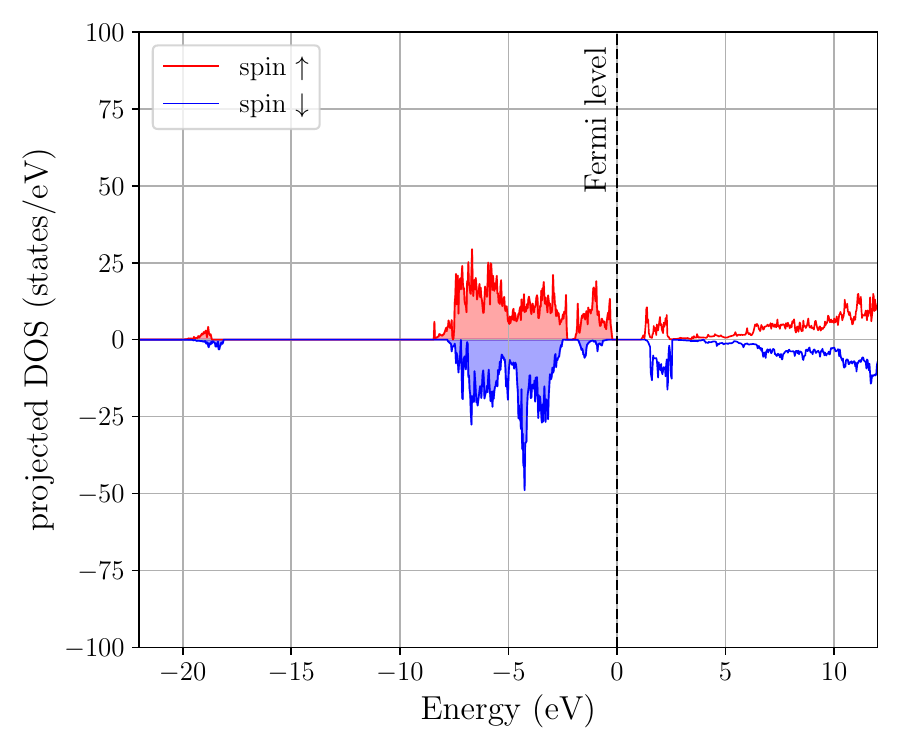}
        \caption{}\label{subfig:MZF_A_p}
    \end{subfigure}\\
    \begin{subfigure}{0.45\textwidth}
        \includegraphics[width=\textwidth]{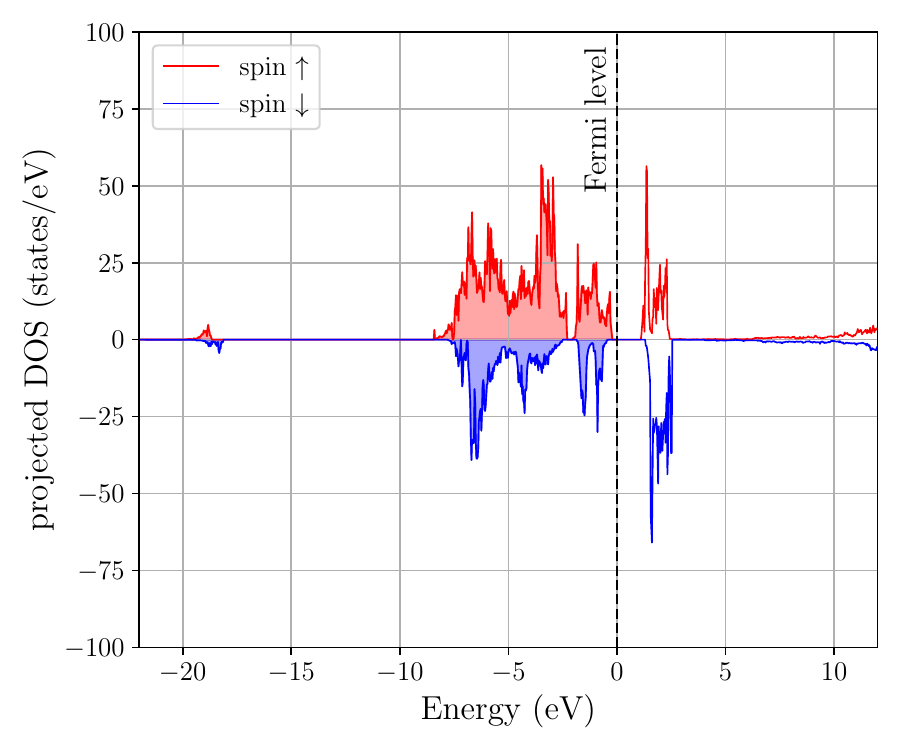}
        \caption{}\label{subfig:MZF_A_d}
    \end{subfigure}
\caption{\small \em Orbital-projected electron DOS for \ce{(Mn_{0.5},Zn_{0.5})Fe2O4}--Config.\,1: (a) \(s\)-projected DOS, (b) \(p\)-projected DOS, and (c) \(d\)-projected DOS.}
\label{fig:figure3_4}
\end{figure*}

\begin{figure*}
    \centering
    \begin{subfigure}{0.45\textwidth}
        \includegraphics[width=\textwidth]{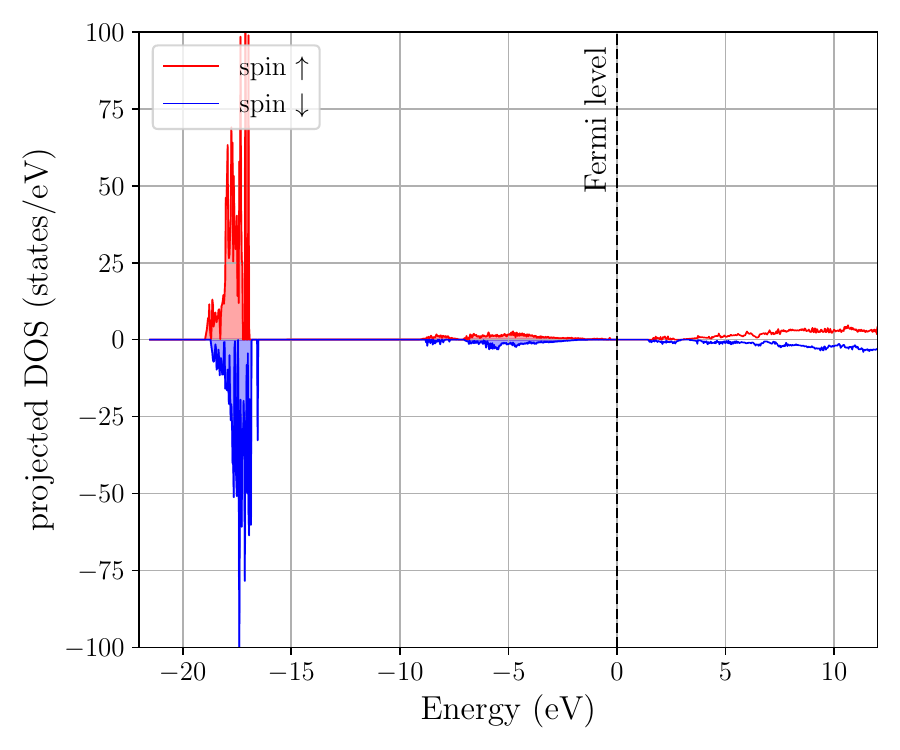}
        \caption{}\label{subfig:MZF_B_s}
    \end{subfigure}\\
    \begin{subfigure}{0.45\textwidth}
        \includegraphics[width=\textwidth]{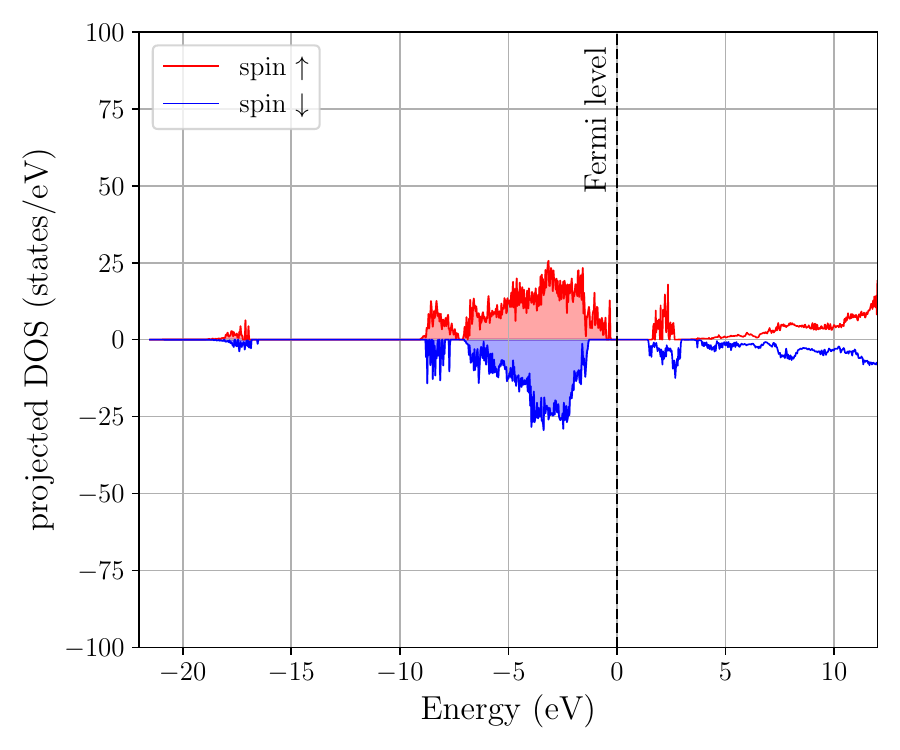}
        \caption{}\label{subfig:MZF_B_p}
    \end{subfigure}\\
    \begin{subfigure}{0.45\textwidth}
        \includegraphics[width=\textwidth]{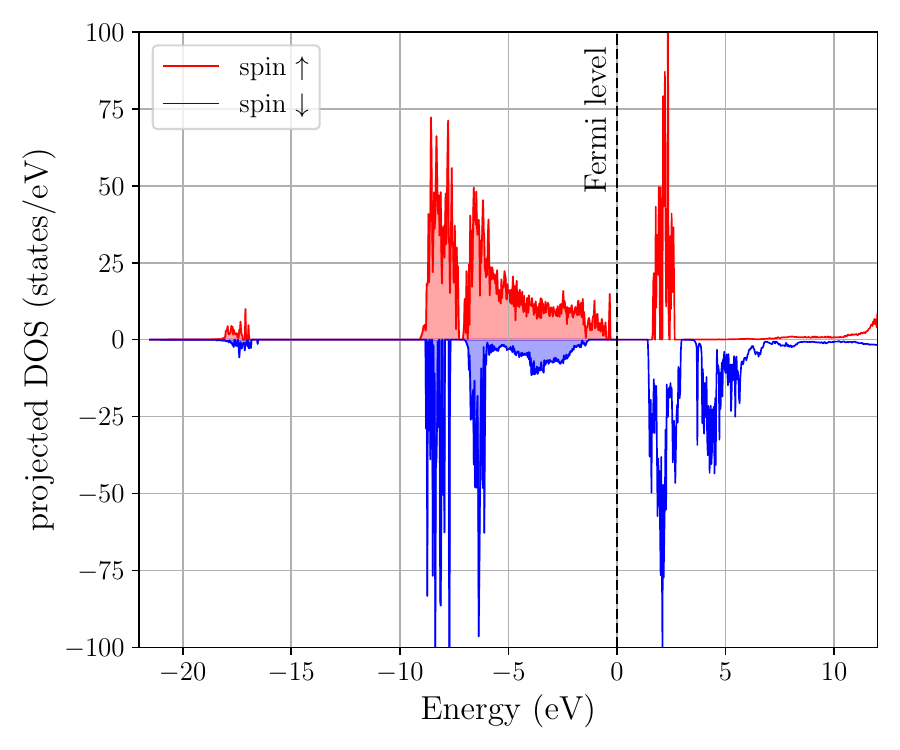}
        \caption{}\label{subfig:MZF_B_d}
    \end{subfigure}
\caption{\small \em Orbital-projected electron DOS for \ce{(Mn_{0.5},Zn_{0.5})Fe2O4}--Config.\,2: (a) \(s\)-projected DOS, (b) \(p\)-projected DOS, and (c) \(d\)-projected DOS.}
\label{fig:figure3_5}
\end{figure*}

\begin{figure*}
    \centering
    \begin{subfigure}{0.45\textwidth}
        \includegraphics[width=\textwidth]{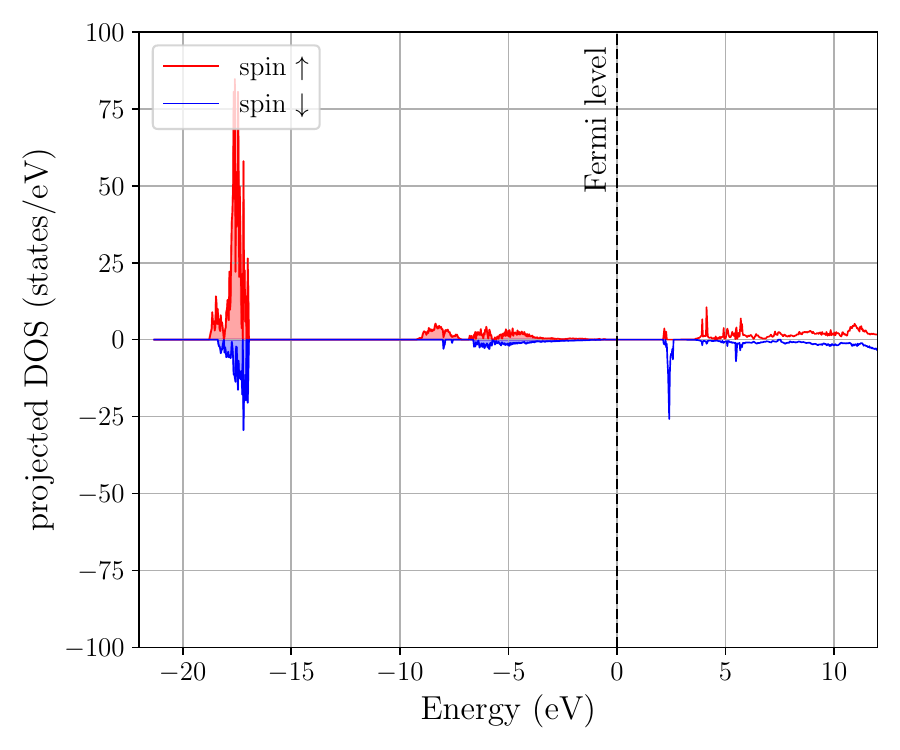}
        \caption{}\label{subfig:MZF_C_s}
    \end{subfigure}\\
    \begin{subfigure}{0.45\textwidth}
        \includegraphics[width=\textwidth]{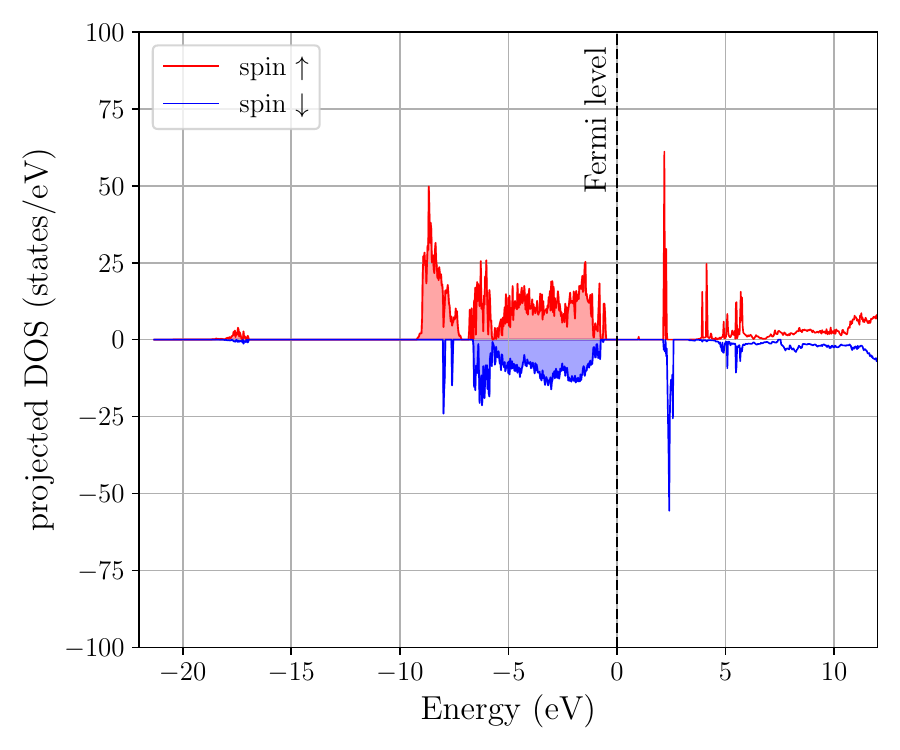}
        \caption{}\label{subfig:MZF_C_p}
    \end{subfigure}\\
    \begin{subfigure}{0.45\textwidth}
        \includegraphics[width=\textwidth]{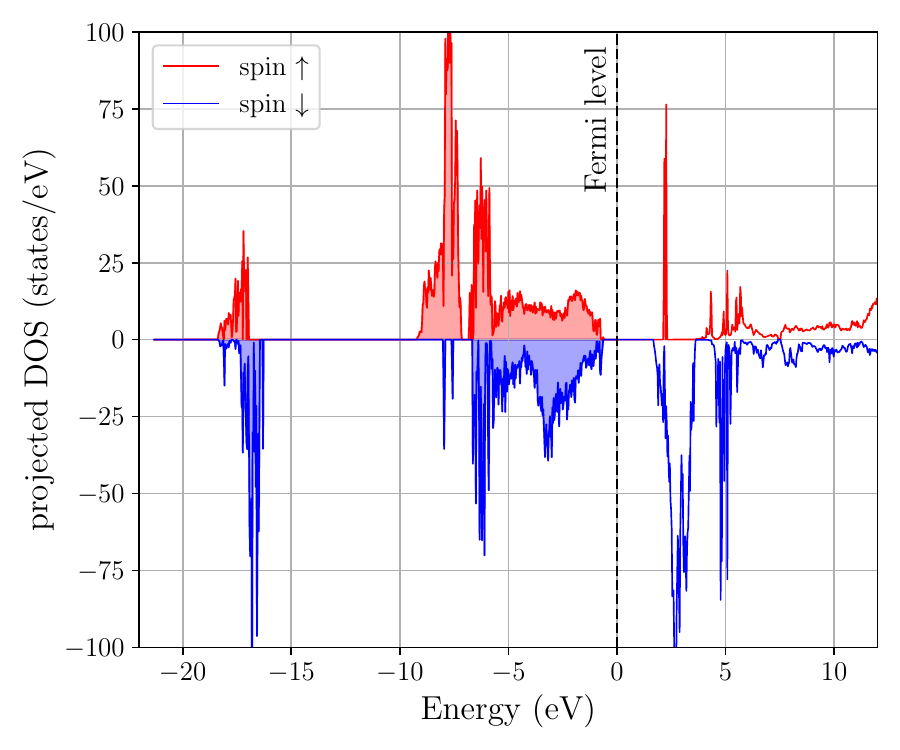}
        \caption{}\label{subfig:MZF_C_d}
    \end{subfigure}
\caption{\small \em Orbital-projected electron DOS for \ce{(Mn_{0.5},Zn_{0.5})Fe2O4}--Config.\,3
: (a) \(s\)-projected DOS, (b) \(p\)-projected DOS, and (c) \(d\)-projected DOS.}
\label{fig:figure3_6}
\end{figure*}

In \autoref{subfig:el_a}, the spin-resolved total eDOS of \ce{Fe3O4} is strongly asymmetric at the Fermi level since the minority channel (blue) retains finite spectral weight at \(E_F\), whereas the majority channel (red) is strongly suppressed, approaching a deep pseudogap. This spin-selective metallicity is consistent with the commonly reported half-metal-like electronic structure of cubic magnetite in DFT+\(U\)-type descriptions. The orbital-projected DOS in \autoref{fig:figure3_1} clarifies the origin of the main features in the spectrum. The deep feature near \(\sim -20\) eV is concentrated almost entirely in the \(s\)-projected DOS and is therefore consistent with the oxygen \(2s\) manifold. The broad occupied complex extending from roughly \(-9\) eV toward the Fermi level is composed mainly of hybridized \(p\)- and \(d\)-type states, while the states that survive at and just above \(E_F\) are predominantly \(d\)-like with a smaller but still visible \(p\) contribution. Thus, the low-energy electronic structure of \ce{Fe3O4} is not only spin-selective, but also clearly governed by transition-metal \(d\) states superimposed on appreciable valence-band \(p\)-\(d\) hybridization.

In contrast, \autoref{subfig:el_b} (\ce{MnFe2O4}, Mn confined to A sites) shows a strong depletion of the eDOS at \(E_F\) in both spin channels, consistent with an insulating or small-gap ferrimagnetic state in this configuration. Relative to \autoref{subfig:el_a}, the states that cross or closely approach the Fermi level are removed, indicating that the spin-selective metallicity of magnetite is not preserved when Mn is forced entirely onto the tetrahedral sub-lattice. The corresponding projected DOS in \autoref{fig:figure3_2} shows that the upper valence region remains predominantly \(p\)-\(d\) hybridized, while the first strong unoccupied features are more clearly \(d\)-dominated. In this sense, the effect of Mn substitution is not merely to suppress the total DOS at \(E_F\), but to increase the separation between the occupied hybridized valence manifold and the low-lying transition-metal \(d\)-derived states above the gap. Allowing Mn to occupy both sub-lattices (\autoref{subfig:el_c}, 50/50 Mn on A and B sites) modifies the near-\(E_F\) spectrum relative to the all-A case, but the Fermi level still lies in a strongly depleted region of the total eDOS. Compared with \autoref{subfig:el_b}, there is some redistribution of spectral weight in the \(\sim -2\) to 0 eV window, most evident in the minority channel, and the projected DOS in \autoref{fig:figure3_3} indicates that this change is associated primarily with a repositioning of the transition-metal \(d\) manifold relative to the valence-band edge. However, \autoref{subfig:el_c} does not recover the clear minority-metallic crossing observed for \ce{Fe3O4} in \autoref{subfig:el_a}. On the basis of the total and orbital-projected eDOS together, this configuration is therefore best described as insulating/small-gap or pseudogapped, with inversion acting to reshape the band-edge \(d\)-weight rather than unambiguously restoring a half-metallic state.

The mixed Mn--Zn ferrite cases in \autoref{subfig:el_d}--\ref{subfig:el_f} all show a pronounced suppression of the eDOS at the Fermi level compared with magnetite, indicating that the spin-selective metallicity of \ce{Fe3O4} is not retained upon partial substitution by Mn/Zn and the associated cation rearrangements. In \autoref{subfig:el_d} (Mn and Zn placed on A sites), both spin channels remain strongly depleted near \(E_F\), with the spectral weight concentrated well below the Fermi level and in unoccupied features above \(\sim 1\)--3 eV. The projected DOS in \autoref{fig:figure3_4} shows that the occupied valence manifold remains largely \(p\)-\(d\) hybridized, whereas the low-lying unoccupied features are predominantly \(d\)-like. Placing Zn on A sites while transferring Mn to the B sub-lattice (\autoref{subfig:el_e}) does not generate an obvious metallic crossing at \(E_F\). Instead, the Fermi level again lies within a deep depletion region for both spins. Compared with \autoref{subfig:el_d}, the minority eDOS shows a stronger redistribution of states in the \(\sim -2\) to 0 eV window, while the projected DOS in \autoref{fig:figure3_5} indicates a more pronounced reshaping of the conduction-side \(d\)-derived features. The intermediate Mn partitioning case (\autoref{subfig:el_f}, Zn fixed on A with Mn shared between A and B) likewise retains a depleted \(E_F\) region with only modest changes in the immediate vicinity of the Fermi level relative to \autoref{subfig:el_d}--\ref{subfig:el_e}. As seen in \autoref{fig:figure3_6}, the same overall orbital hierarchy is preserved, but the detailed distribution of band-edge \(d\)-weight is again altered by the Mn arrangement. Taken together, \autoref{subfig:el_d}--\ref{subfig:el_f} suggest that, within the present DFT+\(U\)+\(J\) description, the Mn--Zn substituted ferrites remain insulating/small-gap or pseudogapped across the tested cation distributions, and that cation arrangement primarily tunes the detailed spectral weight on the valence- and conduction-side of the depletion rather than producing a magnetite-like half-metallic eDOS.

Overall, \autoref{fig:figure3} together with \autoref{fig:figure3_1}--\ref{fig:figure3_6} show that the main electronic effect of cation replacement is not the appearance of entirely new orbital manifolds, but a redistribution of the transition-metal \(d\)-derived states relative to the occupied \(p\)-\(d\) valence complex and the Fermi-level region. The projected DOS therefore makes the role of cation arrangement clearer than the total DOS alone: \ce{Fe3O4} retains a minority-spin, \(d\)-dominated near-\(E_F\) response, whereas Mn substitution suppresses this low-energy spectral weight and the Zn-containing cases remain more strongly depleted at the Fermi level.

\subsection{Magnon density of states}

\begin{figure*}
    \centering
    \begin{subfigure}{0.45\textwidth}
        \includegraphics[width=\textwidth]{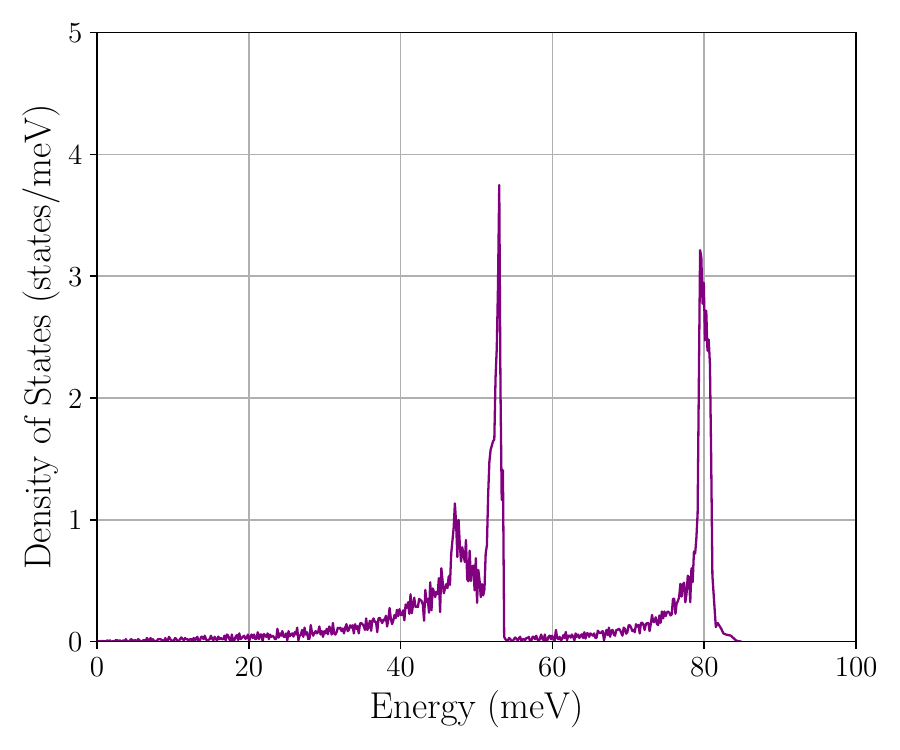}
        \caption{}\label{subfig:mDOS_1a}
    \end{subfigure}\hspace{1em}
    \begin{subfigure}{0.45\textwidth}
        \includegraphics[width=\textwidth]{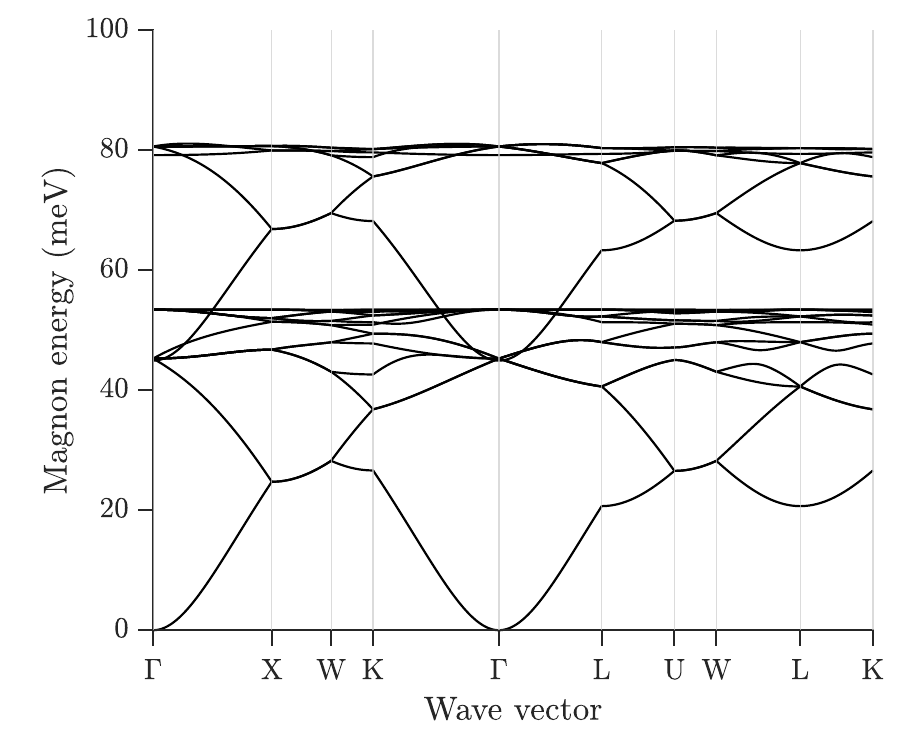}
        \caption{}\label{subfig:mband_1b}
    \end{subfigure}
    \caption{\small \em Magnon density of states and magnon band structure of \ce{Fe3O4}. (a) Total magnon DOS. (b) Corresponding magnon dispersion.}
    \label{fig:magnon_Fe3O4}
\end{figure*}

\begin{figure*}
    \centering
    \begin{subfigure}{0.45\textwidth}
        \includegraphics[width=\textwidth]{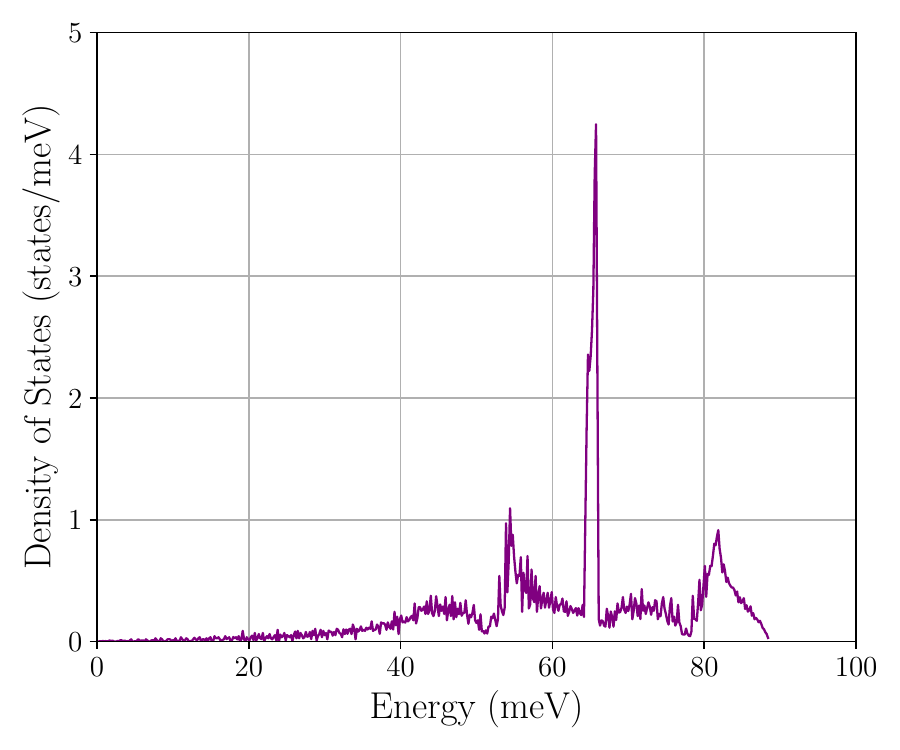}
        \caption{}\label{subfig:mDOS_2a}
    \end{subfigure}\hspace{1em}
    \begin{subfigure}{0.45\textwidth}
        \includegraphics[width=\textwidth]{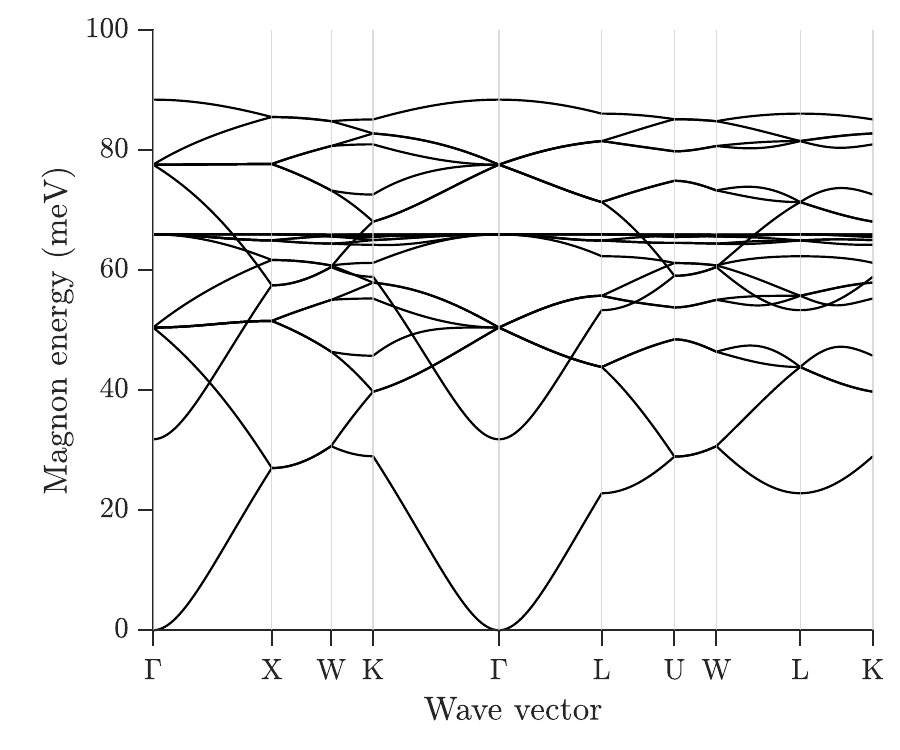}
        \caption{}\label{subfig:mband_2b}
    \end{subfigure}\\
    \begin{subfigure}{0.45\textwidth}
        \includegraphics[width=\textwidth]{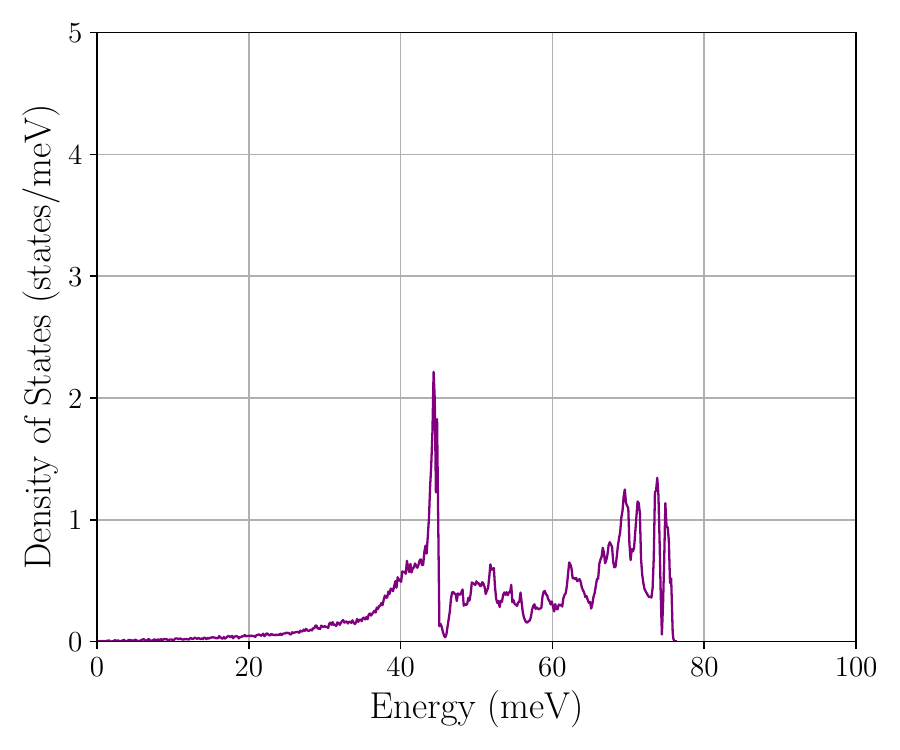}
        \caption{}\label{subfig:mDOS_2c}
    \end{subfigure}\hspace{1em}
    \begin{subfigure}{0.45\textwidth}
        \includegraphics[width=\textwidth]{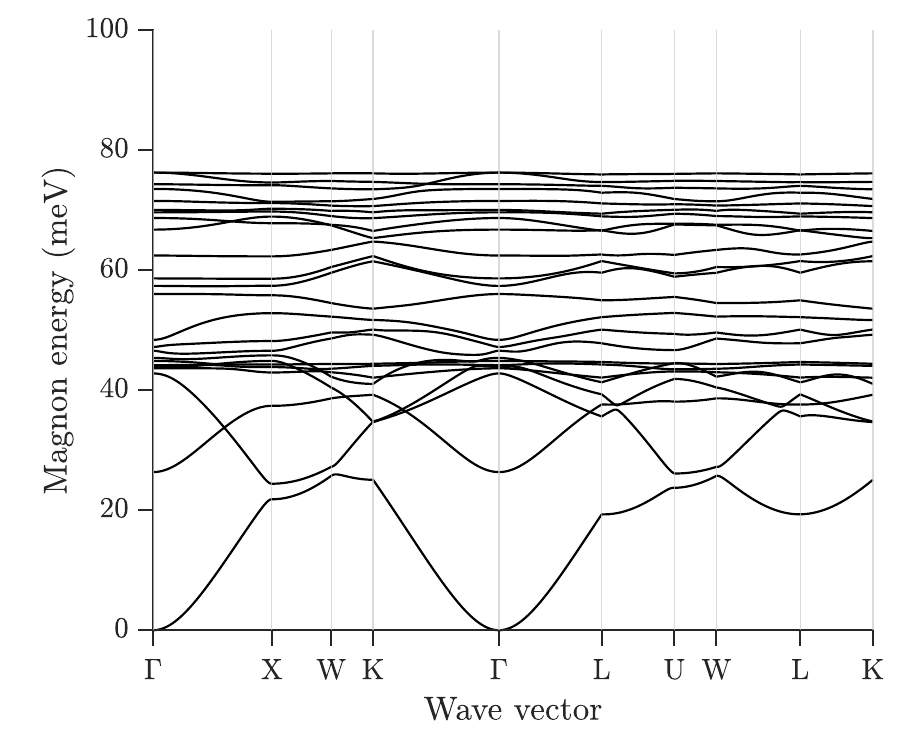}
        \caption{}\label{subfig:mband_2d}
    \end{subfigure}
    \caption{\small \em Magnon density of states and magnon band structures of \ce{MnFe2O4} for two cation arrangements. (a) Total magnon DOS and (b) corresponding magnon dispersion for Config.\,1, where all Mn atoms occupy A sites. (c) Total magnon DOS and (d) corresponding magnon dispersion for Config.\,2, where Mn is distributed equally between A and B sites.}
    \label{fig:magnon_MnFe2O4}
\end{figure*}

\begin{figure*}
    \centering
    \begin{subfigure}{0.45\textwidth}
        \includegraphics[width=\textwidth]{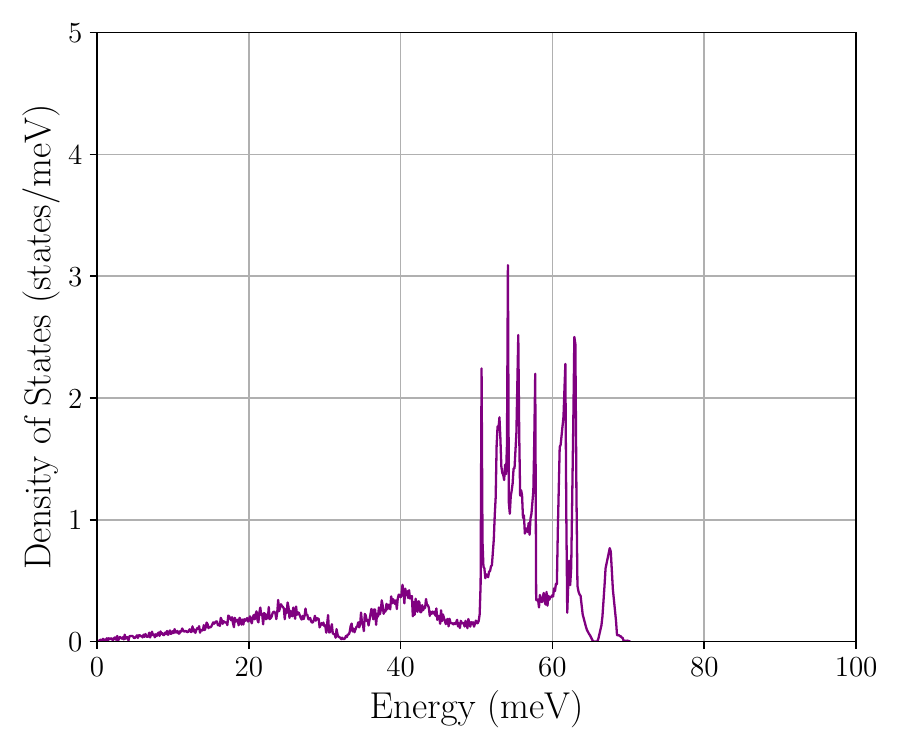}
        \caption{}\label{subfig:mDOS_3a}
    \end{subfigure}\hspace{1em}
    \begin{subfigure}{0.45\textwidth}
        \includegraphics[width=\textwidth]{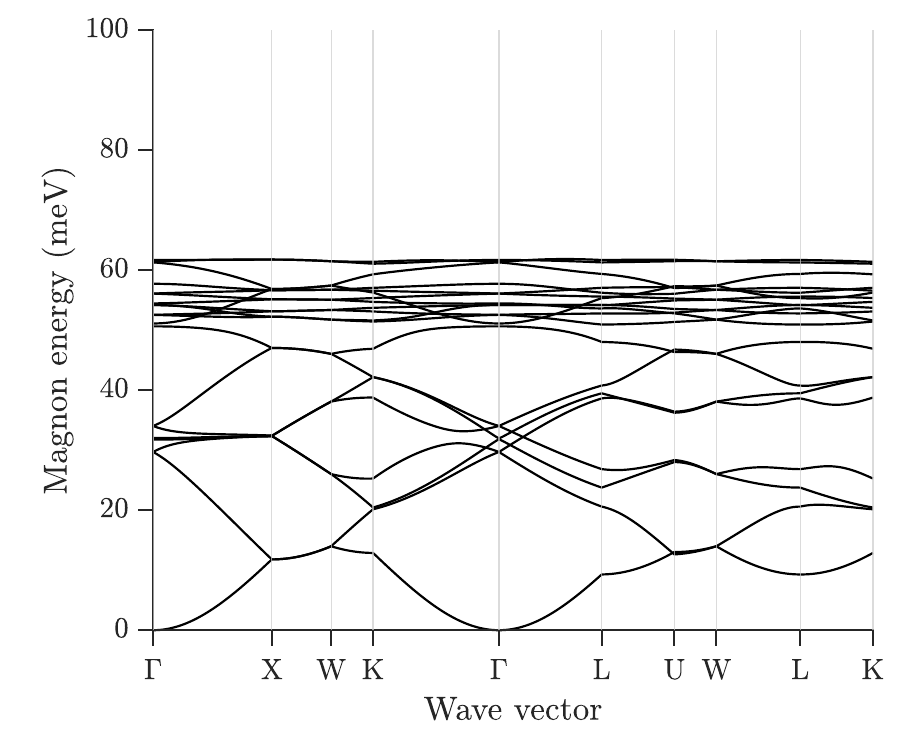}
        \caption{}\label{subfig:mband_3b}
    \end{subfigure}\\
    \begin{subfigure}{0.45\textwidth}
        \includegraphics[width=\textwidth]{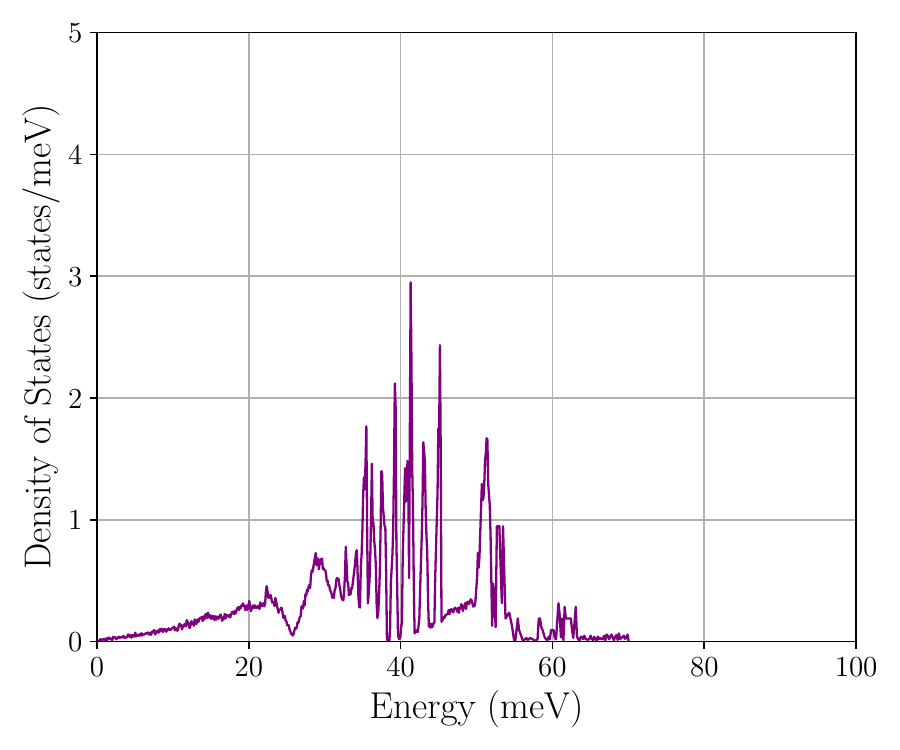}
        \caption{}\label{subfig:mDOS_3c}
    \end{subfigure}\hspace{1em}
    \begin{subfigure}{0.45\textwidth}
        \includegraphics[width=\textwidth]{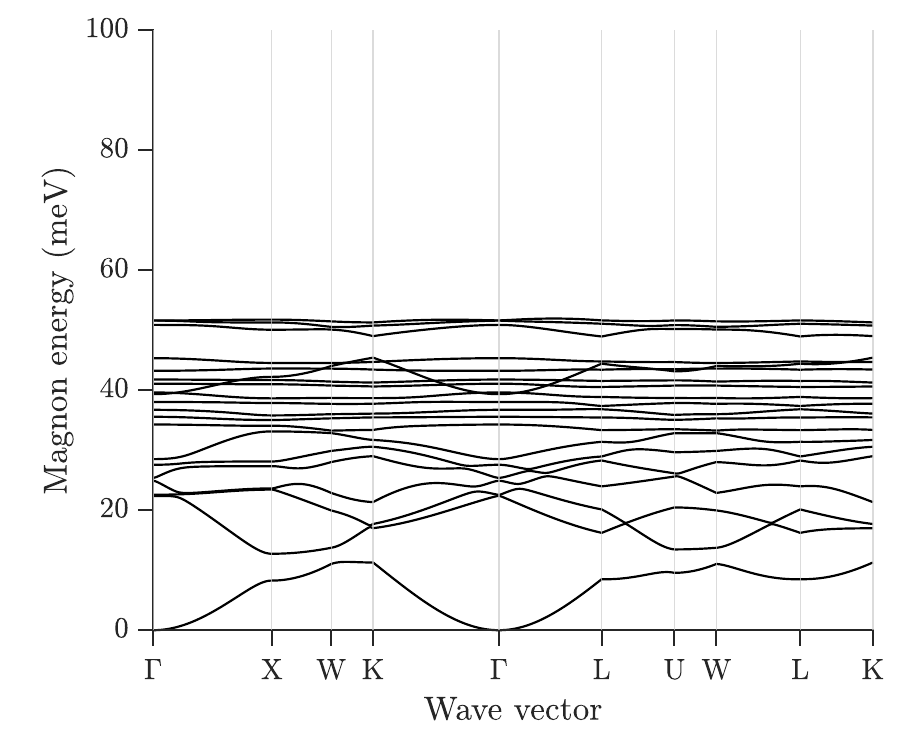}
        \caption{}\label{subfig:mband_3d}
    \end{subfigure}\\
    \begin{subfigure}{0.45\textwidth}
        \includegraphics[width=\textwidth]{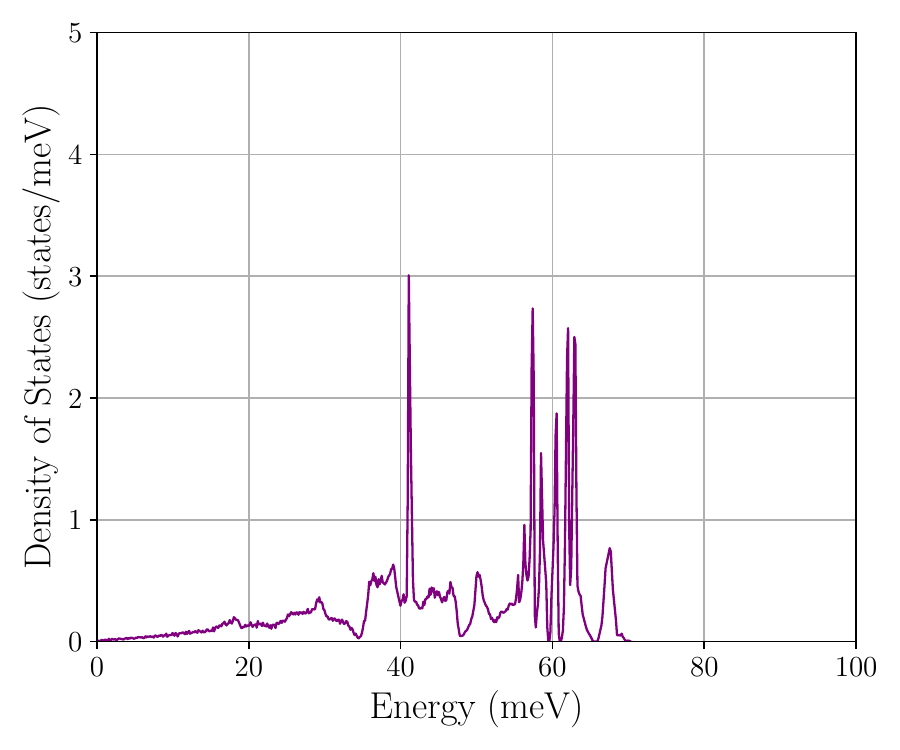}
        \caption{}\label{subfig:mDOS_3e}
    \end{subfigure}\hspace{1em}
    \begin{subfigure}{0.45\textwidth}
        \includegraphics[width=\textwidth]{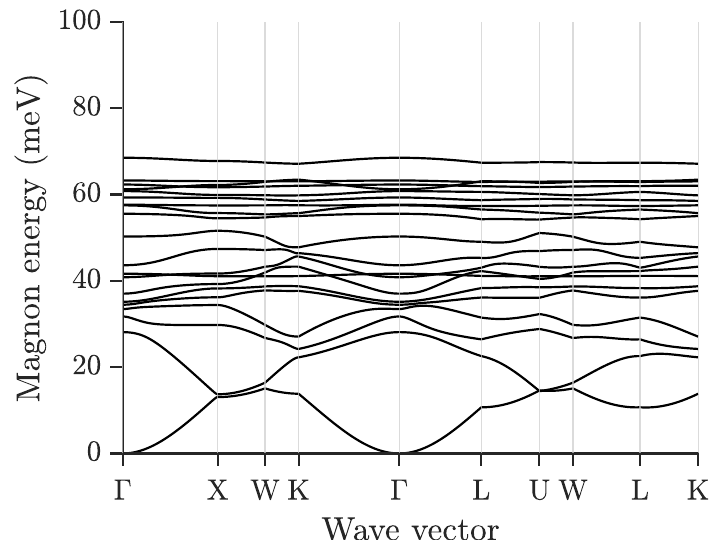}
        \caption{}\label{subfig:mband_3f}
    \end{subfigure}
    \caption{\small \em Magnon density of states and magnon band structures of \ce{(Mn_{0.5},Zn_{0.5})Fe2O4} for three cation arrangements. (a) Total magnon DOS and (b) corresponding magnon dispersion for Config.\,1, where all Zn and all Mn atoms occupy A sites. (c) Total magnon DOS and (d) corresponding magnon dispersion for Config.\,2, where all Zn atoms occupy A sites and all Mn atoms occupy B sites. (e) Total magnon DOS and (f) corresponding magnon dispersion for Config.\,3, where all Zn atoms occupy A sites and Mn is distributed equally between A and B sites.}
    \label{fig:magnon_ZnMnFe2O4}
\end{figure*}

\autoref{fig:magnon_Fe3O4}--\ref{fig:magnon_ZnMnFe2O4} present the magnon DOS together with the corresponding magnon bandstructures for the same six compositions/configurations considered in the electronic analysis. Across all cases, the mDOS turns on from zero energy, consistent with a gapless acoustic branch in the absence of explicit anisotropy terms. The main differences between systems are therefore governed not by the opening of a magnon gap, but by how the spectral weight is distributed over energy and by the maximum excitation energy (the band top), both of which reflect how the exchange network and cation arrangement shape the spin-wave spectrum. The accompanying band structures are particularly useful here because they show whether a given DOS peak arises from a broad family of branches, from a compressed bandwidth, or from weakly dispersive branch segments that generate van-Hove-like accumulations of states.

For \ce{Fe3O4}, the mDOS in \autoref{fig:magnon_Fe3O4}\subref{subfig:mDOS_1a} is characterized by two pronounced accumulations of states separated by a substantial depletion region. The spectrum rises gradually from low energies and develops a strong narrow peak near \(\sim 55\) meV, followed by a region of very low mDOS through much of the upper mid-energy range. A second sharp peak appears near \(\sim 80\) meV before the mDOS rapidly decays to near zero by \(\sim 85\)--90 meV, defining the largest bandwidth among the six cases. The corresponding band structure in \autoref{fig:magnon_Fe3O4}\subref{subfig:mband_1b} indicates that these two dominant DOS accumulations arise from separated families of magnon branches occupying distinct energy ranges rather than from a single broad continuum. In this sense, \ce{Fe3O4} exhibits the stiffest spin-wave spectrum here, with both the highest band top and strong van-Hove-like singularities associated with extrema or weakly dispersive branch segments.\cite{salaVanHoveSingularity2021}

When Mn is constrained to occupy only A sites in \ce{MnFe2O4}, the overall bandwidth remains comparable to that of \ce{Fe3O4}, as seen by comparing \autoref{fig:magnon_MnFe2O4}\subref{subfig:mDOS_2a} with \autoref{fig:magnon_Fe3O4}\subref{subfig:mDOS_1a}, but the dominant spectral features are reorganized. The mDOS remains nonzero from low energy and fills the mid-energy window more continuously than in \ce{Fe3O4}, culminating in a very strong singular peak near \(\sim 65\) meV. In contrast to magnetite, the post-peak region does not exhibit an extended interval of near-zero mDOS; instead, finite weight persists through much of the \(\sim 40\)--80 meV range, and the high-energy tail still extends toward \(\sim 85\)--90 meV. The band structure in \autoref{fig:magnon_MnFe2O4}\subref{subfig:mband_2b} is consistent with this more continuous DOS, since the magnon branches populate the mid-energy range more densely and do not separate into two clearly isolated families as in \ce{Fe3O4}. Thus, the all-A Mn arrangement shifts the dominant accumulation of modes upward, from \(\sim 55\) to \(\sim 65\) meV, while reducing the clear two-lobe partitioning seen in magnetite.

Allowing Mn to occupy the B sub-lattice in \ce{MnFe2O4} produces a clear softening and compression of the magnon spectrum, as seen in \autoref{fig:magnon_MnFe2O4}\subref{subfig:mDOS_2c}. The most prominent mDOS peak shifts downward to \(\sim 42\)--45 meV, and the upper-energy extent contracts, with the mDOS approaching zero by roughly \(\sim 78\)--80 meV. Above the dominant low-energy singularity, the spectrum consists of a broad structured band with multiple smaller peaks extending through the \(\sim 50\)--75 meV region rather than a single dominant high-energy accumulation. The corresponding band structure in \autoref{fig:magnon_MnFe2O4}\subref{subfig:mband_2d} shows this redistribution more directly: the branches are shifted downward overall, the band top is reduced, and the spectral weight is spread over a softer, more compressed energy window. Relative to both \ce{Fe3O4} and the all-A Mn case, this configuration therefore exhibits a softer spin-wave spectrum, as expected when the dominant exchange pathways and sub-lattice-resolved couplings are altered by inversion.

The Mn--Zn ferrite configurations further reduce the accessible magnon energy range and produce a more fragmented mDOS with many narrow spikes, as shown in \autoref{fig:magnon_ZnMnFe2O4}. When Mn and Zn are both placed on A sites, the spectrum in \autoref{fig:magnon_ZnMnFe2O4}\subref{subfig:mDOS_3a} retains a gradual low-energy rise but concentrates much of its weight into a dense cluster of sharp peaks between \(\sim 55\) and 66 meV, with the mDOS essentially vanishing by \(\sim 70\) meV. The corresponding band structure in \autoref{fig:magnon_ZnMnFe2O4}\subref{subfig:mband_3b} indicates a compressed spectrum containing several relatively flat or weakly dispersive branch segments in this same energy range, which explains the strong van-Hove-like structure in the DOS \cite{salaVanHoveSingularity2021,vanhoveOccurrenceSingularitiesElastic1953}.

Placing Zn on A sites while transferring Mn to B sites yields the softest Mn--Zn spectrum of the three mixed cases. In \autoref{fig:magnon_ZnMnFe2O4}\subref{subfig:mDOS_3c}, the mDOS already displays substantial structure and elevated weight in the \(\sim 20\)--45 meV window, including numerous narrow spikes, while the high-energy tail is strongly curtailed and the mDOS dies out by roughly \(\sim 60\)--65 meV. The band structure in \autoref{fig:magnon_ZnMnFe2O4}\subref{subfig:mband_3d} is consistent with this behavior, showing that a substantial fraction of the branches is shifted to lower energies and that the overall magnon bandwidth is further compressed. Compared with the all-(Mn,Zn)-on-A configuration, this case therefore shifts the spectral weight markedly downward and increases the density of low-energy excitations.

In the intermediate Mn partitioning case, with Zn fixed on A and Mn distributed between A and B, the mDOS in \autoref{fig:magnon_ZnMnFe2O4}\subref{subfig:mDOS_3e} exhibits signatures of both preceding Mn--Zn arrangements. A sharp singular feature near \(\sim 40\)--42 meV reappears, while an additional cluster of spikes emerges around \(\sim 58\)--63 meV, and the spectrum again terminates near \(\sim 70\) meV. The corresponding band structure in \autoref{fig:magnon_ZnMnFe2O4}\subref{subfig:mband_3f} indicates that these two DOS accumulations arise from distinct sets of branches within a reduced overall bandwidth, consistent with multiple magnetic environments generated by distributing Mn across both sub-lattices while Zn remains fixed. Taken together, the Mn--Zn results indicate that cation arrangement primarily controls the degree of magnon softening and the redistribution of mDOS toward lower energies, with the Zn-containing configurations generally exhibiting lower band tops and more strongly structured mid-energy spectra than \ce{Fe3O4} and \ce{MnFe2O4}.

\subsection{Phonon density of states}

\begin{figure*}
\centering
\vspace{-2em}
\begin{subfigure}{0.45\textwidth}
    \includegraphics[width=\textwidth]{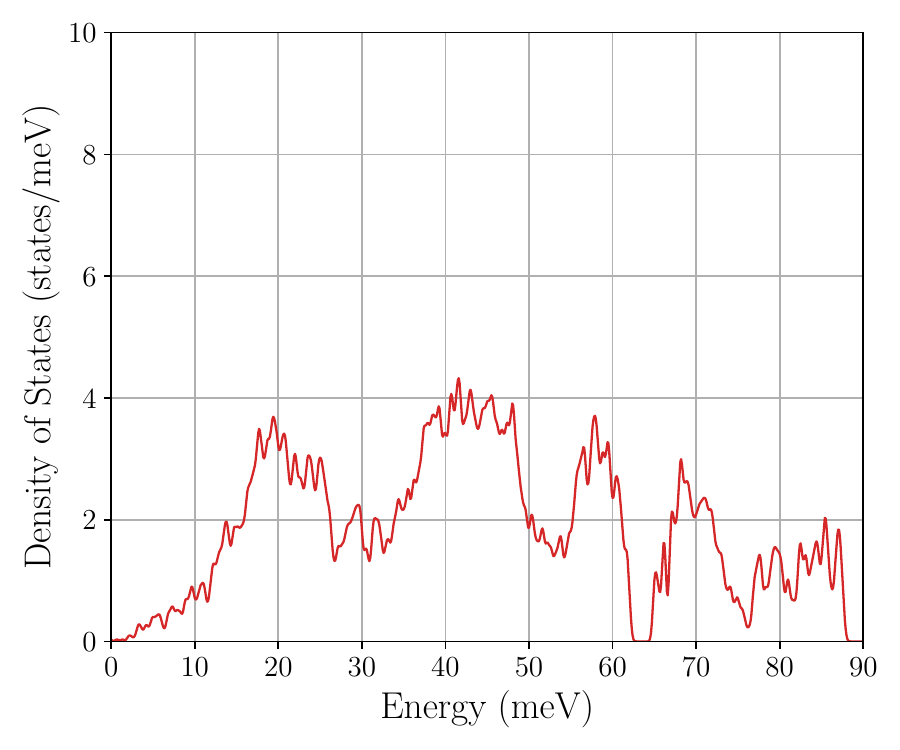}
    \caption{}\label{subfig:ph_a}
\end{subfigure}%\hspace{2em}
\begin{subfigure}{0.45\textwidth}
    \includegraphics[width=\textwidth]{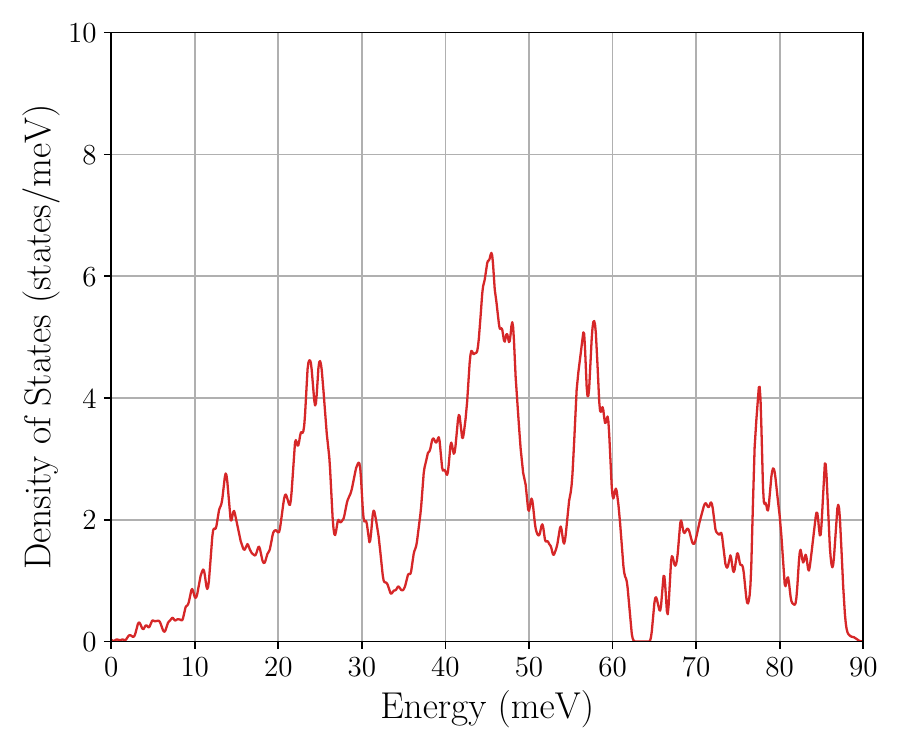}
    \caption{}\label{subfig:ph_b}
\end{subfigure}\\
\vspace{0.5em}
\begin{subfigure}{0.45\textwidth}
    \includegraphics[width=\textwidth]{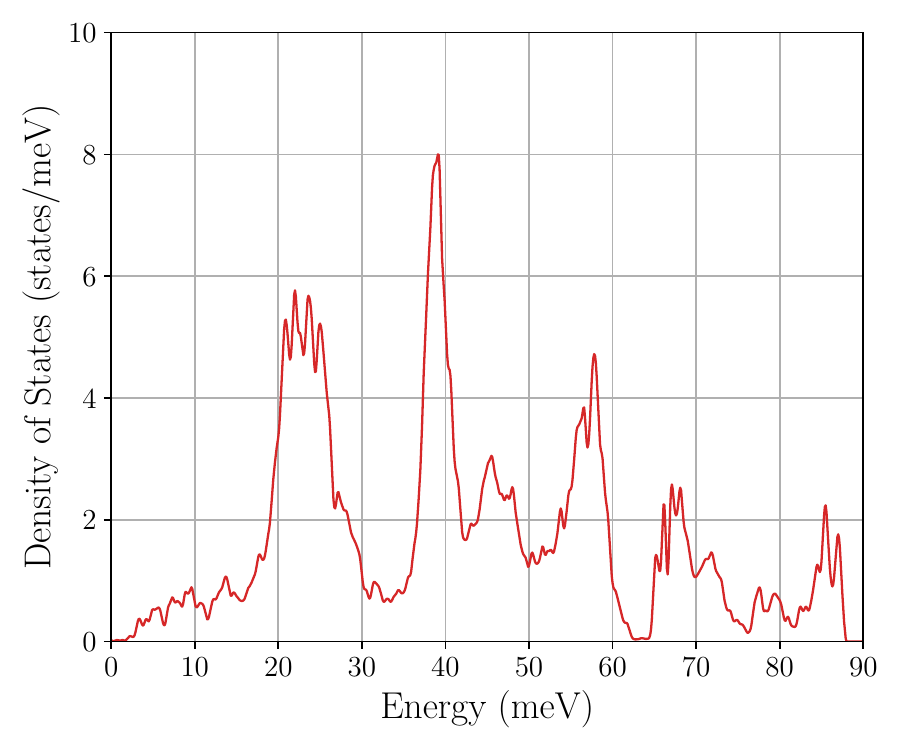}
    \caption{}\label{subfig:ph_c}
\end{subfigure}%\hspace{2em}
\begin{subfigure}{0.45\textwidth}
    \includegraphics[width=\textwidth]{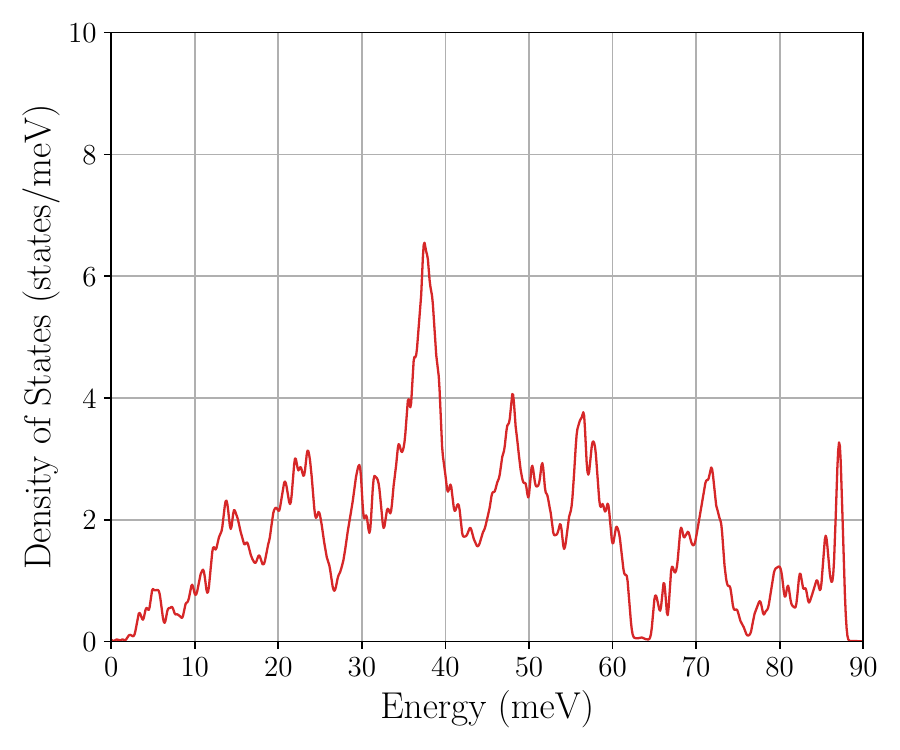}
    \caption{}\label{subfig:ph_d}
\end{subfigure}\\
\vspace{0.5em}
\begin{subfigure}{0.45\textwidth}
    \includegraphics[width=\textwidth]{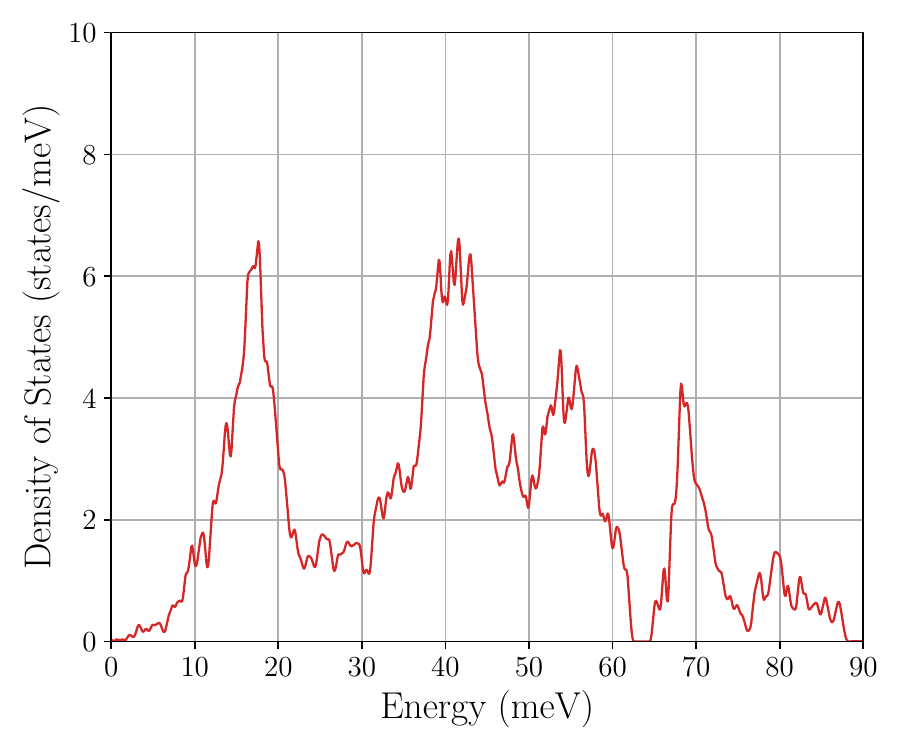}
    \caption{}\label{subfig:ph_e}
\end{subfigure}%\hspace{2em}
\begin{subfigure}{0.45\textwidth}
    \includegraphics[width=\textwidth]{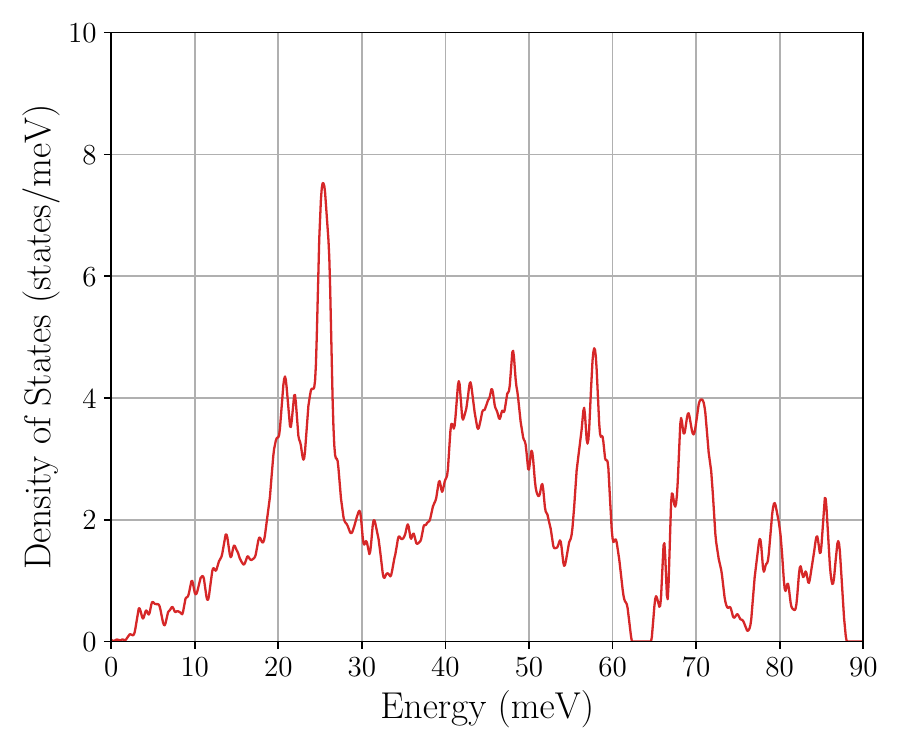}
    \caption{}\label{subfig:ph_f}
\end{subfigure}

    \caption{\small \em Phonon DOS for spinel ferrites with different cationic compositions and configurations. (a) archetypal \ce{Fe3O4}, (b) \ce{MnFe2O4} Config.\,1 where all Mn atoms occupy A-sites of the spinel structure, (c) \ce{MnFe2O4} Config.\,2 where half of the Mn atoms occupy A-sites and the other half occupy B-sites, (d) \ce{(Mn_{0.5},Zn_{0.5})Fe2O4} Config.\,1 where all Zn \& Mn atoms occupy A-sites, (e) \ce{(Mn_{0.5},Zn_{0.5})Fe2O4} Config.\,2 where all Zn atoms occupy A-sites \& all Mn atoms occupy B-sites, (f) \ce{(Mn_{0.5},Zn_{0.5})Fe2O4} Config.\,3 where all Zn atoms occupy A-sites \& the Mn atoms occupy half A-sites and half B-sites.}  
\label{fig:figure5}
\end{figure*}

In \autoref{subfig:ph_a}--\ref{subfig:ph_f}, the total pDOS spans approximately 0--90 meV and turns on smoothly from zero energy, consistent with acoustic modes at long wavelength. Rather than splitting into sharply isolated blocks, each spectrum exhibits multiple broad manifolds with a configuration-dependent peak structure, and all six panels share a prominent depletion (a `mini-gap') centered around roughly 62--65 meV. Above this depletion, a high-energy tail extends to the band top near \(\sim 90\) meV with several relatively sharp features. These common features are physically meaningful. The persistent depletion near 62--65 meV suggests a robust separation between two vibrational manifolds retained across all six cubic-spinel configurations: a lower/mid-frequency manifold dominated by mixed cation--oxygen motions and bending/deformation character, and a higher-frequency manifold with stronger oxygen-dominated stretching character. Likewise, the fact that the phonon band top remains close to \(\sim 90\) meV in all configurations indicates that cation substitution and inversion redistribute spectral weight within the phonon spectrum more strongly than they alter the stiffest metal--oxygen bonding scale in the lattice. While species-resolved assignments require projected pDOS, the higher-energy portion of the spectrum is typically associated with the oxygen-dominated stretching character, whereas the lower- and mid-energy regions contain mixed cation--oxygen bending and cation-participating modes.

For \ce{MnFe2O4} with Mn confined to A sites, the pDOS, shown in \autoref{subfig:ph_b}, rises gradually from 0 meV and develops a sequence of structured peaks throughout the low- and mid-energy range. The dominant spectral weight is concentrated between roughly 35 and 60 meV, with a particularly strong peak near $\sim 50$ meV and an additional pronounced structure approaching $\sim 58-60$ meV. A deep depletion follows at $\sim 62-65$ meV after which only a comparatively weaker high-energy tail persists up to the band top near $\sim 90$ meV. Relative to \ce{Fe3O4}, this configuration shifts and concentrates the mid-frequency weight (especially around 45-60 meV), while retaining the same overall bandwidth, indicating that cation arrangement reshapes the distribution of vibrational states more than it changes the maximum phonon energy. When Mn is distributed between A and B sites (see \autoref{subfig:ph_c}), the phonon DOS becomes more sharply structured with a very prominent peak near $\sim 40$ meV that exceeds the corresponding mid-band amplitudes in \ce{Fe3O4} and the all-A Mn configuration. The lower-energy region also exhibits a strong group of features around $\sim 20-25$ meV. As in the other panels, the pDOS drops into a deep depletion near $\sim 62-65$ meV and then continues as a weaker high-energy tail toward $\sim 90$ meV. Compared with the all-A Mn case, this configuration shifts substantial vibrational weight downward (from the 50-60 meV region toward $\sim 40$ meV) while preserving the overall bandwidth.

For \ce{(Mn_{0.5},Zn_{0.5})Fe2O4} with Mn and Zn placed on A sites, the pDOS shown in \autoref{subfig:ph_d} remains continuous from 0 to $\sim 90$ meV but shows a particularly strong mid-frequency maximum centered near $\sim 38-40$ meV accompanied by additional structure throughout the 45-60 meV region. The same pronounced depletion near $\sim 62-65$ meV is retained after which only modest high-energy features persist. Relative to the Mn-free magnetite panel, the mixed Mn--Zn configuration concentrates more spectral weight into a narrower mid-band peak near $\sim 40$ meV consistent with a cation-configuration-driven reshaping of intermediate-frequency vibrations. Placing Zn on A sites while forcing Mn onto B sites (see \autoref{subfig:ph_e}) produces one of the strongest enhancements of low-energy weight among the Mn--Zn cases, namely, a pronounced peak already emerges near $\sim 18-20$ meV in addition to a broad and intense mid-band centered around $\sim 40-45$ meV. The deep depletion near $\sim 62-65$ meV again appears clearly, and the high-energy tail above it is comparatively weaker and more fragmented. This spectrum, therefore, shifts a notable fraction of vibrational density toward lower energies compared with the all-A (Mn,Zn) arrangement, indicating that which species occupies the octahedral network materially affects the placement and intensity of mid-frequency manifolds.

In the intermediate Mn partitioning case (in \autoref{subfig:ph_f}), the pDOS exhibits both a strong low- and mid-frequency structure: a dominant low-energy peak appears near $\sim 25-27$ meV followed by a broad manifold around $\sim 40-50$ meV and additional peaks approaching $\sim 55-60$ meV. The internal depletion near $\sim 62-65$ meV remains a robust feature, and a reduced but still nonzero high-energy tail persists up to $\sim 90$ meV. Compared with \autoref{subfig:ph_e}, the strongest low-energy accumulation shifts upward (from $\sim 18-20$ meV to $\sim 25-27$ meV), while the mid-band retains significant intensity, underscoring that partial inversion tunes the balance between low-frequency and mid-frequency vibrational populations.

\section{Conclusions}\label{sec:conclusion}

This study delivers a consistent, configuration-aware set of electronic, phonon, and magnon excitation spectra for  spinel ferrites spanning magnetite, jacobsite, and Mn--Zn mixed ferrites. The workflow combines a self-consistent LR determination of Hubbard $U$ and Hund's $J$ parameters within rotationally invariant DFT+$U$+$J$, energy mapping onto a Heisenberg model suitable for mixed-cation sub-lattices, linear spin-wave calculations for magnons, and finite-displacement phonons computed on the same relaxed structures. This alignment across methods ensures that the resulting spectra form a coherent dataset in which differences are attributable to chemistry and A/B-site configurations.

The results underscore that the excitation spectrum is the appropriate microscopic object for comparing ferrites across a composition--configuration landscape. Changes in inversion and site occupation redistribute $d$--$p$ hybridization and re-weight exchange pathways, thereby shifting electronic, vibrational, and spin-wave spectral features in correlated ways. These configuration-resolved DOS are therefore not merely post-processing outputs; they are compact representations of how local chemistry and lattice topology shape the available excitation channels in each material.

Future work can proceed in two complementary directions. On the materials side, the configurational space can be expanded (e.g., with additional inversion levels and disorder realizations) and the magnetic model refined (further-neighbor exchanges and, where necessary, anisotropy terms) to produce more complete, experimentally comparable spectra. On the modeling side, the computed electron/magnon/phonon DOS provide standardized inputs for frameworks that require microscopic excitation content---particularly those aimed at interpreting or predicting material response under time-dependent fields---so that performance changes can be traced back to identifiable spectral shifts rather than treated as changes in empirical constants. The latter is presented in a forthcoming paper that employs the SEAQT formalism.

\section*{Acknowledgments}

D.D. would like to acknowledge helpful discussions with Dr.~L\'{o}rien MacEnulty on linear response calculations and the Department of Materials Science and Engineering for financial support of his doctoral research. The authors thank the Advanced Research Computing (ARC) at Virginia Tech for providing the computational resources and technical support that have contributed to the results reported here.

\section*{Author Contributions}

All authors contributed equally to this work. D.D.: conceptualization; methodology; software; writing original draft; visualization; and validation. M.R.v.S.: conceptualization; methodology; and writing review and editing; supervision. W.T.R. Jr.: conceptualization; methodology; writing review and editing; and supervision.

%\appendix

\begin{appendices}

\section{Second Nearest-Neighbor Exchange constants of \ce{Fe3O4}}\label{app:second-NN-Fe3O4}

In the spinel-ferrite literature, the three commonly quoted ``nearest-neighbor'' exchange constants, namely $J_{AB}$, $J_{AA}$, and $J_{BB}$, do \emph{not} correspond to a single common radial shell of the full cation network. Rather, they correspond to the \emph{shortest} A--B, A--A, and B--B shells, respectively. This distinction is important for inverse-spinel \ce{Fe3O4}, because the shortest B--B, A--B, and A--A cation separations are all different. Accordingly, a strict crystallographic shell labeling and the standard ferrite exchange labeling are not identical.

\paragraph{Pure crystallographic radial-shell ordering:}
Let $\mathcal{R}_m$ denote the $m$th shortest cation--cation shell of the full cubic spinel cation network, ordered only by distance. Using the ideal cubic spinel geometry, the shortest cation--cation distances are (see \autoref{tab:spinel-pair-family-shells})
\begin{equation}\label{eq:radial-shell-order-spinel}
    d_{BB}^{(1)}=\frac{\sqrt{2}}{4}\,a,
    \quad
    d_{AB}^{(1)}=\frac{\sqrt{11}}{8}\,a,
    \quad
    d_{AA}^{(1)}=\frac{\sqrt{3}}{4}\,a,
\end{equation}
so that
\begin{equation}\label{eq:radial-shell-inequality-spinel}
    d_{BB}^{(1)}\, <\, d_{AB}^{(1)}\, <\, d_{AA}^{(1)}
\end{equation}
Therefore, in \emph{purely crystallographic} terms, the shortest B--B shell is the first radial shell of the cation network, the shortest A--B shell is the second radial shell, and the shortest A--A shell is the third radial shell. The standard ferrite NN constants $J_{BB}$, $J_{AB}$, and $J_{AA}$ therefore belong to \emph{different radial shells} even though they are all conventionally called ``nearest-neighbor'' exchange constants.

\paragraph{Pair-family shell convention:}
Because the ferrite literature distinguishes exchange couplings primarily by sub-lattice pair type rather than by global radial rank, it is more precise to organize the exchange model in terms of \emph{pair-family shells}. Let
\begin{equation}\label{eq:pair-family-shell-definition}
    \mu \in \{AB,\,AA,\,BB\},
\end{equation}
and for each pair family $\mu$, let
\begin{equation}\label{eq:pair-family-shells}
    \mathcal{N}_{\mu}^{(n)}
    \equiv
    \left\{
    \langle i,j\rangle \in \mu
    \;:\;
    d_{ij}=d_{\mu}^{(n)}
    \right\},
    \qquad
    d_{\mu}^{(1)} < d_{\mu}^{(2)} < d_{\mu}^{(3)} < \cdots ,
\end{equation}
where $d_{\mu}^{(n)}$ is the $n$th shortest distance \emph{within the fixed pair family} $\mu$. In this notation,
\begin{equation}\label{eq:first-shell-identification}
    J_{AB}\equiv J_{AB}^{(1)},
    \qquad
    J_{AA}\equiv J_{AA}^{(1)},
    \qquad
    J_{BB}\equiv J_{BB}^{(1)}.
\end{equation}
The corresponding second-neighbor exchange constants are then defined consistently as
\begin{equation}\label{eq:second-shell-identification}
    J_{AB}^{(2)},
    \qquad
    J_{AA}^{(2)},
    \qquad
    J_{BB}^{(2)}.
\end{equation}
This convention preserves the standard ferrite language while removing the ambiguity associated with a bare radial-shell labeling such as $J_1$, $J_2$, and $J_3$. In particular, ``second nearest neighbor'' in the present sense means \emph{the second shell within a given pair family}, not the second shortest cation--cation distance of the entire crystal.

\begin{table*}[t]
\centering
\caption{First and second pair-family shells for the ideal cubic spinel cation network. Distances are given in terms of the lattice parameter $a$. The coordination numbers $z_{\mu}^{(n)}$ are given per site of the relevant sub-lattice; for A--B shells, the notation ``A/B'' indicates the counts seen from an A site and from a B site, respectively. The numbers of bonds per formula unit are denoted by $N_{\mu}^{(n)}$.}
\label{tab:spinel-pair-family-shells}
\renewcommand{\arraystretch}{1.15}
\setlength{\tabcolsep}{5pt}
\begin{tabular}{c|ccc|ccc}
\toprule
\textbf{Pair family} &
\textbf{$d_{\mu}^{(1)}$} &
\textbf{$z_{\mu}^{(1)}$} &
\textbf{$N_{\mu}^{(1)}$} &
\textbf{$d_{\mu}^{(2)}$} &
\textbf{$z_{\mu}^{(2)}$} &
\textbf{$N_{\mu}^{(2)}$} \\
\midrule
$AA$ &
$\dfrac{\sqrt{3}}{4}\,a$ &
$4$ &
$2$ &
$\dfrac{\sqrt{2}}{2}\,a$ &
$12$ &
$6$ \\[6pt]

$BB$ &
$\dfrac{\sqrt{2}}{4}\,a$ &
$6$ &
$6$ &
$\dfrac{\sqrt{6}}{4}\,a$ &
$12$ &
$12$ \\[6pt]

$AB$ &
$\dfrac{\sqrt{11}}{8}\,a$ &
$12/6$ &
$12$ &
$\dfrac{3\sqrt{3}}{8}\,a$ &
$16/8$ &
$16$ \\
\bottomrule
\end{tabular}
\end{table*}

\paragraph{Special cases for the A and B sub-lattices:}
The A-site cations form a diamond sub-lattice. Hence $J_{AA}^{(1)}$ and $J_{AA}^{(2)}$ coincide with the usual first- and second-neighbor couplings of the diamond lattice. Likewise, the B-site cations form a pyrochlore sub-lattice, so $J_{BB}^{(1)}$ and $J_{BB}^{(2)}$ coincide with the usual first- and second-neighbor couplings of the pyrochlore lattice. By contrast, the A--B family does not reduce to a standard one-sublattice notation, and it is therefore preferable to retain the explicit symbols $J_{AB}^{(1)}$ and $J_{AB}^{(2)}$.

\paragraph{Heisenberg Hamiltonian in pair-family form:}
With this convention, the collinear mapping Hamiltonian truncated after the second shell in each pair family is
\begin{equation}\label{eq:Heis-pair-family-2shell}
    \begin{aligned}
    H
    = &- \sum_{n=1}^{2} J_{AB}^{(n)}
    \sum_{\langle i\in A,\,j\in B\rangle \in \mathcal{N}_{AB}^{(n)}}
    \bm{S}_i^{\top}\bm{S}_j\,\sigma_i\sigma_j \\
    &- \sum_{n=1}^{2} J_{AA}^{(n)}
    \sum_{\langle i,i'\in A\rangle \in \mathcal{N}_{AA}^{(n)}}
    \bm{S}_i^{\top}\bm{S}_{i'}\,\sigma_i\sigma_{i'} \\
    &- \sum_{n=1}^{2} J_{BB}^{(n)}
    \sum_{\langle j,j'\in B\rangle \in \mathcal{N}_{BB}^{(n)}}
    \bm{S}_j^{\top}\bm{S}_{j'}\,\sigma_j\sigma_{j'} .
    \end{aligned}
\end{equation}
The sign convention is unchanged from \autoref{eq:NN-Heis-spinel}: $J>0$ favors ferromagnetic alignment and $J<0$ favors antiferromagnetic alignment.

\paragraph{Bond-weighted sums:}
In direct analogy with the first-shell quantities defined in \autoref{eq:Wdefs} and \autoref{eq:Cdefs}, define for each pair family and shell
\begin{equation}\label{eq:Wdefs-pair-family}
    W_{\mu}^{(n)}
    =
    \sum_{\langle i,j\rangle \in \mathcal{N}_{\mu}^{(n)}} S_iS_j,
    \qquad
    \mu\in\{AB,AA,BB\}, \quad n\in\{1,2\},
\end{equation}
and for any collinear reference state $k$,
\begin{equation}\label{eq:Cdefs-pair-family}
    C_{\mu}^{(n,k)}
    =
    \sum_{\langle i,j\rangle \in \mathcal{N}_{\mu}^{(n)}} S_iS_j\,
    \sigma_i^{(k)}\sigma_j^{(k)}.
\end{equation}
The DFT energy differences relative to the ferrimagnetic reference state then satisfy
\begin{equation}\label{eq:DeltaEgeneral-pair-family}
    \Delta E_k
    \equiv E_k-E_{\mathrm{FiM}}
    =
    -\sum_{\mu\in\{AB,AA,BB\}}
    \sum_{n=1}^{2}
    J_{\mu}^{(n)}
    \left[
    C_{\mu}^{(n,k)}-C_{\mu}^{(n,\mathrm{FiM})}
    \right].
\end{equation}

\paragraph{Determining the NN and NNN exchange constants:}
For the original NN model, the unknown exchange vector is
\begin{equation}\label{eq:Jvector-three}
    \bm{J}_{\mathrm{NN}}
    =
    \begin{bmatrix}
    J_{AB}^{(1)} &
    J_{AA}^{(1)} &
    J_{BB}^{(1)}
    \end{bmatrix}^{\!\top},
\end{equation}
and three linearly independent energy differences are sufficient. This is why the four-state set
\[
\{\mathrm{FiM},\ \mathrm{FM},\ \mathrm{A_{AF}},\ \mathrm{B_{AF}}\}
\]
is enough to determine the NN constants, as done in the main text. For the extended first+second-shell model, however, the unknown exchange vector becomes
\begin{equation}\label{eq:Jvector-six}
    \bm{J}
    =
    \begin{bmatrix}
    J_{AB}^{(1)} &
    J_{AA}^{(1)} &
    J_{BB}^{(1)} &
    J_{AB}^{(2)} &
    J_{AA}^{(2)} &
    J_{BB}^{(2)}
    \end{bmatrix}^{\!\top},
\end{equation}
so that \emph{six linearly independent energy differences} are required. If $\mathrm{FiM}$ is used as the reference state, this means that at least \emph{seven total collinear spin states} are needed: the reference $\mathrm{FiM}$ plus six additional linearly independent configurations.\\

Let $k=1,\dots,M$ label the $M$ non-reference configurations and define
\begin{equation}\label{eq:deltaC-compact}
    \Delta C_{\mu}^{(n,k)}
    \equiv
    C_{\mu}^{(n,k)}-C_{\mu}^{(n,\mathrm{FiM})}.
\end{equation}
Then the energy mapping can be written compactly as $\Delta \bm{E} = -\mathbf{A}\bm{J}$, where the columns of $\mathbf{A}$ correspond to the six shell-resolved configurations,
\[
(AB,1),\ (AA,1),\ (BB,1),\ (AB,2),\ (AA,2),\ (BB,2),
\]
and the matrix entries are
\begin{equation}\label{eq:Amatrix-six}
    A_{k,\alpha} = \Delta C_{\alpha}^{(k)}.
\end{equation}

If $M=6$ and $\mathbf{A}$ has full rank, the six exchange constants follow from the exact solution

\begin{equation}\label{eq:exact-solution-six}
    \bm{J} = -\,\mathbf{A}^{-1}\Delta\bm{E}.
\end{equation}

In practice, an overdetermined fit with $M>6$ is preferable, in which case the least-squares estimate is $\bm{J} = -(\mathbf{A}^{\top}\mathbf{A})^{-1}\mathbf{A}^{\top}\Delta\bm{E}$. The essential requirement is that the chosen set of spin configurations produce a coefficient matrix $\mathbf{A}$ of rank six.

\paragraph{Choice of spin configurations:}
The four spin-configurations already used for the NN mapping,
\[
\mathrm{FiM},\qquad
\mathrm{FM},\qquad
\mathrm{A_{AF}},\qquad
\mathrm{B_{AF}},
\]
remain a natural starting point for the six-parameter fit, but they provide only three independent energy differences and are therefore insufficient by themselves. At least three additional collinear spin configurations must be introduced. There is, however, no unique universal choice for the last three configurations. Their detailed spin patterns depend on the supercell adopted and on the actual shell enumeration in that cell. In practice, the selected states should be accepted only after verifying explicitly that the resulting matrix $\mathbf{A}$ has rank six. A convenient practical strategy adopted here is to enlarge the magnetic cell and add
\begin{enumerate}[(i)]
    \item a second inequivalent A-sublattice antiferromagnetic configuration, $\mathrm{A_{AF}^{\prime}}$, chosen so that the first- and second-shell A--A exchanges differ from those of $\mathrm{A_{AF}}$;
    \item a second inequivalent B-sublattice antiferromagnetic configuration, $\mathrm{B_{AF}^{\prime}}$, chosen so that the first- and second-shell B--B exchanges differ from those of $\mathrm{B_{AF}}$;
    \item a mixed A/B collinear configuration, denoted here by $\mathrm{AB_{mix}}$, chosen so that the first- and second-shell A--B exchanges are not changed in the same proportion as in the simple $\mathrm{FiM}\leftrightarrow\mathrm{FM}$ reversal.
\end{enumerate}
A minimal seven-state set may therefore be written schematically as
\[
\{\mathrm{FiM},\ \mathrm{FM},\ \mathrm{A_{AF}},\ \mathrm{B_{AF}},\ \mathrm{A_{AF}^{\prime}},\ \mathrm{B_{AF}^{\prime}},\ \mathrm{AB_{mix}}\}.
\]

\paragraph{The case of \ce{Fe3O4}:}
For cubic \ce{Fe3O4}, the conventional three-parameter model $\{J_{AB},J_{AA},J_{BB}\}$ should therefore be interpreted as the set of \emph{first pair-family shells}
\[
\{J_{AB}^{(1)},\,J_{AA}^{(1)},\,J_{BB}^{(1)}\},
\]
not as the first three radial shells of the full cation network. Following the same logic, the natural second-neighbor extension is the three-parameter set
\[
\{J_{AB}^{(2)},\,J_{AA}^{(2)},\,J_{BB}^{(2)}\},
\]
where each term denotes the second shell within its own pair family. This formulation preserves the standard ferrite notation while remaining fully consistent with the underlying crystallographic geometry.

\paragraph{Second nearest-neighbor exchange constants for \ce{Fe3O4}:}
The shell-resolved second-neighbor exchange constants obtained for cubic \ce{Fe3O4} are summarized in \autoref{tab:nnn-exchange-fe3o4}. Here, ``second nearest neighbor'' is understood in the pair-family sense established above; that is, $J_{AB}^{(2)}$, $J_{BB}^{(2)}$, and $J_{AA}^{(2)}$ denote the second shell within the A--B, B--B, and A--A families, respectively, rather than the second radial shell of the full cation network. All values are extracted from the same shell-resolved collinear energy-mapping procedure, with the same sign convention as in the main NN analysis: $J>0$ favors ferromagnetic alignment and $J<0$ favors antiferromagnetic alignment.

A point worth noting is that once the exchange model is extended from the three-parameter NN fit to the six-parameter NN+NNN fit, the fitted first-shell couplings are no longer numerically identical to the NN-only values. In other words, the NN-only constants $\{J_{AB},J_{BB},J_{AA}\}$ should be regarded as effective first-shell parameters of the truncated model, whereas the quantities $\{J_{AB}^{(1)},J_{BB}^{(1)},J_{AA}^{(1)}\}$ obtained from the NN+NNN fit are the renormalized first-shell values obtained when second-shell terms are included explicitly.

\begin{table}[h]
    \centering
    \caption{Comparison of NN-only and NN+NNN exchange constants for cubic \ce{Fe3O4}. All values are reported in meV and follow the same sign convention as the NN constants in \autoref{tab:nn-exchange}.}
    \label{tab:nnn-exchange-fe3o4}
    \renewcommand\arraystretch{1.35}
    \setlength{\tabcolsep}{10pt}
    \begin{tabular}{|c|c|c|}
    \hline
    \textbf{Exchange constant} & \textbf{NN only} & \textbf{NN + NNN} \\
    \hline
    $J_{AB}$ or $J_{AB}^{(1)}$ & $-2.38$ & $-2.20$ \\
    $J_{BB}$ or $J_{BB}^{(1)}$ & $+0.56$ & $+0.49$ \\
    $J_{AA}$ or $J_{AA}^{(1)}$ & $-0.26$ & $-0.53$ \\
    \hline
    $J_{AB}^{(2)}$ & --- & $-0.213$ \\
    $J_{BB}^{(2)}$ & --- & $-0.071$ \\
    $J_{AA}^{(2)}$ & --- & $-0.013$ \\
    \hline
    \end{tabular}
\end{table}

\autoref{tab:nnn-exchange-fe3o4} shows that including the second pair-family shells leaves the dominant first-shell A--B antiferromagnetic interaction intact, but also renormalizes the fitted first-shell values relative to the NN-only model. In particular, $|J_{AB}^{(1)}|$ decreases slightly from $2.38$ to $2.20$ meV, $J_{BB}^{(1)}$ decreases from $0.56$ to $0.49$ meV, and $|J_{AA}^{(1)}|$ increases from $0.26$ to $0.53$ meV when the second-shell terms are fitted simultaneously.

A useful normalization is to compare each second-shell term to its corresponding first-shell value from the same NN+NNN fit,
\begin{equation*}
    \left|\frac{J_{AB}^{(2)}}{J_{AB}^{(1)}}\right| = 0.097, \qquad
    \left|\frac{J_{BB}^{(2)}}{J_{BB}^{(1)}}\right| = 0.145, \qquad
    \left|\frac{J_{AA}^{(2)}}{J_{AA}^{(1)}}\right| = 0.025 .
\end{equation*}
These ratios show that the second-shell terms remain smaller than the leading first-shell couplings, but they are not strictly negligible within a refined exchange fit. The negative sign of $J_{AB}^{(2)}$ indicates that the longer-range A--B channel reinforces, rather than competes with, the ferrimagnetic A--B coupling already established at first shell. By contrast, the B--B channel changes sign between first and second shell, with $J_{BB}^{(1)}>0$ but $J_{BB}^{(2)}<0$, indicating that the longer-range interactions on the octahedral network are weaker and more mixed than the leading NN term. The A--A second-shell coupling is also antiferromagnetic, but its very small magnitude indicates that the tetrahedral sub-lattice remains only weakly coupled beyond the first shell.

\begin{figure}[h]
    \centering
    \begin{subfigure}{0.49\linewidth}
        \includegraphics[width=\linewidth]{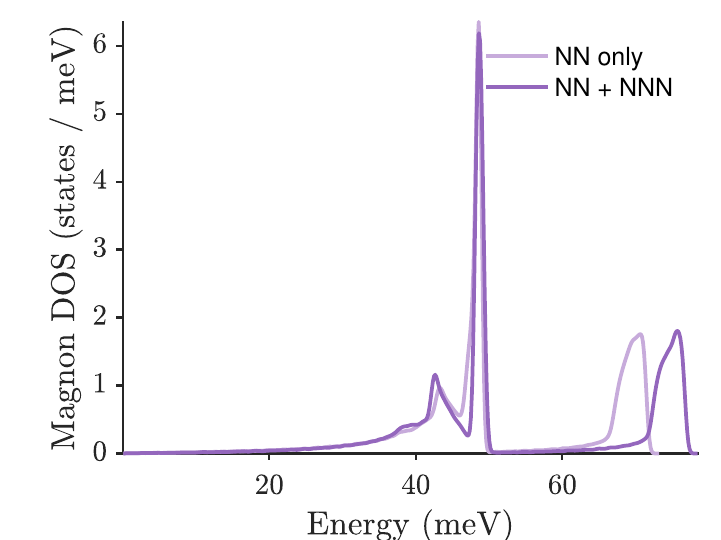}
        \caption{}
        \label{subfig:fe3o4_compare_mdos}
    \end{subfigure}\hfill
    \begin{subfigure}{0.49\linewidth}
        \includegraphics[width=\linewidth]{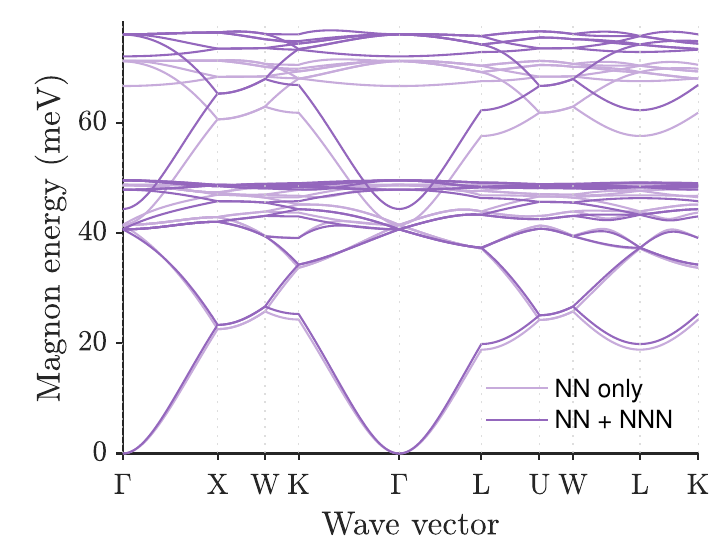}
        \caption{}
        \label{subfig:fe3o4_compare_mband}
    \end{subfigure}
    \caption{\small \em Comparison of the magnon spectrum of cubic \ce{Fe3O4} obtained from the NN-only and NN+NNN exchange models. (a) Magnon DOS. (b) Magnon band structure.}
    \label{fig:fe3o4_compare_nn_nnn}
\end{figure}

The effect of including the second-shell exchange terms on the spin-wave spectrum is shown in \autoref{fig:fe3o4_compare_nn_nnn}. The comparison indicates that the NN+NNN model does not qualitatively reconstruct the magnon spectrum of cubic \ce{Fe3O4}: the overall bandwidth, the principal branch groupings, and the main DOS features are all retained. However, visible quantitative shifts do appear in both the band structure and the DOS, which is consistent with the renormalization of the first-shell constants when the second-shell terms are fitted explicitly.

Overall, the shell-resolved NN+NNN constants provide a modest but physically meaningful refinement of the exchange landscape of cubic \ce{Fe3O4}. The dominant magnetic backbone still comes from the first-shell A--B antiferromagnetic interaction, but the extended model captures weak longer-range corrections and shows explicitly that the effective first-shell constants depend on whether those longer-range terms are omitted or included in the fit.

\end{appendices}

%%%%%%%%%%%%%%%%%%%%%%%%%%%%%%%%%%%%%%%%%%%%%%%%%%%%%%%%%%%%%%%%%%%%%
%% If you are using classical BibTeX rather than biblatex,
%% remove the \printbibliography and uncomment the \bibliograpy one
%%%%%%%%%%%%%%%%%%%%%%%%%%%%%%%%%%%%%%%%%%%%%%%%%%%%%%%%%%%%%%%%%%%%%
\printbibliography

@book{harrisModernFerritesVolume2023,
  title = {Modern Ferrites: {{Volume}} 2: {{Emerging}} Technologies and Applications},
  shorttitle = {Modern Ferrites},
  editor = {Harris, Vincent G.},
  year = 2023,
  publisher = {IEEE Press, Wiley},
  address = {Hoboken, NJ},
  isbn = {978-1-394-15613-9},
  langid = {english}
}

@book{somiyaHandbookAdvancedCeramics2013,
  title = {Handbook of {{Advanced Ceramics}}: {{Materials}}, {{Applications}}, {{Processing}}, and {{Properties}}},
  shorttitle = {Handbook of {{Advanced Ceramics}}},
  author = {Somiya, Shigeyuki},
  year = 2013,
  edition = {2nd ed},
  publisher = {Elsevier Science \& Technology},
  address = {San Diego},
  isbn = {978-0-12-385470-4},
  langid = {english}
}

@article{harrisModernMicrowaveFerrites2012,
  title = {Modern {{Microwave Ferrites}}},
  author = {Harris, Vincent G.},
  year = 2012,
  month = mar,
  journal = {IEEE Transactions on Magnetics},
  volume = 48,
  number = 3,
  pages = {1075--1104},
  issn = {0018-9464, 1941-0069},
}

@book{cullityIntroductionMagneticMaterials2008,
  title = {Introduction to {{Magnetic Materials}}},
  author = {Cullity, B. D. and Graham, C. D.},
  year = 2008,
  month = nov,
  edition = {1},
  publisher = {Wiley},
  isbn = {978-0-471-47741-9 978-0-470-38632-3},
  langid = {english}
}

@book{smitFerritesPhysicalProperties1959,
  title = {Ferrites; Physical Properties of Ferrimagnetic Oxides in Relation to Their Technical Applications. [{{Translated}} by {{G}}.{{E}}. {{Luton}}]},
  author = {Smit, J. and Wijn, Henricus Petrus Johannes},
  year = 1959,
  publisher = {Philips' Technical Library},
  address = {Eindhoven},
  langid = {english}
}

@techreport{JFEFerrite2022,
  title        = {High Resistivity High Initial Permeability Mn--Zn Ferrite Applied for 10 MHz Range},
  institution  = {JFE Steel Corporation / JFE Ferrite Co., Ltd.},
  type         = {JFE Technical Report},
  number       = {27},
  year         = {2022},
  month        = {March},
  url          = {https://www.jfe-steel.co.jp/en/research/report/027/pdf/027-05.pdf},
}

@article{thakurReviewMnZnFerrites2020,
  title = {A Review on {{MnZn}} Ferrites: {{Synthesis}}, Characterization and Applications},
  shorttitle = {A Review on {{MnZn}} Ferrites},
  author = {Thakur, Preeti and Chahar, Deepika and Taneja, Shilpa and Bhalla, Nikhil and Thakur, Atul},
  year = 2020,
  month = jul,
  journal = {Ceramics International},
  volume = 46,
  number = 10,
  pages = {15740--15763},
  issn = {02728842},
  langid = {english},
}

@article{balajiMagneticPropertiesMnFe2O42002,
  title = {Magnetic Properties of {{MnFe2O4}} Nanoparticles},
  author = {Balaji, G and Gajbhiye, N.S and Wilde, G and Weissm{\"u}ller, J},
  year = 2002,
  month = apr,
  journal = {Journal of Magnetism and Magnetic Materials},
  volume = 242,
  pages = {617--620},
  issn = {03048853},
  doi = {10.1016/S0304-8853(01)01043-5},
  urldate = {2025-09-12},
  copyright = {https://www.elsevier.com/tdm/userlicense/1.0/},
  langid = {english}
}

@article{heSoftMagneticMaterials2023,
  title = {Soft Magnetic Materials for Power Inductors: {{State}} of Art and Future Development},
  shorttitle = {Soft Magnetic Materials for Power Inductors},
  author = {He, Jiayi and Yuan, Han and Nie, Min and Guo, Hai and Yu, Hongya and Liu, Zhongwu and Sun, Rong},
  year = 2023,
  month = dec,
  journal = {Materials Today Electronics},
  volume = 6,
  pages = {100066},
  issn = {27729494},
  langid = {english}
}

@article{berettaSteepestEntropyAscent2014,
  title = {Steepest Entropy Ascent Model for Far-Nonequilibrium Thermodynamics: {{Unified}} Implementation of the Maximum Entropy Production Principle},
  shorttitle = {Steepest Entropy Ascent Model for Far-Nonequilibrium Thermodynamics},
  author = {Beretta, Gian Paolo},
  year = 2014,
  month = oct,
  journal = {Physical Review E},
  volume = 90,
  number = 4,
  pages = {042113},
  issn = {1539-3755, 1550-2376},
  doi = {10.1103/PhysRevE.90.042113},
  urldate = {2023-06-05},
  langid = {english}
}

@article{liSteepestEntropyAscent2018,
  title = {Steepest Entropy Ascent Quantum Thermodynamic Model of Electron and Phonon Transport},
  author = {Li, Guanchen and {von Spakovsky}, Michael R. and Hin, Celine},
  year = 2018,
  month = jan,
  journal = {Physical Review B},
  volume = 97,
  number = 2,
  pages = {024308},
  issn = {2469-9950, 2469-9969},
  doi = {10.1103/PhysRevB.97.024308},
  urldate = {2023-06-05},
  langid = {english}
}

@article{liGeneralizedThermodynamicRelations2016,
  title = {Generalized Thermodynamic Relations for a System Experiencing Heat and Mass Diffusion in the Far-from-Equilibrium Realm Based on Steepest Entropy Ascent},
  author = {Li, Guanchen and {von Spakovsky}, Michael R.},
  year = 2016,
  month = sep,
  journal = {Physical Review E},
  volume = 94,
  number = 3,
  pages = {032117},
  issn = {2470-0045, 2470-0053},
  doi = {10.1103/PhysRevE.94.032117},
  urldate = {2023-06-05},
  langid = {english}
}

@article{liSteepestentropyascentQuantumThermodynamic2016,
  title = {Steepest-Entropy-Ascent Quantum Thermodynamic Modeling of the Relaxation Process of Isolated Chemically Reactive Systems Using Density of States and the Concept of Hypoequilibrium State},
  author = {Li, Guanchen and {von Spakovsky}, Michael R.},
  year = 2016,
  month = jan,
  journal = {Physical Review E},
  volume = 93,
  number = 1,
  pages = {012137},
  issn = {2470-0045, 2470-0053},
  doi = {10.1103/PhysRevE.93.012137},
  urldate = {2023-06-05},
  langid = {english}
}

@article{liSteepestentropyascentModelMesoscopic2018,
  title = {Steepest-Entropy-Ascent Model of Mesoscopic Quantum Systems Far from Equilibrium along with Generalized Thermodynamic Definitions of Measurement and Reservoir},
  author = {Li, Guanchen and von Spakovsky, Michael R.},
  year = 2018,
  month = oct,
  journal = {Physical Review E},
  volume = 98,
  number = 4,
  pages = {042113},
  issn = {2470-0045, 2470-0053},
  doi = {10.1103/PhysRevE.98.042113},
  urldate = {2023-06-05},
  langid = {english}
}

@misc{wordenPredictingCoupledElectron2024,
  title = {Predicting {{Coupled Electron}} and {{Phonon Transport Using Steepest-Entropy-Ascent Quantum Thermodynamics}}},
  author = {Worden, J. A. and von Spakovsky, M. R. and Hin, C.},
  year = 2024,
  month = nov,
  number = {arXiv:2307.12478},
  eprint = {2307.12478},
  primaryclass = {physics},
  publisher = {arXiv},
  doi = {10.48550/arXiv.2307.12478},
  urldate = {2025-01-27},
  archiveprefix = {arXiv},
}

@article{kresseEfficiencyAbinitioTotal1996,
  title = {Efficiency of Ab-Initio Total Energy Calculations for Metals and Semiconductors Using a Plane-Wave Basis Set},
  author = {Kresse, G. and Furthm{\"u}ller, J.},
  year = 1996,
  month = jul,
  journal = {Computational Materials Science},
  volume = {6},
  number = {1},
  pages = {15--50},
  issn = {09270256},
  doi = {10.1016/0927-0256(96)00008-0},
  urldate = {2025-04-09},
  copyright = {https://www.elsevier.com/tdm/userlicense/1.0/},
  langid = {english}
}

@article{kresseEfficientIterativeSchemes1996,
  title = {Efficient Iterative Schemes for {\emph{Ab Initio}} Total-Energy Calculations Using a Plane-Wave Basis Set},
  author = {Kresse, G. and Furthm{\"u}ller, J.},
  year = 1996,
  month = oct,
  journal = {Physical Review B},
  volume = {54},
  number = {16},
  pages = {11169--11186},
  issn = {0163-1829, 1095-3795},
  doi = {10.1103/PhysRevB.54.11169},
  urldate = {2025-04-09},
  copyright = {http://link.aps.org/licenses/aps-default-license},
  langid = {english}
}

@article{kresseInitioMolecularDynamics1993,
  title = {{\emph{Ab Initio}} Molecular Dynamics for Liquid Metals},
  author = {Kresse, G. and Hafner, J.},
  year = 1993,
  month = jan,
  journal = {Physical Review B},
  volume = {47},
  number = {1},
  pages = {558--561},
  issn = {0163-1829, 1095-3795},
  doi = {10.1103/PhysRevB.47.558},
  urldate = {2025-04-09},
  copyright = {http://link.aps.org/licenses/aps-default-license},
  langid = {english}
}

@article{kresseTheoryCrystalStructures1994,
  title = {Theory of the Crystal Structures of Selenium and Tellurium: {{The}} Effect of Generalized-Gradient Corrections to the Local-Density Approximation},
  shorttitle = {Theory of the Crystal Structures of Selenium and Tellurium},
  author = {Kresse, G. and Furthm{\"u}ller, J. and Hafner, J.},
  year = 1994,
  month = nov,
  journal = {Physical Review B},
  volume = {50},
  number = {18},
  pages = {13181--13185},
  issn = {0163-1829, 1095-3795},
  doi = {10.1103/PhysRevB.50.13181},
  urldate = {2025-04-09},
  copyright = {http://link.aps.org/licenses/aps-default-license},
  langid = {english}
}

@article{kresseUltrasoftPseudopotentialsProjector1999,
  title = {From Ultrasoft Pseudopotentials to the Projector Augmented-Wave Method},
  author = {Kresse, G. and Joubert, D.},
  year = 1999,
  month = jan,
  journal = {Physical Review B},
  volume = {59},
  number = {3},
  pages = {1758--1775},
  issn = {0163-1829, 1095-3795},
  doi = {10.1103/PhysRevB.59.1758},
  urldate = {2025-04-09},
  copyright = {http://link.aps.org/licenses/aps-default-license},
  langid = {english}
}

@article{beckePerspectiveFiftyYears2014,
  title = {Perspective: {{Fifty}} Years of Density-Functional Theory in Chemical Physics},
  shorttitle = {Perspective},
  author = {Becke, Axel D.},
  year = 2014,
  month = may,
  journal = {The Journal of Chemical Physics},
  volume = 140,
  number = 18,
  pages = {18A301},
  issn = {0021-9606, 1089-7690},
  doi = {10.1063/1.4869598},
  langid = {english},
}

@article{mengWhenDensityFunctional2016,
  title = {When {{Density Functional Approximations Meet Iron Oxides}}},
  author = {Meng, Yu and Liu, Xing-Wu and Huo, Chun-Fang and Guo, Wen-Ping and Cao, Dong-Bo and Peng, Qing and Dearden, Albert and Gonze, Xavier and Yang, Yong and Wang, Jianguo and Jiao, Haijun and Li, Yongwang and Wen, Xiao-Dong},
  year = 2016,
  month = oct,
  journal = {Journal of Chemical Theory and Computation},
  volume = 12,
  number = 10,
  pages = {5132--5144},
  issn = {1549-9618, 1549-9626},
  doi = {10.1021/acs.jctc.6b00640},
  langid = {english},
}

@book{pavariniCorrelatedElectronsModels2012,
  title = {Correlated Electrons: From Models to Materials: Lecture Notes of the {{Autumn School Correlated Electrons}} 2012: At {{Forschungszentrum J{\"u}lich}}, 3-7 {{September}} 2012},
  shorttitle = {Correlated Electrons},
  editor = {Pavarini, Eva and Koch, Erik and Anders, Frithjof and Jarrell, Mark and {Institute for Advanced Simulation} and {German Research School for Simulation Sciences}},
  year = 2012,
  series = {Schriften Des {{Forschungszentrums J{\"u}lich}}. {{Reihe Modeling}} and {{Simulation}}},
  number = {Band 2},
  publisher = {Forschungszentrum J{\"u}lich, Zentralbibliothek, Verl},
  address = {J{\"u}lich},
  isbn = {978-3-89336-796-2},
  langid = {english},
}

@article{anisimovFirstprinciplesCalculationsElectronic1997,
  title = {First-Principles Calculations of the Electronic Structure and Spectra of Strongly Correlated Systems: The {{{\textbf{LDA}}}} + {{{\emph{U}}}} Method},
  shorttitle = {First-Principles Calculations of the Electronic Structure and Spectra of Strongly Correlated Systems},
  author = {Anisimov, Vladimir I and Aryasetiawan, F and Lichtenstein, A I},
  year = 1997,
  month = jan,
  journal = {Journal of Physics: Condensed Matter},
  volume = 9,
  number = 4,
  pages = {767--808},
  issn = {0953-8984, 1361-648X},
  doi = {10.1088/0953-8984/9/4/002},
}

@article{bajajMolecularDFT+UTransferable2021,
  title = {Molecular {{DFT}}+{{U}}: {{A Transferable}}, {{Low-Cost Approach}} to {{Eliminate Delocalization Error}}},
  shorttitle = {Molecular {{DFT}}+{{U}}},
  author = {Bajaj, Akash and Kulik, Heather J.},
  year = 2021,
  month = apr,
  journal = {The Journal of Physical Chemistry Letters},
  volume = 12,
  number = 14,
  pages = {3633--3640},
  issn = {1948-7185, 1948-7185},
  doi = {10.1021/acs.jpclett.1c00796},
  urldate = {2025-09-12},
  copyright = {https://doi.org/10.15223/policy-029},
  langid = {english},
}

@article{streltsovOrbitalPhysicsTransition2017,
  title = {Orbital Physics in Transition Metal Compounds: New Trends},
  shorttitle = {Orbital Physics in Transition Metal Compounds},
  author = {Streltsov, S V and Khomskii, D I},
  year = 2017,
  month = nov,
  journal = {Physics-Uspekhi},
  volume = 60,
  number = 11,
  pages = {1121--1146},
  issn = {1063-7869, 1468-4780},
  doi = {10.3367/UFNe.2017.08.038196},
}

@article{linscottRoleSpinCalculation2018,
  title = {Role of Spin in the Calculation of {{Hubbard U}} and {{Hund}}'s {{J}} Parameters from First Principles},
  author = {Linscott, Edward B. and Cole, Daniel J. and Payne, Michael C. and O'Regan, David D.},
  year = 2018,
  month = dec,
  journal = {Physical Review B},
  volume = 98,
  number = 23,
  pages = {235157},
  issn = {2469-9950, 2469-9969},
  doi = {10.1103/PhysRevB.98.235157},
  urldate = {2025-09-12},
  langid = {english},
}

@article{georgesStrongCorrelationsHunds2013,
  title = {Strong {{Correlations}} from {{Hund}}'s {{Coupling}}},
  author = {Georges, Antoine and Medici, Luca De' and Mravlje, Jernej},
  year = 2013,
  month = apr,
  journal = {Annual Review of Condensed Matter Physics},
  volume = 4,
  number = 1,
  pages = {137--178},
  issn = {1947-5454, 1947-5462},
  doi = {10.1146/annurev-conmatphys-020911-125045},
  langid = {english},
}

@article{leiriacampojrExtendedDFTMethod2010,
  title = {Extended {{DFT}} + {{{\emph{U}}}} + {{{\emph{V}}}} Method with On-Site and Inter-Site Electronic Interactions},
  author = {Leiria Campo Jr, Vivaldo and Cococcioni, Matteo},
  year = 2010,
  month = feb,
  journal = {Journal of Physics: Condensed Matter},
  volume = 22,
  number = 5,
  pages = {055602},
  issn = {0953-8984, 1361-648X},
  doi = {10.1088/0953-8984/22/5/055602},
}

@article{wangLocalProjectionDensity2016,
  title = {The Local Projection in the Density Functional Theory plus {{{\emph{U}}}} Approach: {{A}} Critical Assessment},
  shorttitle = {The Local Projection in the Density Functional Theory plus {{{\emph{U}}}} Approach},
  author = {Wang, Yue-Chao and Chen, Ze-Hua and Jiang, Hong},
  year = 2016,
  month = apr,
  journal = {The Journal of Chemical Physics},
  volume = 144,
  number = 14,
  pages = {144106},
  issn = {0021-9606, 1089-7690},
  doi = {10.1063/1.4945608},
  langid = {english}
}

@article{mooreHighthroughputDeterminationHubbard2024,
  title = {High-Throughput Determination of {{Hubbard U}} and {{Hund J}} Values for Transition Metal Oxides via the Linear Response Formalism},
  author = {Moore, Guy C. and Horton, Matthew K. and Linscott, Edward and Ganose, Alexander M. and Siron, Martin and O'Regan, David D. and Persson, Kristin A.},
  year = 2024,
  month = jan,
  journal = {Physical Review Materials},
  volume = 8,
  number = 1,
  pages = {014409},
  issn = {2475-9953},
  doi = {10.1103/PhysRevMaterials.8.014409},
  urldate = {2025-09-12},
  langid = {english}
}

@article{cococcioniLinearResponseApproach2005a,
  title = {Linear Response Approach to the Calculation of the Effective Interaction Parameters in the {{LDA}} + {{U}} Method},
  author = {Cococcioni, Matteo and De Gironcoli, Stefano},
  year = 2005,
  month = jan,
  journal = {Physical Review B},
  volume = 71,
  number = 3,
  pages = {035105},
  issn = {1098-0121, 1550-235X},
  doi = {10.1103/PhysRevB.71.035105},
  urldate = {2025-09-12},
  copyright = {http://link.aps.org/licenses/aps-default-license},
  langid = {english},
}

@article{vaugierHubbardHundExchange2012,
  title = {Hubbard {{U}} and {{Hund}} Exchange {{J}} in Transition Metal Oxides: {{Screening}} versus Localization Trends from Constrained Random Phase Approximation},
  author = {Vaugier, Lo{\"i}g and Jiang, Hong and Biermann, Silke},
  year = 2012,
  month = oct,
  journal = {Physical Review B},
  volume = 86,
  number = 16,
  pages = {165105},
  issn = {1098-0121, 1550-235X},
  doi = {10.1103/PhysRevB.86.165105},
  urldate = {2025-09-12},
  copyright = {http://link.aps.org/licenses/aps-default-license},
  langid = {english},
}

@article{liechtensteinDensityfunctionalTheoryStrong1995a,
  title = {Density-Functional Theory and Strong Interactions: {{Orbital}} Ordering in {{Mott-Hubbard}} Insulators},
  shorttitle = {Density-Functional Theory and Strong Interactions},
  author = {Liechtenstein, A. I. and Anisimov, V. I. and Zaanen, J.},
  year = 1995,
  month = aug,
  journal = {Physical Review B},
  volume = 52,
  number = 8,
  pages = {R5467-R5470},
  issn = {0163-1829, 1095-3795},
  doi = {10.1103/PhysRevB.52.R5467},
  urldate = {2025-09-12},
  copyright = {http://link.aps.org/licenses/aps-default-license},
  langid = {english}
}

@article{himmetogluHubbardcorrectedDFTEnergy2014,
  title = {Hubbard-Corrected {{DFT}} Energy Functionals: {{The LDA}}+{{U}} Description of Correlated Systems},
  shorttitle = {Hubbard-Corrected {{DFT}} Energy Functionals},
  author = {Himmetoglu, Burak and Floris, Andrea and De Gironcoli, Stefano and Cococcioni, Matteo},
  year = 2014,
  month = jan,
  journal = {International Journal of Quantum Chemistry},
  volume = 114,
  number = 1,
  pages = {14--49},
  issn = {00207608},
  doi = {10.1002/qua.24521},
  urldate = {2025-09-12},
  copyright = {http://doi.wiley.com/10.1002/tdm\_license\_1.1},
  langid = {english},
}

@article{glasser1963spin,
  title={Spin wave spectra of magnetite},
  author={Glasser, M Lawrence and Milford, Frederick J},
  journal={Physical Review},
  volume=130,
  number={5},
  pages={1783},
  year={1963},
  publisher={APS}
}

@article{tothLinearSpinWave2015a,
  title = {Linear Spin Wave Theory for Single-{{Q}} Incommensurate Magnetic Structures},
  author = {Toth, S and Lake, B},
  year = 2015,
  month = apr,
  journal = {Journal of Physics: Condensed Matter},
  volume = {27},
  number = {16},
  pages = {166002},
  issn = {0953-8984, 1361-648X},
  doi = {10.1088/0953-8984/27/16/166002},
  urldate = {2025-04-09}
}

@article{mcqueeneyInvestigationPresenceCharge2006,
  title = {Investigation of the Presence of Charge Order in Magnetite by Measurement of the Spin Wave Spectrum},
  author = {McQueeney, R. J. and Yethiraj, M. and Montfrooij, W. and Gardner, J. S. and Metcalf, P. and Honig, J. M.},
  year = 2006,
  month = may,
  journal = {Physical Review B},
  volume = {73},
  number = {17},
  pages = {174409},
  issn = {1098-0121, 1550-235X},
  doi = {10.1103/PhysRevB.73.174409},
  urldate = {2025-12-19},
  copyright = {http://link.aps.org/licenses/aps-default-license},
  langid = {english}
}

@article{sunTemperatureFrequencyCharacteristics2011,
  title = {Temperature and Frequency Characteristics of Low-Loss {{MnZn}} Ferrite in a Wide Temperature Range},
  author = {Sun, Ke and Lan, Zhongwen and Yu, Zhong and Xu, Zhiyong and Jiang, Xiaona and Wang, Zihui and Liu, Zhi and Luo, Ming},
  year = 2011,
  month = may,
  journal = {Journal of Applied Physics},
  volume = {109},
  number = {10},
  pages = {106103},
  issn = {0021-8979, 1089-7550},
  doi = {10.1063/1.3583551},
  langid = {english}
}

@article{srivastavaAngleDependenceFerromagnetic1999,
  title = {Angle Dependence of the Ferromagnetic Resonance Linewidth and Two Magnon Losses in Pulsed Laser Deposited Films of Yttrium Iron Garnet, {{MnZn}} Ferrite, and {{NiZn}} Ferrite},
  author = {Srivastava, Anuj K. and Hurben, Michael J. and Wittenauer, Michael A. and Kabos, Pavel and Patton, Carl E. and Ramesh, R. and Dorsey, Paul C. and Chrisey, Douglas B.},
  year = 1999,
  month = jun,
  journal = {Journal of Applied Physics},
  volume = {85},
  number = {11},
  pages = {7838--7848},
  issn = {0021-8979, 1089-7550},
  doi = {10.1063/1.370595},
  langid = {english}
}

@article{santos-carballalFirstprinciplesStudyInversion2015,
  title = {First-Principles Study of the Inversion Thermodynamics and Electronic Structure of {{Fe M}} 2 {{X}} 4 (Thio)Spinels ( {{M}} = {{Cr}} , {{Mn}}, {{Co}}, {{Ni}}; {{X}} = {{O}} , {{S}})},
  author = {{Santos-Carballal}, David and Roldan, Alberto and {Grau-Crespo}, Ricardo and De Leeuw, Nora H.},
  year = 2015,
  month = may,
  journal = {Physical Review B},
  volume = {91},
  number = {19},
  pages = {195106},
  issn = {1098-0121, 1550-235X},
  doi = {10.1103/PhysRevB.91.195106},
  langid = {english},
}

@article{naveasFirstPrinciplesCalculationsMagnetite2023a,
  title = {First-{{Principles Calculations}} of {{Magnetite}} ({{Fe}}{\textsubscript{3}} {{O}}{\textsubscript{4}} ) above the {{Verwey Temperature}} by {{Using Self-Consistent DFT}} + {{{\emph{U}}}} + {{{\emph{V}}}}},
  author = {Naveas, Nelson and Pulido, Ruth and Marini, Carlo and Gargiani, Pierluigi and {Hernandez-Montelongo}, Jacobo and Brito, Ivan and {Manso-Silv{\'a}n}, Miguel},
  year = 2023,
  month = dec,
  journal = {Journal of Chemical Theory and Computation},
  volume = {19},
  number = {23},
  pages = {8610--8623},
  issn = {1549-9618, 1549-9626},
  doi = {10.1021/acs.jctc.3c00860},
  langid = {english}
}

@article{sakuraiCationDistributionValence2008,
  title = {Cation Distribution and Valence State in {{Mn}}--{{Zn}} Ferrite Examined by Synchrotron {{X-rays}}},
  author = {Sakurai, Syoichi and Sasaki, Satoshi and Okube, Maki and Ohara, Hiroki and Toyoda, Takeshi},
  year = 2008,
  month = oct,
  journal = {Physica B: Condensed Matter},
  volume = {403},
  number = {19-20},
  pages = {3589--3595},
  issn = {09214526},
  doi = {10.1016/j.physb.2008.05.035},
  langid = {english}
}

@article{dudarevElectronenergylossSpectraStructural1998,
  title = {Electron-Energy-Loss Spectra and the Structural Stability of Nickel Oxide: {{An LSDA}}+{{U}} Study},
  shorttitle = {Electron-Energy-Loss Spectra and the Structural Stability of Nickel Oxide},
  author = {Dudarev, S. L. and Botton, G. A. and Savrasov, S. Y. and Humphreys, C. J. and Sutton, A. P.},
  year = 1998,
  month = jan,
  journal = {Physical Review B},
  volume = {57},
  number = {3},
  pages = {1505--1509},
  issn = {0163-1829, 1095-3795},
  doi = {10.1103/PhysRevB.57.1505},
  copyright = {http://link.aps.org/licenses/aps-default-license},
  langid = {english}
}

@article{perdewGeneralizedGradientApproximation1996a,
  title = {Generalized {{Gradient Approximation Made Simple}}},
  author = {Perdew, John P. and Burke, Kieron and Ernzerhof, Matthias},
  year = 1996,
  month = oct,
  journal = {Physical Review Letters},
  volume = {77},
  number = {18},
  pages = {3865--3868},
  issn = {0031-9007, 1079-7114},
  doi = {10.1103/PhysRevLett.77.3865},
  copyright = {http://link.aps.org/licenses/aps-default-license},
  langid = {english}
}

@article{monkhorstSpecialPointsBrillouinzone1976,
  title = {Special Points for {{Brillouin-zone}} Integrations},
  author = {Monkhorst, Hendrik J. and Pack, James D.},
  year = 1976,
  month = jun,
  journal = {Physical Review B},
  volume = {13},
  number = {12},
  pages = {5188--5192},
  issn = {0556-2805},
  doi = {10.1103/PhysRevB.13.5188},
  copyright = {http://link.aps.org/licenses/aps-default-license},
  langid = {english}
}

@book{baderAtomsMoleculesQuantum1994,
  title = {Atoms in Molecules: A Quantum Theory},
  shorttitle = {Atoms in Molecules},
  author = {Bader, Richard F. W.},
  year = 1994,
  series = {The {{International}} Series of Monographs on Chemistry},
  number = {22},
  publisher = {Clarendon Press ; Oxford University Press},
  address = {Oxford [England] : New York},
  isbn = {978-0-19-855865-1},
  lccn = {QD462 .B33 1994}
}

@article{henkelmanFastRobustAlgorithm2006,
  title = {A Fast and Robust Algorithm for {{Bader}} Decomposition of Charge Density},
  author = {Henkelman, Graeme and Arnaldsson, Andri and J{\'o}nsson, Hannes},
  year = 2006,
  month = jun,
  journal = {Computational Materials Science},
  volume = {36},
  number = {3},
  pages = {354--360},
  issn = {09270256},
  doi = {10.1016/j.commatsci.2005.04.010},
  copyright = {https://www.elsevier.com/tdm/userlicense/1.0/},
  langid = {english}
}

@article{yuAccurateEfficientAlgorithm2011,
  title = {Accurate and Efficient Algorithm for {{Bader}} Charge Integration},
  author = {Yu, Min and Trinkle, Dallas R.},
  year = 2011,
  month = feb,
  journal = {The Journal of Chemical Physics},
  volume = {134},
  number = {6},
  pages = {064111},
  issn = {0021-9606, 1089-7690},
  doi = {10.1063/1.3553716},
  langid = {english},
}

@article{tangGridbasedBaderAnalysis2009,
  title = {A Grid-Based {{Bader}} Analysis Algorithm without Lattice Bias},
  author = {Tang, W and Sanville, E and Henkelman, G},
  year = 2009,
  month = feb,
  journal = {Journal of Physics: Condensed Matter},
  volume = {21},
  number = {8},
  pages = {084204},
  issn = {0953-8984, 1361-648X},
  doi = {10.1088/0953-8984/21/8/084204},
}

@article{lambertEvaluationFirstprinciplesHubbard2024,
  title = {Evaluation of First-Principles {{Hubbard}} and {{Hund}} Corrected {{DFT}} for Defect Formation Energies in Non-Magnetic Transition Metal Oxides},
  author = {Lambert, Daniel S. and O'Regan, David D.},
  year = 2024,
  journal = {RSC Advances},
  volume = {14},
  number = {52},
  pages = {38645--38659},
  issn = {2046-2069},
  doi = {10.1039/D4RA07774A},
  langid = {english},
}

@article{heisenbergZurTheorieFerromagnetismus1928,
  title = {{Zur Theorie des Ferromagnetismus}},
  author = {Heisenberg, W.},
  year = 1928,
  month = sep,
  journal = {Zeitschrift f{\"u}r Physik},
  volume = {49},
  number = {9-10},
  pages = {619--636},
  issn = {1434-6001, 1434-601X},
  doi = {10.1007/BF01328601},
  copyright = {http://www.springer.com/tdm},
  langid = {ngerman}
}

@article{liechtensteinLocalSpinDensity1987,
  title = {Local Spin Density Functional Approach to the Theory of Exchange Interactions in Ferromagnetic Metals and Alloys},
  author = {Liechtenstein, A.I. and Katsnelson, M.I. and Antropov, V.P. and Gubanov, V.A.},
  year = 1987,
  month = may,
  journal = {Journal of Magnetism and Magnetic Materials},
  volume = {67},
  number = {1},
  pages = {65--74},
  issn = {03048853},
  doi = {10.1016/0304-8853(87)90721-9},
  langid = {english}
}

@article{xiangMagneticPropertiesEnergymapping2013,
  title = {Magnetic Properties and Energy-Mapping Analysis},
  author = {Xiang, Hongjun and Lee, Changhoon and Koo, Hyun-Joo and Gong, Xingao and Whangbo, Myung-Hwan},
  year = 2013,
  journal = {Dalton Trans.},
  volume = {42},
  number = {4},
  pages = {823--853},
  issn = {1477-9226, 1477-9234},
  doi = {10.1039/C2DT31662E},
  langid = {english}
}

@article{srivastavaExchangeConstantsSpinel1979,
  title = {Exchange Constants in Spinel Ferrites},
  author = {Srivastava, C. M. and Srinivasan, G. and Nanadikar, N. G.},
  year = 1979,
  month = jan,
  journal = {Physical Review B},
  volume = {19},
  number = {1},
  pages = {499--508},
  issn = {0163-1829},
  doi = {10.1103/PhysRevB.19.499},
  copyright = {http://link.aps.org/licenses/aps-default-license},
  langid = {english}
}

@incollection{harrisGoodenoughKanamoriAnderson2022,
  title = {Goodenough--{{Kanamori}}--{{Anderson Rules}} of {{Superexchange Applied}} to {{Ferrite Systems}}},
  booktitle = {Modern {{Ferrites}}},
  author = {Harris, Vincent G. and Andalib, Parisa},
  editor = {Harris, Vincent G.},
  year = 2022,
  month = nov,
  edition = {1},
  pages = {31--67},
  publisher = {Wiley},
  doi = {10.1002/9781394156146.ch2},
  isbn = {978-1-394-15613-9 978-1-394-15614-6},
  langid = {english}
}

@misc{singhStoryMagnetismHeisenberg2018,
  title = {The Story of Magnetism: From {{Heisenberg}}, {{Slater}}, and {{Stoner}} to {{Van Vleck}}, and the Issues of Exchange and Correlation},
  shorttitle = {The Story of Magnetism},
  author = {Singh, Navinder},
  year = 2018,
  publisher = {arXiv},
  doi = {10.48550/ARXIV.1807.11291},
}

@article{bercoffExchangeConstantsTransfer1997,
  title = {Exchange Constants and Transfer Integrals of Spinel Ferrites},
  author = {Bercoff, P.G. and Bertorello, H.R.},
  year = 1997,
  month = may,
  journal = {Journal of Magnetism and Magnetic Materials},
  volume = {169},
  number = {3},
  pages = {314--322},
  issn = {03048853},
  doi = {10.1016/S0304-8853(96)00748-2},
  langid = {english}
}

@article{ciofiniMappingManyelectronGeneralised2005,
  title = {Mapping the Many-Electron Generalised Spin-Exchange {{Hamiltonian}} to Accurate Post-{{HF}} Calculations},
  author = {Ciofini, Ilaria and Adamo, Carlo and Barone, Vincenzo and Berthier, Gaston and Rassat, Andr{\'e}},
  year = 2005,
  month = mar,
  journal = {Chemical Physics},
  volume = {309},
  number = {2-3},
  pages = {133--141},
  issn = {03010104},
  doi = {10.1016/j.chemphys.2004.09.001},
  langid = {english}
}

@article{katsnelsonFirstprinciplesCalculationsMagnetic2000,
  title = {First-Principles Calculations of Magnetic Interactions in Correlated Systems},
  author = {Katsnelson, M. I. and Lichtenstein, A. I.},
  year = 2000,
  month = apr,
  journal = {Physical Review B},
  volume = {61},
  number = {13},
  pages = {8906--8912},
  issn = {0163-1829, 1095-3795},
  doi = {10.1103/PhysRevB.61.8906},
  langid = {english},
}

@article{holsteinFieldDependenceIntrinsic1940,
  title = {Field {{Dependence}} of the {{Intrinsic Domain Magnetization}} of a {{Ferromagnet}}},
  author = {Holstein, T. and Primakoff, H.},
  year = 1940,
  month = dec,
  journal = {Physical Review},
  volume = {58},
  number = {12},
  pages = {1098--1113},
  issn = {0031-899X},
  doi = {10.1103/PhysRev.58.1098},
  langid = {english}
}

@book{auerbachInteractingElectronsQuantum1994,
  title = {Interacting {{Electrons}} and {{Quantum Magnetism}}},
  author = {Auerbach, Assa},
  editor = {Birman, Joseph L. and Lynn, Jeffrey W. and Silverman, Mark P. and Stanley, H. Eugene and Voloshin, Mikhail},
  year = 1994,
  series = {Graduate {{Texts}} in {{Contemporary Physics}}},
  publisher = {Springer New York},
  address = {New York, NY},
  doi = {10.1007/978-1-4612-0869-3},
  isbn = {978-1-4612-6928-1 978-1-4612-0869-3},
}

@article{colpaDiagonalizationQuadraticBoson1978,
  title = {Diagonalization of the Quadratic Boson Hamiltonian},
  author = {Colpa, J.H.P.},
  year = 1978,
  month = sep,
  journal = {Physica A: Statistical Mechanics and its Applications},
  volume = {93},
  number = {3-4},
  pages = {327--353},
  issn = {03784371},
  doi = {10.1016/0378-4371(78)90160-7},
  langid = {english}
}

@article{camleyConsequencesDzyaloshinskiiMoriyaInteraction2023,
  title = {Consequences of the {{Dzyaloshinskii-Moriya}} Interaction},
  author = {Camley, Robert E. and Livesey, Karen L.},
  year = 2023,
  month = aug,
  journal = {Surface Science Reports},
  volume = {78},
  number = {3},
  pages = {100605},
  issn = {01675729},
  doi = {10.1016/j.surfrep.2023.100605},
  langid = {english}
}

@article{togoFirstprinciplesCalculationsFerroelastic2008,
  title = {First-Principles Calculations of the Ferroelastic Transition between Rutile-Type and {{CaCl}} 2 -Type {{SiO}} 2 at High Pressures},
  author = {Togo, Atsushi and Oba, Fumiyasu and Tanaka, Isao},
  year = 2008,
  month = oct,
  journal = {Physical Review B},
  volume = {78},
  number = {13},
  pages = {134106},
  issn = {1098-0121, 1550-235X},
  doi = {10.1103/PhysRevB.78.134106},
  langid = {english}
}

@article{togoFirstPrinciplesPhonon2015a,
  title = {First Principles Phonon Calculations in Materials Science},
  author = {Togo, Atsushi and Tanaka, Isao},
  year = 2015,
  month = nov,
  journal = {Scripta Materialia},
  volume = {108},
  pages = {1--5},
  issn = {13596462},
  doi = {10.1016/j.scriptamat.2015.07.021},
  langid = {english},
}

@article{parlinskiFirstPrinciplesDeterminationSoft1997,
  title = {First-{{Principles Determination}} of the {{Soft Mode}} in {{Cubic ZrO}} 2},
  author = {Parlinski, K. and Li, Z. Q. and Kawazoe, Y.},
  year = 1997,
  month = may,
  journal = {Physical Review Letters},
  volume = {78},
  number = {21},
  pages = {4063--4066},
  issn = {0031-9007, 1079-7114},
  doi = {10.1103/PhysRevLett.78.4063},
  langid = {english}
}

@article{salaVanHoveSingularity2021,
  title = {Van {{Hove}} Singularity in the Magnon Spectrum of the Antiferromagnetic Quantum Honeycomb Lattice},
  author = {Sala, G. and Stone, M. B. and Rai, Binod K. and May, A. F. and Laurell, Pontus and Garlea, V. O. and Butch, N. P. and Lumsden, M. D. and Ehlers, G. and Pokharel, G. and Podlesnyak, A. and Mandrus, D. and Parker, D. S. and Okamoto, S. and Hal{\'a}sz, G{\'a}bor B. and Christianson, A. D.},
  year = 2021,
  month = jan,
  journal = {Nature Communications},
  volume = {12},
  number = {1},
  pages = {171},
  issn = {2041-1723},
  doi = {10.1038/s41467-020-20335-5},
  langid = {english},
}

@article{vanhoveOccurrenceSingularitiesElastic1953,
  title = {The {{Occurrence}} of {{Singularities}} in the {{Elastic Frequency Distribution}} of a {{Crystal}}},
  author = {Van Hove, L{\'e}on},
  year = 1953,
  month = mar,
  journal = {Physical Review},
  volume = {89},
  number = {6},
  pages = {1189--1193},
  issn = {0031-899X},
  doi = {10.1103/PhysRev.89.1189},
  langid = {english}
}

@article{bockDelithiationSpinelFerrites2020,
  title = {({{De}})Lithiation of Spinel Ferrites {{Fe}}{\textsubscript{3}} {{O}}{\textsubscript{4}} , {{MgFe}}{\textsubscript{2}} {{O}}{\textsubscript{4}} , and {{ZnFe}}{\textsubscript{2}} {{O}}{\textsubscript{4}} : A Combined Spectroscopic, Diffraction and Theory Study},
  shorttitle = {({{De}})Lithiation of Spinel Ferrites {{Fe}}{\textsubscript{3}} {{O}}{\textsubscript{4}} , {{MgFe}}{\textsubscript{2}} {{O}}{\textsubscript{4}} , and {{ZnFe}}{\textsubscript{2}} {{O}}{\textsubscript{4}}},
  author = {Bock, David C. and Tallman, Killian R. and Guo, Haoyue and Quilty, Calvin and Yan, Shan and Smith, Paul F. and Zhang, Bingjie and Lutz, Diana M. and McCarthy, Alison H. and Huie, Matthew M. and Burnett, Veronica and Bruck, Andrea M. and Marschilok, Amy C. and Takeuchi, Esther S. and Liu, Ping and Takeuchi, Kenneth J.},
  year = 2020,
  journal = {Physical Chemistry Chemical Physics},
  volume = {22},
  number = {45},
  pages = {26200--26215},
  issn = {1463-9076, 1463-9084},
  doi = {10.1039/D0CP02322A},
  langid = {english}
}

@article{anantharamanMagneticPropertiesUltrafine1998,
  title = {On the Magnetic Properties of Ultra-Fine Zinc Ferrites},
  author = {Anantharaman, M.R. and Jagatheesan, S. and Malini, K.A. and Sindhu, S. and Narayanasamy, A. and Chinnasamy, C.N. and Jacobs, J.P. and Reijne, S. and Seshan, K. and Smits, R.H.H. and Brongersma, H.H.},
  year = 1998,
  month = oct,
  journal = {Journal of Magnetism and Magnetic Materials},
  volume = {189},
  number = {1},
  pages = {83--88},
  issn = {03048853},
  doi = {10.1016/S0304-8853(98)00171-1},
  langid = {english}
}

@article{bohraNanostructuredZnFe2O4Exotic2021,
  title = {Nanostructured {{ZnFe2O4}}: {{An Exotic Energy Material}}},
  shorttitle = {Nanostructured {{ZnFe2O4}}},
  author = {Bohra, Murtaza and Alman, Vidya and Arras, R{\'e}mi},
  year = 2021,
  month = may,
  journal = {Nanomaterials},
  volume = {11},
  number = {5},
  pages = {1286},
  issn = {2079-4991},
  doi = {10.3390/nano11051286},
  langid = {english},
}

@article{rodrigueztorresEvidenceDefectinducedFerromagnetism2011,
  title = {Evidence of Defect-Induced Ferromagnetism in {{ZnFe}} 2 {{O}} 4 Thin Films},
  author = {Rodr{\'i}guez Torres, C. E. and Golmar, F. and Ziese, M. and Esquinazi, P. and Heluani, S. P.},
  year = 2011,
  month = aug,
  journal = {Physical Review B},
  volume = {84},
  number = {6},
  pages = {064404},
  issn = {1098-0121, 1550-235X},
  doi = {10.1103/PhysRevB.84.064404},
  langid = {english},
}

@article{gebauerOxygenVacanciesZirconia2023,
  title = {Oxygen {{Vacancies}} in {{Zirconia}} and {{Their Migration}}: {{The Role}} of {{Hubbard-U Parameters}} in {{Density Functional Theory}}},
  shorttitle = {Oxygen {{Vacancies}} in {{Zirconia}} and {{Their Migration}}},
  author = {Gebauer, Ralph},
  year = 2023,
  month = mar,
  journal = {Crystals},
  volume = {13},
  number = {4},
  pages = {574},
  issn = {2073-4352},
  doi = {10.3390/cryst13040574},
  langid = {english},
}

@article{chaiMinimumTrackingLinear2024,
  title = {Minimum {{Tracking Linear Response Hubbard}} and {{Hund Corrected Density Functional Theory}} in {{CP2K}}},
  author = {Chai, Ziwei and Si, Rutong and Chen, Mingyang and Teobaldi, Gilberto and O'Regan, David D. and Liu, Li-Min},
  year = 2024,
  month = oct,
  journal = {Journal of Chemical Theory and Computation},
  volume = {20},
  number = {20},
  pages = {8984--9002},
  issn = {1549-9618, 1549-9626},
  doi = {10.1021/acs.jctc.4c00964},
  langid = {english},
}

@article{zieseMagnetoresistanceGrainBoundaries2002,
  title = {Magnetoresistance at Grain Boundaries Artificially Introduced into Magnetite Films},
  author = {Ziese, M and H{\"o}hne, R and Hong, N.H and Dienelt, J and Zimmer, K and Esquinazi, P},
  year = 2002,
  month = apr,
  journal = {Journal of Magnetism and Magnetic Materials},
  volume = {242--245},
  pages = {450--452},
  issn = {03048853},
  doi = {10.1016/S0304-8853(01)01338-5},
  urldate = {2026-01-13},
  copyright = {https://www.elsevier.com/tdm/userlicense/1.0/},
  langid = {english}
}

@book{cornellIronOxidesStructure2003,
  title = {The {{Iron Oxides}}: {{Structure}}, {{Properties}}, {{Reactions}}, {{Occurences}} and {{Uses}}},
  shorttitle = {The {{Iron Oxides}}},
  author = {Cornell, R. M. and Schwertmann, U.},
  year = 2003,
  month = jul,
  edition = {1},
  publisher = {Wiley},
  doi = {10.1002/3527602097},
  urldate = {2026-01-13},
  copyright = {http://doi.wiley.com/10.1002/tdm\_license\_1.1},
  isbn = {978-3-527-30274-1 978-3-527-60209-4},
  langid = {english}
}

@article{hakamiStructuralDielectricMagnetic2024,
  title = {Structural, Dielectric and Magnetic Properties of {{MnFe2O4}}/ {{MWCNTs}} Based Nanocomposites for Technological Applications},
  author = {Hakami, Othman},
  year = 2024,
  month = jun,
  journal = {Surfaces and Interfaces},
  volume = {49},
  pages = {104387},
  issn = {24680230},
  doi = {10.1016/j.surfin.2024.104387},
  urldate = {2026-01-13},
  langid = {english}
}

@article{macenultyOptimizationStrategiesDeveloped2023,
  title = {Optimization Strategies Developed on {{NiO}} for {{Heisenberg}} Exchange Coupling Calculations Using Projector Augmented Wave Based First-Principles {{DFT}}+{{U}}+{{J}}},
  author = {MacEnulty, L{\'o}rien and O'Regan, David D.},
  year = 2023,
  month = dec,
  journal = {Physical Review B},
  volume = {108},
  number = {24},
  pages = {245137},
  issn = {2469-9950, 2469-9969},
  doi = {10.1103/PhysRevB.108.245137},
  urldate = {2026-01-20},
  langid = {english},
}

@article{yuFe3O4SurfaceElectronic2012,
  title = {{{Fe3O4}} Surface Electronic Structures and Stability from {{GGA}}+{{U}}},
  author = {Yu, Xiaohu and Huo, Chun-Fang and Li, Yong-Wang and Wang, Jianguo and Jiao, Haijun},
  year = 2012,
  month = may,
  journal = {Surface Science},
  volume = {606},
  number = {9-10},
  pages = {872--879},
  issn = {00396028},
  doi = {10.1016/j.susc.2012.02.003},
  urldate = {2026-03-19},
  copyright = {https://www.elsevier.com/tdm/userlicense/1.0/},
  langid = {english}
}

@article{huangCationMagneticOrders2013,
  title = {Cation and Magnetic Orders in {{MnFe2O4}} from Density Functional Calculations},
  author = {Huang, Jhih-Rong and Cheng, Ching},
  year = 2013,
  month = jan,
  journal = {Journal of Applied Physics},
  volume = {113},
  number = {3},
  pages = {033912},
  issn = {0021-8979, 1089-7550},
  doi = {10.1063/1.4776771},
  urldate = {2026-03-19},
  langid = {english}
}

@article{liTailoringStructuralElectronic2025,
  title = {Tailoring {{Structural}}, {{Electronic}}, and {{Magnetic Properties}} of {{Fe}}{\textsubscript{3}} {{O}}{\textsubscript{4}} and {{MnZn-Ferrites}} through {{Metal}}/{{Non-metal Doping}}: {{A DFT}}+{{U Study}}},
  year = 2025,
  month = jul,
  journal = {The Journal of Physical Chemistry C},
  volume = {129},
  number = {26},
  pages = {12124--12135},
  issn = {1932-7447, 1932-7455},
  doi = {10.1021/acs.jpcc.5c02045},
  urldate = {2026-03-19},
  copyright = {https://doi.org/10.15223/policy-029},
  langid = {english}
}

@techreport{osti_4763367,
  author       = {Brockhouse, Bertram N. and Watanabe, H.},
  title        = {Spin waves in magnetite from neutron scattering},
  institution  = {Atomic Energy of Canada Ltd., Chalk River, Ontario (Canada)},
  url          = {https://www.osti.gov/biblio/4763367},
  place        = {Canada},
  year         = {1962},
  month        = {09}}

@article{wegenerInelasticNeutronScattering1974,
  title = {Inelastic Neutron Scattering Study of Acoustical Magnons in {{MnFe2O4}}},
  author = {Wegener, W. and Scheerlinck, D. and Legrand, E. and Hautecler, S. and Brabers, V.},
  year = 1974,
  month = jul,
  journal = {Solid State Communications},
  volume = {15},
  number = {2},
  pages = {345--349},
  issn = {00381098},
  doi = {10.1016/0038-1098(74)90773-X},
  urldate = {2026-03-19},
  copyright = {https://www.elsevier.com/tdm/userlicense/1.0/},
  langid = {english}
}
%\bibliography{DD_references}

\end{document}